\documentclass[11pt]{article}
\usepackage{epsf}
\usepackage{axodraw}
\usepackage{epsfig}                            
\usepackage{graphicx}
\usepackage{rotate}
\usepackage{latexsym}
\renewcommand{\cal}{\mathcal}
\begin{document}
%
%
%
%
\newcommand{\nl}{\nonumber\\}
\newcommand{\nn}{\nonumber}
\newcommand{\ds}{\displaystyle}
\newcommand{\mpar}[1]{{\marginpar{\hbadness10000%
                      \sloppy\hfuzz10pt\boldmath\bf#1}}%
                      \typeout{marginpar: #1}\ignorespaces}
\def\mnew{\mpar{\hfil NEW \hfil}\ignorespaces}
\newcommand{\lpar}{\left(}                            
\newcommand{\rpar}{\right)} 
\newcommand{\lrbr}{\left[}
\newcommand{\rrbr}{\right]}
\newcommand{\lcbr}{\left\{}
\newcommand{\rcbr}{\right\}} 
\newcommand{\rbrak}[1]{\lrbr#1\rrbr}
\newcommand{\bq}{\begin{equation}}                    
\newcommand{\eq}{\end{equation}}
\newcommand{\bqa}{\begin{eqnarray}}
\newcommand{\eqa}{\end{eqnarray}}
\newcommand{\ba}[1]{\begin{array}{#1}}
\newcommand{\ea}{\end{array}}
\newcommand{\ben}{\begin{enumerate}}
\newcommand{\een}{\end{enumerate}}
\newcommand{\bei}{\begin{itemize}}
\newcommand{\eei}{\end{itemize}}
\newcommand{\bec}{\begin{center}}
\newcommand{\eec}{\end{center}}
\newcommand{\eqn}[1]{Eq.(\ref{#1})}
\newcommand{\eqns}[2]{Eqs.(\ref{#1}--\ref{#2})}
\newcommand{\eqnss}[1]{Eqs.(\ref{#1})}
\newcommand{\eqnsc}[2]{Eqs.(\ref{#1},~\ref{#2})}
\newcommand{\tbn}[1]{Tab.~\ref{#1}}
\newcommand{\tbns}[2]{Tabs.~\ref{#1}--\ref{#2}}
\newcommand{\tbnsc}[2]{Tabs.~\ref{#1},~\ref{#2}}
\newcommand{\fig}[1]{Fig.~\ref{#1}}
\newcommand{\figs}[2]{Figs.~\ref{#1}--\ref{#2}}
\newcommand{\sect}[1]{Sect.~\ref{#1}}
\newcommand{\subsect}[1]{Sub-Sect.~\ref{#1}}
%
%
\newcommand{\TeV}{\;\mathrm{TeV}}                     
\newcommand{\GeV}{\;\mathrm{GeV}}
\newcommand{\MeV}{\;\mathrm{MeV}}
\newcommand{\nb}{\;\mathrm{nb}}
\newcommand{\pb}{\;\mathrm{pb}}
\newcommand{\fb}{\;\mathrm{fb}}
\def\Re{\mathop{\operator@font Re}\nolimits}
\def\Im{\mathop{\operator@font Im}\nolimits}
\newcommand{\ord}[1]{{\cal O}\lpar#1\rpar}
\newcommand{\group}{SU(2)\otimes U(1)}
\newcommand{\ib}{i}
\newcommand{\asums}[1]{\sum_{#1}}
\newcommand{\asumt}[2]{\sum_{#1}^{#2}}
\newcommand{\asum}[3]{\sum_{#1=#2}^{#3}}
%
%
\newcommand{\tmi}{\times 10^{-1}}
\newcommand{\tmii}{\times 10^{-2}}
\newcommand{\tmiii}{\times 10^{-3}}
\newcommand{\tmiv}{\times 10^{-4}}
\newcommand{\tmfv}{\times 10^{-5}}
\newcommand{\tmfvi}{\times 10^{-6}}
\newcommand{\tmfvii}{\times 10^{-7}}
\newcommand{\tmfviii}{\times 10^{-8}}
\newcommand{\tmfix}{\times 10^{-9}}
\newcommand{\tmfx}{\times 10^{-10}}
%
%
\newcommand{\fer}{{\rm{fer}}}
\newcommand{\bos}{{\rm{bos}}}
\newcommand{\lep}{{l}}
\newcommand{\had}{{h}}
\newcommand{\gen}{\rm{g}}
\newcommand{\dbl}{\rm{d}}
\newcommand{\philone}{\phi}
\newcommand{\philoneb}{\phi_{0}}
\newcommand{\phiind}[1]{\phi_{#1}}
\newcommand{\gBi}[2]{B_{#1}^{#2}}
\newcommand{\gBn}[1]{B_{#1}}
%
%
\newcommand{\ph}{\gamma}
\newcommand{\ab}{A}
\newcommand{\abr}{A^r}
\newcommand{\abb}{A^{0}}
\newcommand{\abi}[1]{A_{#1}}
\newcommand{\abri}[1]{A^r_{#1}}
\newcommand{\abbi}[1]{A^{0}_{#1}}
\newcommand{\wb}{W}            
\newcommand{\wbi}[1]{W_{#1}}           
\newcommand{\wbp}{W^{+}}
\newcommand{\wbm}{W^{-}}
\newcommand{\wbpm}{W^{\pm}}
\newcommand{\wbpi}[1]{W^{+}_{#1}}
\newcommand{\wbmi}[1]{W^{-}_{#1}}
\newcommand{\wbpmi}[1]{W^{\pm}_{#1}}
\newcommand{\wbli}[1]{W^{[+}_{#1}}
\newcommand{\wbri}[1]{W^{-]}_{#1}}
\newcommand{\zb}{Z}
\newcommand{\zbi}[1]{Z_{#1}}
\newcommand{\vb}{V}
\newcommand{\vbi}[1]{V_{#1}}      
\newcommand{\vbiv}[1]{V^{*}_{#1}}      
\newcommand{\Pb}{P}
\newcommand{\Sb}{S}
\newcommand{\Bb}{B}
%
%
\newcommand{\hk}{K}
\newcommand{\hKi}[1]{K_{#1}}
\newcommand{\hkg}{\phi}
\newcommand{\hkn}{\phi^{0}}                 
\newcommand{\hkp}{\phi^{+}}
\newcommand{\hkm}{\phi^{-}}
\newcommand{\hkpm}{\phi^{\pm}}
\newcommand{\hkmp}{\phi^{\mp}}
\newcommand{\hki}[1]{\phi^{#1}}
\newcommand{\hb}{H}
\newcommand{\hbi}[1]{H_{#1}}
\newcommand{\hkl}{\phi^{[+\cgfi\cgfi}}
\newcommand{\hkr}{\phi^{-]}}
%
%
\newcommand{\fpx}{X}
\newcommand{\fpy}{Y}
\newcommand{\fpxp}{X^+}
\newcommand{\fpxm}{X^-}
\newcommand{\fpxpm}{X^{\pm}}
\newcommand{\fpxi}[1]{X^{#1}}
\newcommand{\fpyZ}{Y^{\ssZ}}
\newcommand{\fpyA}{Y^{\ssA}}
\newcommand{\fpyZA}{Y_{\ssZ,\ssA}}
\newcommand{\fpbxi}[1]{{\overline{X}}^{#1}}
\newcommand{\fpbyZ}{{\overline{Y}}^{\ssZ}}
\newcommand{\fpbyA}{{\overline{Y}}^{\ssA}}
\newcommand{\fpbyZA}{{\overline{Y}}^{\ssZ,\ssA}}
%
%
\newcommand{\Flone}{F}
\newcommand{\fpsi}{\psi}
\newcommand{\fpsii}[1]{\psi^{#1}}
\newcommand{\fpsib}{\psi^{0}}
\newcommand{\fpsir}{\psi^r}
\newcommand{\fpsiL}{\psi_{_L}}
\newcommand{\fpsiR}{\psi_{_R}}
\newcommand{\fpsiLi}[1]{\psi_{_L}^{#1}}
\newcommand{\fpsiRi}[1]{\psi_{_R}^{#1}}
\newcommand{\fpsiLbi}[1]{\psi_{_{0L}}^{#1}}
\newcommand{\fpsiRbi}[1]{\psi_{_{0R}}^{#1}}
\newcommand{\fpsiLR}{\psi_{_{L,R}}}
\newcommand{\fbpsi}{{\overline{\psi}}}
\newcommand{\fbpsii}[1]{{\overline{\psi}}^{#1}}
\newcommand{\fbpsir}{{\overline{\psi}}^r}
\newcommand{\fbpsiL}{{\overline{\psi}}_{_L}}
\newcommand{\fbpsiR}{{\overline{\psi}}_{_R}}
\newcommand{\fbpsiLi}[1]{\overline{\psi_{_L}}^{#1}}
\newcommand{\fbpsiRi}[1]{\overline{\psi_{_R}}^{#1}}
\newcommand{\fe}{e}
\newcommand{\ff}{f}
\newcommand{\fep}{e^{+}}
\newcommand{\fem}{e^{-}}
\newcommand{\fepm}{e^{\pm}}
\newcommand{\fp}{f^{+}}
\newcommand{\fm}{f^{-}}
\newcommand{\fhp}{h^{+}}
\newcommand{\fhm}{h^{-}}
\newcommand{\fh}{h}
\newcommand{\flm}{\mu}
\newcommand{\flmp}{\mu^{+}}
\newcommand{\flmm}{\mu^{-}}
\newcommand{\fll}{l}
\newcommand{\fllp}{l^{+}}
\newcommand{\fllm}{l^{-}}
\newcommand{\flt}{\tau}
\newcommand{\fltp}{\tau^{+}}
\newcommand{\fltm}{\tau^{-}}
\newcommand{\fq}{q}
\newcommand{\fqi}[1]{\fq_{#1}}
\newcommand{\bfqi}[1]{\barq_{#1}}
\newcommand{\ffQ}{Q}
\newcommand{\fu}{u}
\newcommand{\fd}{d}
\newcommand{\fc}{c}
\newcommand{\fs}{s}
\newcommand{\fqp}{q'}
\newcommand{\fup}{u'}
\newcommand{\fdp}{d'}
\newcommand{\fcp}{c'}
\newcommand{\fsp}{s'}
\newcommand{\fdpp}{d''}
\newcommand{\ffi}[1]{f_{#1}}
\newcommand{\bffi}[1]{{\overline{f}}_{#1}}
\newcommand{\ffpi}[1]{f'_{#1}}
\newcommand{\bffpi}[1]{{\overline{f}}'_{#1}}
\newcommand{\ft}{t}
\newcommand{\ffb}{b}
\newcommand{\ffp}{f'}
\newcommand{\fft}{{\tilde{f}}}
\newcommand{\fl}{l}
\newcommand{\fli}[1]{\fl_{#1}}
\newcommand{\fnu}{\nu}
\newcommand{\fU}{U}
\newcommand{\fD}{D}
\newcommand{\fUc}{\overline{U}}
\newcommand{\fDc}{\overline{D}}
\newcommand{\fnul}{\nu_l}
\newcommand{\fnue}{\nu_e}
\newcommand{\fnum}{\nu_{\mu}}
\newcommand{\fnut}{\nu_{\tau}}
\newcommand{\fbe}{{\overline{e}}}
\newcommand{\fbu}{{\overline{u}}}
\newcommand{\fbd}{{\overline{d}}}
\newcommand{\fbf}{{\overline{f}}}
\newcommand{\fbfp}{{\overline{f}}'}
\newcommand{\fbl}{{\overline{l}}}
\newcommand{\fbnu}{{\overline{\nu}}}
\newcommand{\fbnul}{{\overline{\nu}}_{\fl}}
\newcommand{\fbnue}{{\overline{\nu}}_{\fe}}
\newcommand{\fbnum}{{\overline{\nu}}_{\flm}}
\newcommand{\fbnut}{{\overline{\nu}}_{\flt}}
\newcommand{\fuL}{u_{_L}}
\newcommand{\fdL}{d_{_L}}
\newcommand{\ffL}{f_{_L}}
\newcommand{\fbuL}{{\overline{u}}_{_L}}
\newcommand{\fbdL}{{\overline{d}}_{_L}}
\newcommand{\fbfL}{{\overline{f}}_{_L}}
\newcommand{\fuR}{u_{_R}}
\newcommand{\fdR}{d_{_R}}
\newcommand{\ffR}{f_{_R}}
\newcommand{\fbuR}{{\overline{u}}_{_R}}
\newcommand{\fbdR}{{\overline{d}}_{_R}}
\newcommand{\fbfR}{{\overline{f}}_{_R}}
%
%
\newcommand{\barf}{\overline f}                
\newcommand{\barl}{\overline l}
\newcommand{\barq}{\overline q}
\newcommand{\barqp}{\overline{q}'}
\newcommand{\barb}{\overline b}
\newcommand{\bart}{\overline t}
\newcommand{\barc}{\overline c}
\newcommand{\baru}{\overline u}
\newcommand{\bard}{\overline d}
\newcommand{\bars}{\overline s}
\newcommand{\barv}{\overline v}
\newcommand{\barnu}{\overline{\nu}}
\newcommand{\barne}{\overline{\nu}_{\fe}}
\newcommand{\barnm}{\overline{\nu}_{\flm}}
\newcommand{\barnt}{\overline{\nu}_{\flt}}
%
%
\newcommand{\glu}{g}
%
%
\newcommand{\prot}{p}
\newcommand{\aprot}{{\bar{p}}}
\newcommand{\Nucln}{N}
%
%
\newcommand{\tM}{{\tilde M}}
\newcommand{\tMs}{{\tilde M}^2}
\newcommand{\tW}{{\tilde \Gamma}}
\newcommand{\tWs}{{\tilde\Gamma}^2}
\newcommand{\fphi}{\phi}
\newcommand{\fJpsi}{J/\psi}
\newcommand{\fgpsi}{\psi}
\newcommand{\Glone}{\Gamma}
\newcommand{\Gloni}[1]{\Gamma_{#1}}
\newcommand{\Glones}{\Gamma^2}
\newcommand{\Glonec}{\Gamma^3}
\newcommand{\glone}{\gamma}
\newcommand{\glones}{\gamma^2}
\newcommand{\gloneq}{\gamma^4}
\newcommand{\gloni}[1]{\gamma_{#1}}
\newcommand{\glonis}[1]{\gamma^2_{#1}}
\newcommand{\Grest}[2]{\Gamma_{#1}^{#2}}
\newcommand{\grest}[2]{\gamma_{#1}^{#2}}
\newcommand{\resampl}{A_{_R}}
\newcommand{\resasyi}[1]{{\cal{A}}_{#1}}
\newcommand{\sSrest}[1]{\sigma_{#1}}
\newcommand{\Srest}[2]{\sigma_{#1}\lpar{#2}\rpar}
\newcommand{\Gdist}[1]{{\cal{G}}\lpar{#1}\rpar}
\newcommand{\sGdist}{{\cal{G}}}
\newcommand{\Aarea}{A_{0}}
\newcommand{\Aareai}[1]{{\cal{A}}\lpar{#1}\rpar}
\newcommand{\sAarea}{{\cal{A}}}
\newcommand{\resolw}{\sigma_{\Energ}}
\newcommand{\chizer}{\chi_{_0}}
\newcommand{\ini}{\rm{in}}
\newcommand{\fin}{\rm{fin}}
\newcommand{\ifi}{\rm{if}}
\newcommand{\ipf}{\rm{i+f}}
\newcommand{\tot}{\rm{tot}}
\newcommand{\Bac}{Q}
\newcommand{\Res}{R}
\newcommand{\Int}{I}
\newcommand{\NRe}{NR}
\newcommand{\ratoe}{\delta}
\newcommand{\ratoes}{\delta^2}
%
%
\newcommand{\Fbox}[2]{f^{\rm{box}}_{#1}\lpar{#2}\rpar}
\newcommand{\Dbox}[2]{\delta^{\rm{box}}_{#1}\lpar{#2}\rpar}
\newcommand{\Bbox}[3]{{\cal{B}}_{#1}^{#2}\lpar{#3}\rpar}
%
%
\newcommand{\phm}{\lambda}
\newcommand{\phms}{\lambda^2}
\newcommand{\mV}{M_{_V}}
\newcommand{\mw}{M_{_W}}
\newcommand{\mX}{M_{_X}}
\newcommand{\mY}{M_{_Y}}
\newcommand{\LM}{M}
\newcommand{\mz}{M_{_Z}}
\newcommand{\bzm}{M_{_0}}
\newcommand{\mh}{M_{_H}}
\newcommand{\bhm}{M_{_{0H}}}
\newcommand{\mf}{m_f}
\newcommand{\mfp}{m_{f'}}
\newcommand{\mfh}{m_{h}}
\newcommand{\mt}{m_t}
\newcommand{\me}{m_e}
\newcommand{\mm}{m_{\mu}}
\newcommand{\mtau}{m_{\tau}}
\newcommand{\muq}{m_u}
\newcommand{\md}{m_d}
\newcommand{\muqp}{m'_u}
\newcommand{\mdqp}{m'_d}
\newcommand{\mc}{m_c}
\newcommand{\ms}{m_s}
\newcommand{\mb}{m_b}
\newcommand{\mup}{M_u}                              
\newcommand{\mdp}{M_d}
\newcommand{\mcp}{M_c}
\newcommand{\msp}{M_s}
\newcommand{\mbp}{M_b}
%
%
\newcommand{\mls}{m^2_l}
\newcommand{\mVs}{M^2_{_V}}
\newcommand{\mws}{M^2_{_W}}
\newcommand{\mwc}{M^3_{_W}}
\newcommand{\LMs}{M^2}
\newcommand{\LMc}{M^3}
\newcommand{\mzs}{M^2_{_Z}}
\newcommand{\mzc}{M^3_{_Z}}
\newcommand{\bzms}{M^2_{_0}}
\newcommand{\bzmc}{M^3_{_0}}
\newcommand{\bhms}{M^2_{_{0H}}}
\newcommand{\mhs}{M^2_{_H}}
\newcommand{\mfs}{m^2_f}
\newcommand{\mfc}{m^3_f}
\newcommand{\mfps}{m^2_{f'}}
\newcommand{\mfhs}{m^2_{h}}
\newcommand{\mfpc}{m^3_{f'}}
\newcommand{\mts}{m^2_t}
\newcommand{\mes}{m^2_e}
\newcommand{\mms}{m^2_{\mu}}
\newcommand{\mmc}{m^3_{\mu}}
\newcommand{\mmfour}{m^4_{\mu}}
\newcommand{\mmf}{m^5_{\mu}}
\newcommand{\mmfive}{m^5_{\mu}}
\newcommand{\mmsix}{m^6_{\mu}}
\newcommand{\mminv}{\frac{1}{m_{\mu}}}
\newcommand{\mtaus}{m^2_{\tau}}
\newcommand{\mus}{m^2_u}
\newcommand{\mds}{m^2_d}
\newcommand{\muqps}{m'^2_u}
\newcommand{\mdqps}{m'^2_d}
\newcommand{\mcs}{m^2_c}
\newcommand{\mss}{m^2_s}
\newcommand{\mbs}{m^2_b}
\newcommand{\mups}{M^2_u}
\newcommand{\mdps}{M^2_d}
\newcommand{\mcps}{M^2_c}
\newcommand{\msps}{M^2_s}
\newcommand{\mbps}{M^2_b}
%
%
\newcommand{\muf}{\mu_{\ff}}
\newcommand{\mufs}{\mu^2_{\ff}}
\newcommand{\mufq}{\mu^4_{\ff}}
\newcommand{\mufx}{\mu^6_{\ff}}
\newcommand{\muz}{\mu_{_{\zb}}}
\newcommand{\muw}{\mu_{_{\wb}}}
\newcommand{\mut}{\mu_{\ft}}
\newcommand{\muzs}{\mu^2_{_{\zb}}}
\newcommand{\muws}{\mu^2_{_{\wb}}}
\newcommand{\muts}{\mu^2_{\ft}}
\newcommand{\muSW}{\mu^2_{_{\wb}}}
\newcommand{\muwq}{\mu^4_{_{\wb}}}
\newcommand{\muwsx}{\mu^6_{_{\wb}}}
\newcommand{\muwms}{\mu^{-2}_{_{\wb}}}
\newcommand{\muhs}{\mu^2_{_{\hb}}}
\newcommand{\muhq}{\mu^4_{_{\hb}}}
\newcommand{\muhsx}{\mu^6_{_{\hb}}}
\newcommand{\mutq}{\mu^4_{_{\hb}}}   
\newcommand{\mutsx}{\mu^6_{_{\hb}}}  
\newcommand{\muL}{\mu}
\newcommand{\muS}{\mu^2}
\newcommand{\muQ}{\mu^4}
\newcommand{\muizs}{\mu^2_{0}}
\newcommand{\muizq}{\mu^4_{0}}
\newcommand{\muis}{\mu^2_{1}}
\newcommand{\muiis}{\mu^2_{2}}
\newcommand{\muiiis}{\mu^2_{3}}
\newcommand{\muii}[1]{\mu_{#1}}
\newcommand{\muisi}[1]{\mu^2_{#1}}
\newcommand{\muiqi}[1]{\mu^4_{#1}}
\newcommand{\muixi}[1]{\mu^6_{#1}}
\newcommand{\zm}{z_m}
\newcommand{\ri}[1]{r_{#1}}
\newcommand{\xw}{x_w}
\newcommand{\xws}{x^2_w}
\newcommand{\xwc}{x^3_w}
\newcommand{\xth}{x_t}
\newcommand{\xths}{x^2_t}
\newcommand{\xthc}{x^3_t}
\newcommand{\xthf}{x^4_t}
\newcommand{\xthv}{x^5_t}
\newcommand{\xthx}{x^6_t}
\newcommand{\xh}{x_h}
\newcommand{\xhs}{x^2_h}
\newcommand{\xhc}{x^3_h}
\newcommand{\Rl}{R_{\fl}}
\newcommand{\Rb}{R_{\ffb}}
\newcommand{\Rc}{R_{\fc}}
%
%
\newcommand{\mwq}{M^4_{_\wb}}
\newcommand{\mwf}{M^4_{_\wb}}
\newcommand{\LMq}{M^4}
\newcommand{\mzq}{M^4_{_Z}}
\newcommand{\bzmq}{M^4_{_0}}
\newcommand{\mhq}{M^4_{_H}}
\newcommand{\mfq}{m^4_f}
\newcommand{\mfpq}{m^4_{f'}}
\newcommand{\mtq}{m^4_t}
\newcommand{\meq}{m^4_e}
\newcommand{\mmq}{m^4_{\mu}}
\newcommand{\mtauq}{m^4_{\tau}}
\newcommand{\muqq}{m^4_u}
\newcommand{\mdq}{m^4_d}
\newcommand{\mcq}{m^4_c}
\newcommand{\msq}{m^4_s}
\newcommand{\mbq}{m^4_b}
\newcommand{\mupq}{M^4_u}
\newcommand{\mdpq}{M^4_d}
\newcommand{\mcpq}{M^4_c}
\newcommand{\mspq}{M^4_s}
\newcommand{\mbpq}{M^4_b}
%
%
\newcommand{\mwx}{M^6_{_W}}
\newcommand{\mzx}{M^6_{_Z}}
\newcommand{\mfx}{m^6_f}
\newcommand{\mfpx}{m^6_{f'}}
\newcommand{\LMx}{M^6}
%
%
\newcommand{\mer}{m_{er}}
\newcommand{\mlep}{m_l}
\newcommand{\mleps}{m^2_l}
\newcommand{\mone}{m_1}
\newcommand{\mtwo}{m_2}
\newcommand{\mtre}{m_3}
\newcommand{\mfor}{m_4}
\newcommand{\mlone}{m}
\newcommand{\mloneb}{\bar{m}}
\newcommand{\mind}[1]{m_{#1}}
\newcommand{\mones}{m^2_1}
\newcommand{\mtwos}{m^2_2}
\newcommand{\mtres}{m^2_3}
\newcommand{\mfors}{m^2_4}
\newcommand{\mlones}{m^2}
\newcommand{\minds}[1]{m^2_{#1}}
\newcommand{\moneq}{m^4_1}
\newcommand{\mtwoq}{m^4_2}
\newcommand{\mtreq}{m^4_3}
\newcommand{\mforq}{m^4_4}
\newcommand{\mloneq}{m^4}
\newcommand{\mindq}[1]{m^4_{#1}}
\newcommand{\mlonev}{m^5}
\newcommand{\mindv}[1]{m^5_{#1}}
\newcommand{\monex}{m^6_1}
\newcommand{\mtwox}{m^6_2}
\newcommand{\mtrex}{m^6_3}
\newcommand{\mforx}{m^6_4}
\newcommand{\mlonex}{m^6}
\newcommand{\mindx}[1]{m^6_{#1}}
\newcommand{\Mone}{M_1}
\newcommand{\Mtwo}{M_2}
\newcommand{\Mtre}{M_3}
\newcommand{\Mfor}{M_4}
\newcommand{\Mlone}{M}
\newcommand{\Mlonep}{M'}
\newcommand{\Miind}{M_i}
\newcommand{\Mind}[1]{M_{#1}}
\newcommand{\Minds}[1]{M^2_{#1}}
\newcommand{\Mindc}[1]{M^3_{#1}}
\newcommand{\Mindf}[1]{M^4_{#1}}
\newcommand{\Mones}{M^2_1}
\newcommand{\Mtwos}{M^2_2}
\newcommand{\Mtres}{M^2_3}
\newcommand{\Mfors}{M^2_4}
\newcommand{\Mlones}{M^2}
\newcommand{\Mloneps}{M'^2}
\newcommand{\Miinds}{M^2_i}
\newcommand{\Mlonec}{M^3}
\newcommand{\Monec}{M^3_1}
\newcommand{\Mtwoc}{M^3_2}
\newcommand{\Moneq}{M^4_1}
\newcommand{\Mtwoq}{M^4_2}
\newcommand{\Mtreq}{M^4_3}
\newcommand{\Mforq}{M^4_4}
\newcommand{\Mloneq}{M^4}
\newcommand{\Miindq}{M^4_i}
\newcommand{\Monex}{M^6_1}
\newcommand{\Mtwox}{M^6_2}
\newcommand{\Mtrex}{M^6_3}
\newcommand{\Mforx}{M^6_4}
\newcommand{\Mlonex}{M^6}
\newcommand{\Miindx}{M^6_i}
\newcommand{\meb}{m_0}
\newcommand{\mebs}{m^2_0}
%
%
\newcommand{\Mq }{M_q  }
\newcommand{\MqS}{M^2_q}
\newcommand{\Ms }{M_s  }
\newcommand{\MsS}{M^2_s}
\newcommand{\Mc }{M_c  }
\newcommand{\McS}{M^2_c}
\newcommand{\Mb }{M_b  }
\newcommand{\MbS}{M^2_b}
\newcommand{\Mt }{M_t  }
\newcommand{\MtS}{M^2_t}
%
%
\newcommand{\mq}{m_q}
\newcommand{\mqs}{m^2_q}
\newcommand{\mqS}{m^2_q}
\newcommand{\mqQ}{m^4_q}
\newcommand{\mqX}{m^6_q}
\newcommand{\mqp}{m'_q }
\newcommand{\mqpS}{m'^2_q}
\newcommand{\mqpQ}{m'^4_q}
%
%
\newcommand{\lL}{l}
\newcommand{\ls}{l^2}
\newcommand{\LL}{L}
\newcommand{\LcalL}{\cal{L}}
\newcommand{\LS}{L^2}
\newcommand{\LC}{L^3}
\newcommand{\LQ}{L^4}
\newcommand{\lw}{l_w}
\newcommand{\Lw}{L_w}
\newcommand{\Lws}{L^2_w}
\newcommand{\Lz}{L_z}
\newcommand{\Lzs}{L^2_z}
\newcommand{\Li}[1]{L_{#1}}
\newcommand{\Lis}[1]{L^2_{#1}}
\newcommand{\Lic}[1]{L^3_{#1}}
%
%
\newcommand{\sman}{s}
\newcommand{\tman}{t}
\newcommand{\uman}{u}
\newcommand{\smani}[1]{s_{#1}}
\newcommand{\bsmani}[1]{{\bar{s}}_{#1}}
\newcommand{\smans}{s^2}
\newcommand{\tmans}{t^2}
\newcommand{\umans}{u^2}
\newcommand{\shat}{{\hat s}}
\newcommand{\that}{{\hat t}}
\newcommand{\uhat}{{\hat u}}
\newcommand{\hq}{{\hat Q}}
%
%
\newcommand{\smanp}{s'}
\newcommand{\smanpi}[1]{s'_{#1}}
\newcommand{\tmanp}{t'}
\newcommand{\umanp}{u'}
\newcommand{\kappi}[1]{\kappa_{#1}}
\newcommand{\zetai}[1]{\zeta_{#1}}
%
%
%
\newcommand{\Phaspi}[1]{\Gamma_{#1}}
\newcommand{\rbetai}[1]{\beta_{#1}}
\newcommand{\ralphai}[1]{\alpha_{#1}}
\newcommand{\rbetais}[1]{\beta^2_{#1}}
\newcommand{\Lambdi}[1]{\Lambda_{#1}}
\newcommand{\Nomini}[1]{N_{#1}}
\newcommand{\smlone}{\frac{-\sman-\ib\ep}{\mlones}}
%
%
\newcommand{\theti}[1]{\theta_{#1}}
\newcommand{\delti}[1]{\delta_{#1}}
\newcommand{\phigi}[1]{\phi_{#1}}
\newcommand{\acoli}[1]{\xi_{#1}}
\newcommand{\scats}{s}
\newcommand{\scatss}{s^2}
\newcommand{\scatsi}[1]{s_{#1}}
\newcommand{\scatsis}[1]{s^2_{#1}}
\newcommand{\scatst}[2]{s_{#1}^{#2}}
\newcommand{\scatc}{c}
\newcommand{\scatcs}{c^2}
\newcommand{\scatci}[1]{c_{#1}}
\newcommand{\scatcis}[1]{c^2_{#1}}
\newcommand{\scatct}[2]{c_{#1}^{#2}}
\newcommand{\angamt}[2]{\gamma_{#1}^{#2}}
%
%
\newcommand{\Regia}{{\cal{R}}}
\newcommand{\Iconi}[2]{{\cal{I}}_{#1}\lpar{#2}\rpar}
\newcommand{\sIcon}[1]{{\cal{I}}_{#1}}
\newcommand{\betaf}{\beta_{\ff}}
\newcommand{\betafs}{\beta^2_{\ff}}
\newcommand{\Kfact}[2]{{\cal{K}}_{#1}\lpar{#2}\rpar}
%
%
\newcommand{\Struf}[4]{{\cal D}^{#1}_{#2}\lpar{#3;#4}\rpar}
\newcommand{\sStruf}[2]{{\cal D}^{#1}_{#2}}
\newcommand{\Fluxf}[2]{H\lpar{#1;#2}\rpar}
\newcommand{\Fluxfi}[4]{H_{#1}^{#2}\lpar{#3;#4}\rpar}
\newcommand{\sFluxf}{H}
\newcommand{\Bflux}[2]{{\cal{B}}_{#1}\lpar{#2}\rpar}
\newcommand{\bflux}[2]{{\cal{B}}_{#1}\lpar{#2}\rpar}
\newcommand{\Fluxd}[2]{D_{#1}\lpar{#2}\rpar}
\newcommand{\fluxd}[2]{C_{#1}\lpar{#2}\rpar}
\newcommand{\Fluxh}[4]{{\cal{H}}_{#1}^{#2}\lpar{#3;#4}\rpar}
\newcommand{\Sluxh}[4]{{\cal{S}}_{#1}^{#2}\lpar{#3;#4}\rpar}
\newcommand{\Fluxhb}[4]{{\overline{{\cal{H}}}}_{#1}^{#2}\lpar{#3;#4}\rpar}
\newcommand{\sFluxhb}{{\overline{{\cal{H}}}}}
\newcommand{\Sluxhb}[4]{{\overline{{\cal{S}}}}_{#1}^{#2}\lpar{#3;#4}\rpar}
\newcommand{\sSluxhb}[2]{{\overline{{\cal{S}}}}_{#1}^{#2}}
\newcommand{\fluxh}[4]{h_{#1}^{#2}\lpar{#3;#4}\rpar}
\newcommand{\fluxhs}[3]{h_{#1}^{#2}\lpar{#3}\rpar}
\newcommand{\sfluxhs}[2]{h_{#1}^{#2}}
\newcommand{\fluxhb}[4]{{\overline{h}}_{#1}^{#2}\lpar{#3;#4}\rpar}
\newcommand{\Strufd}[2]{D\lpar{#1;#2}\rpar}
%
%
\newcommand{\rMQ}[1]{r^2_{#1}}
\newcommand{\rMQs}[1]{r^4_{#1}}
\newcommand{\rf}{w_{\ff}}
\newcommand{\zf}{z_{\ff}}
\newcommand{\rfs}{w^2_{\ff}}
\newcommand{\zfs}{z^2_{\ff}}
\newcommand{\rfc}{w^3_{\ff}}
\newcommand{\zfc}{z^3_{\ff}}
\newcommand{\df}{d_{\ff}}
\newcommand{\rfp}{w_{\ffp}}
\newcommand{\rfps}{w^2_{\ffp}}
\newcommand{\rfpc}{w^3_{\ffp}}
\newcommand{\rt}{w_{\ft}}
\newcommand{\rts}{w^2_{\ft}}
\newcommand{\dt}{d_{\ft}}
\newcommand{\dts}{d^2_{\ft}}
\newcommand{\rh}{r_{h}}
\newcommand{\Lnrt}{\ln{\rt}}
\newcommand{\Rw}{R_{_{\wb}}}
\newcommand{\Rws}{R^2_{_{\wb}}}
\newcommand{\Rz}{R_{_{\zb}}}
\newcommand{\Rzp}{R^{+}_{_{\zb}}}
\newcommand{\Rzm}{R^{-}_{_{\zb}}}
\newcommand{\Rzs}{R^2_{_{\zb}}}
\newcommand{\Rzc}{R^3_{_{\zb}}}
\newcommand{\Rv}{R_{_{\vb}}}
\newcommand{\rhw}{r_{_{\wb}}}
\newcommand{\rhz}{r_{_{\zb}}}
\newcommand{\rhws}{r^2_{_{\wb}}}
\newcommand{\rhzs}{r^2_{_{\zb}}}
%
%
\newcommand{\vqrato}{z}
\newcommand{\vqrats}{w}
\newcommand{\vqratq}{w^2}
\newcommand{\seyrat}{z}
\newcommand{\sexrat}{w}
\newcommand{\sehrat}{h}
\newcommand{\sewrat}{w}
\newcommand{\sezrat}{z}
\newcommand{\zetav}{\zeta}
\newcommand{\zetavi}[1]{\zeta_{#1}}
\newcommand{\bpo}{\beta^2}
\newcommand{\bpos}{\beta^4}
\newcommand{\bpt}{{\tilde\beta}^2}
\newcommand{\lap}{\kappa}
\newcommand{\hw}{h_{_{\wb}}}
\newcommand{\hz}{h_{_{\zb}}}
%
%
\newcommand{\ec}{e}
\newcommand{\ecs}{e^2}
\newcommand{\ect}{e^3}
\newcommand{\ecq}{e^4}
\newcommand{\ecb}{e_{_0}}
\newcommand{\ecbs}{e^2_{_0}}
\newcommand{\ecbq}{e^4_{_0}}
\newcommand{\eci}[1]{e_{#1}}
\newcommand{\ecis}[1]{e^2_{#1}}
\newcommand{\hate}{{\hat e}}
\newcommand{\gss}{g_{_S}}
\newcommand{\gsss}{g^2_{_S}}
\newcommand{\gssb}{g^2_{_{S_0}}}
\newcommand{\als}{\alpha_{_S}}
\newcommand{\as}{a_{_S}}
\newcommand{\ass}{a^2_{_S}}
\newcommand{\gf}{G_{\ssF}}
\newcommand{\gfs}{G^2_{\ssF}}
\newcommand{\gb}{g} 
\newcommand{\gbi}[1]{g_{#1}}
\newcommand{\gbb}{g_{0}}
\newcommand{\gbs}{g^2}
\newcommand{\gbc}{g^3}
\newcommand{\gbf}{g^4}
\newcommand{\gpb}{g'}
\newcommand{\gpbs}{g'^2}
\newcommand{\vc}[1]{v_{#1}}
\newcommand{\ac}[1]{a_{#1}}
\newcommand{\vcc}[1]{v^*_{#1}}
\newcommand{\acc}[1]{a^*_{#1}}
\newcommand{\hatv}[1]{{\hat v}_{#1}}
\newcommand{\vcs}[1]{v^2_{#1}}
\newcommand{\acs}[1]{a^2_{#1}}
\newcommand{\gcv}[1]{g^{#1}_{\ssV}}
\newcommand{\gca}[1]{g^{#1}_{\ssA}}
\newcommand{\gcp}[1]{g^{+}_{#1}}
\newcommand{\gcm}[1]{g^{-}_{#1}}
\newcommand{\gcpm}[1]{g^{\pm}_{#1}}
\newcommand{\vci}[2]{v^{#2}_{#1}}
\newcommand{\aci}[2]{a^{#2}_{#1}}
\newcommand{\vceff}[1]{v^{#1}_{\rm{eff}}}
\newcommand{\hvc}[1]{\hat{v}_{#1}}
\newcommand{\hvcs}[1]{\hat{v}^2_{#1}}
\newcommand{\Vc}[1]{V_{#1}}
\newcommand{\Ac}[1]{A_{#1}}
\newcommand{\Vcs}[1]{V^2_{#1}}
\newcommand{\Acs}[1]{A^2_{#1}}
\newcommand{\vpa}[2]{\sigma_{#1}^{#2}}
\newcommand{\vma}[2]{\delta_{#1}^{#2}}
\newcommand{\vfw}{\sigma^{a}_{\ff}}
\newcommand{\vfpw}{\sigma^{a}_{\ffp}}
\newcommand{\vfwi}[1]{\sigma^{a}_{#1}}
\newcommand{\vfwsi}[1]{\lpar\sigma^{a}_{#1}\rpar^2}
\newcommand{\vvfw}{\sigma^{a}_{\ff}}
\newcommand{\vvew}{\sigma^{a}_{\fe}}
\newcommand{\gv}{g_{_V}}
\newcommand{\ga}{g_{_A}}
\newcommand{\gve}{g^{\fe}_{_{V}}}
\newcommand{\gae}{g^{\fe}_{_{A}}}
\newcommand{\gvf}{g^{\ff}_{_{V}}}
\newcommand{\gaf}{g^{\ff}_{_{A}}}
\newcommand{\gva}{g_{_{V,A}}}
\newcommand{\gvae}{g^{\fe}_{_{V,A}}}
\newcommand{\gvaf}{g^{\ff}_{_{V,A}}}
\newcommand{\sGv}{{\cal{G}}_{_V}}
\newcommand{\cGa}{{\cal{G}}^{*}_{_A}}
\newcommand{\cGv}{{\cal{G}}^{*}_{_V}}
\newcommand{\sGa}{{\cal{G}}_{_A}}
\newcommand{\Gvf}{{\cal{G}}^{\ff}_{_{V}}}
\newcommand{\Gaf}{{\cal{G}}^{\ff}_{_{A}}}
\newcommand{\Gvaf}{{\cal{G}}^{\ff}_{_{V,A}}}
\newcommand{\Gve}{{\cal{G}}^{\fe}_{_{V}}}
\newcommand{\Gae}{{\cal{G}}^{\fe}_{_{A}}}
\newcommand{\Gvae}{{\cal{G}}^{\fe}_{_{V,A}}}
\newcommand{\gvl}{g^{\fl}_{_{V}}}
\newcommand{\gal}{g^{\fl}_{_{A}}}
\newcommand{\gval}{g^{\fl}_{_{V,A}}}
\newcommand{\gvb}{g^{\ffb}_{_{V}}}
\newcommand{\gab}{g^{\ffb}_{_{A}}}
\newcommand{\fvf}{F_{_V}^{\ff}}
\newcommand{\faf}{F_{_A}^{\ff}}
\newcommand{\fvl}{F_{_V}^{\fl}}
\newcommand{\fal}{F_{_A}^{\fl}}
\newcommand{\corat}{\kappa}
\newcommand{\corats}{\kappa^2}
%
%
\newcommand{\dr}{\Delta r}
\newcommand{\drl}{\Delta r_{_L}}
\newcommand{\drh}{\Delta{\hat r}}
\newcommand{\drhw}{\Delta{\hat r}_{_W}}
\newcommand{\rhou}{\rho_{_U}}
\newcommand{\rhoz}{\rho_{_\zb}}
\newcommand{\rZ}{\rho_{_\zb}}
\newcommand{\rhob}{\rho_{_0}}
\newcommand{\rZf}{\rho^{\ff}_{_\zb}}
\newcommand{\rhoe}{\rho_{\fe}}
\newcommand{\rhof}{\rho_{\ff}}
\newcommand{\rhoi}[1]{\rho_{#1}}
\newcommand{\kZf}{\kappa^{\ff}_{_\zb}}
\newcommand{\rWf}{\rho^{\ff}_{_\wb}}
\newcommand{\brWf}{{\bar{\rho}}^{\ff}_{_\wb}}
\newcommand{\rHf}{\rho^{\ff}_{_\hb}}
\newcommand{\brHf}{{\bar{\rho}}^{\ff}_{_\hb}}
\newcommand{\rhoR}{\rho^R_{_{\zb}}}
\newcommand{\hatrh}{{\hat\rho}}
\newcommand{\ku}{\kappa_{_U}}
\newcommand{\rZdf}[1]{\rho^{#1}_{_\zb}}
\newcommand{\kZdf}[1]{\kappa^{#1}_{_\zb}}
\newcommand{\rdfL}[1]{\rho^{#1}_{_L}}
\newcommand{\kdfL}[1]{\kappa^{#1}_{_L}}
\newcommand{\rdfR}[1]{\rho^{#1}_{\rm{rem}}}
\newcommand{\kdfR}[1]{\kappa^{#1}_{\rm{rem}}}
\newcommand{\bark}{\overline\kappa}
\newcommand{\kbar}{\overline k}
%
%
\newcommand{\stw}{s_{\theta}}             
\newcommand{\ctw}{c_{\theta}}
\newcommand{\stws}{s_{\theta}^2}
\newcommand{\stwc}{s_{\theta}^3}
\newcommand{\stwf}{s_{\theta}^4}
\newcommand{\stwx}{s_{\theta}^6}
\newcommand{\ctws}{c_{\theta}^2}
\newcommand{\ctwc}{c_{\theta}^3}
\newcommand{\ctwf}{c_{\theta}^4}
\newcommand{\ctwx}{c_{\theta}^6}
\newcommand{\stwfiv}{s_{\theta}^5}
\newcommand{\ctwfiv}{c_{\theta}^5}
\newcommand{\stwsix}{s_{\theta}^6}
\newcommand{\ctwsix}{c_{\theta}^6}
%
%
\newcommand{\siw}{s_{_W}}           
\newcommand{\cow}{c_{_W}}
\newcommand{\siws}{s^2_{_W}}
\newcommand{\cows}{c^2_{_W}}
\newcommand{\siwc}{s^3_{_W}}
\newcommand{\cowc}{c^3_{_W}}
\newcommand{\siwf}{s^4_{_W}}
\newcommand{\cowf}{c^4_{_W}}
\newcommand{\siwx}{s^6_{_W}}
\newcommand{\cowx}{c^6_{_W}}
\newcommand{\sons}{s_{_W}}
\newcommand{\sonss}{s^2_{_W}}
\newcommand{\cons}{c_{_W}}
\newcommand{\cooss}{c^2_{_W}}
%
%
\newcommand{\szs}{{\overline s}^2}
\newcommand{\szq}{{\overline s}^4}
\newcommand{\czs}{{\overline c}^2}
\newcommand{\sbs}{s_{_0}^2}
\newcommand{\cbs}{c_{_0}^2}
\newcommand{\dss}{\Delta s^2}
\newcommand{\snes}{s_{\nu e}^2}
\newcommand{\cnes}{c_{\nu e}^2}
\newcommand{\shs}{{\hat s}^2}
\newcommand{\chs}{{\hat c}^2}
\newcommand{\chl}{{\hat c}}
\newcommand{\seffs}{s^2_{\rm{eff}}}
\newcommand{\seffsf}[1]{\sin^2\theta^{#1}_{\rm{eff}}}
\newcommand{\sress}{s^2_{\rm res}}                
\newcommand{\sR}{s_{_R}}
\newcommand{\sRs}{s^2_{_R}}
\newcommand{\ctwe}{c_{\theta}^6}
\newcommand{\sany}{s}
\newcommand{\cany}{c}
\newcommand{\sanys}{s^2}
\newcommand{\canys}{c^2}
%
%
\newcommand{\sip}{u}                             
\newcommand{\siap}{{\bar{v}}}                    
\newcommand{\sop}{{\bar{u}}}                     
\newcommand{\soap}{v}                            
\newcommand{\ip}[1]{u\lpar{#1}\rpar}             
\newcommand{\iap}[1]{{\bar{v}}\lpar{#1}\rpar}    
\newcommand{\op}[1]{{\bar{u}}\lpar{#1}\rpar}     
\newcommand{\oap}[1]{v\lpar{#1}\rpar}            
%
%
\newcommand{\ipp}[2]{u\lpar{#1,#2}\rpar}         
\newcommand{\ipap}[2]{{\bar v}\lpar{#1,#2}\rpar} 
\newcommand{\opp}[2]{{\bar u}\lpar{#1,#2}\rpar}  
\newcommand{\opap}[2]{v\lpar{#1,#2}\rpar}        
\newcommand{\upspi}[1]{u\lpar{#1}\rpar}
\newcommand{\vpspi}[1]{v\lpar{#1}\rpar}
\newcommand{\wpspi}[1]{w\lpar{#1}\rpar}
\newcommand{\ubpspi}[1]{{\bar{u}}\lpar{#1}\rpar}
\newcommand{\vbpspi}[1]{{\bar{v}}\lpar{#1}\rpar}
\newcommand{\wbpspi}[1]{{\bar{w}}\lpar{#1}\rpar}
\newcommand{\udpspi}[1]{u^{\dagger}\lpar{#1}\rpar}
\newcommand{\vdpspi}[1]{v^{\dagger}\lpar{#1}\rpar}
\newcommand{\wdpspi}[1]{w^{\dagger}\lpar{#1}\rpar}
\newcommand{\Ubilin}[1]{U\lpar{#1}\rpar}
\newcommand{\Vbilin}[1]{V\lpar{#1}\rpar}
\newcommand{\Xbilin}[1]{X\lpar{#1}\rpar}
\newcommand{\Ybilin}[1]{Y\lpar{#1}\rpar}
\newcommand{\up}[2]{u_{#1}\lpar #2\rpar}
\newcommand{\vp}[2]{v_{#1}\lpar #2\rpar}
\newcommand{\ubp}[2]{{\overline u}_{#1}\lpar #2\rpar}
\newcommand{\vbp}[2]{{\overline v}_{#1}\lpar #2\rpar}
\newcommand{\Pje}[1]{\frac{1}{2}\lpar 1 + #1\,\gfd\rpar}
\newcommand{\Pj}[1]{\Pi_{#1}}
\newcommand{\trace}{\mbox{Tr}}
%
%
\newcommand{\Poper}[2]{P_{#1}\lpar{#2}\rpar}
\newcommand{\Loper}[2]{\Lambda_{#1}\lpar{#2}\rpar}
\newcommand{\proj}[3]{P_{#1}\lpar{#2,#3}\rpar}
\newcommand{\sproj}[1]{P_{#1}}
\newcommand{\Nden}[3]{N_{#1}^{#2}\lpar{#3}\rpar}
\newcommand{\sNden}[1]{N_{#1}}
\newcommand{\nden}[2]{n_{#1}^{#2}}
%
%
\newcommand{\vwf}[2]{e_{#1}\lpar#2\rpar}             
\newcommand{\vwfb}[2]{{\overline e}_{#1}\lpar#2\rpar}
\newcommand{\pwf}[2]{\epsilon_{#1}\lpar#2\rpar}      
\newcommand{\sla}[1]{/\!\!\!#1}
\newcommand{\slac}[1]{/\!\!\!\!#1}
%
%
\newcommand{\iemom}{p_{_-}}                    
\newcommand{\ipmom}{p_{_+}}
\newcommand{\oemom}{q_{_-}}                    
\newcommand{\opmom}{q_{_+}}
%
%
\newcommand{\spro}[2]{{#1}\cdot{#2}}
%
%
\newcommand{\gfour}{\gamma_4}                    
\newcommand{\gfd}{\gamma_5}                    
\newcommand{\gap}{\lpar 1+\gamma_5\rpar}
\newcommand{\gam}{\lpar 1-\gamma_5\rpar}
\newcommand{\gdp}{\gamma_+}
\newcommand{\gdm}{\gamma_-}
\newcommand{\gdpm}{\gamma_{\pm}}
\newcommand{\gad}{\gamma}
\newcommand{\gapm}{\lpar 1\pm\gamma_5\rpar}
\newcommand{\gadi}[1]{\gamma_{#1}}
\newcommand{\gadu}[1]{\gamma_{#1}}
\newcommand{\gapu}[1]{\gamma^{#1}}
\newcommand{\sigd}[2]{\sigma_{#1#2}}
\newcommand{\sumsp}{\overline{\sum_{\mbox{spins}}}}
%
%
\newcommand{\li}[2]{\mathrm{Li}_{#1}\lpar\displaystyle{#2}\rpar} 
\newcommand{\etaf}[2]{\eta\lpar#1,#2\rpar}
\newcommand{\lkall}[3]{\lambda\lpar#1,#2,#3\rpar}       
\newcommand{\slkall}[3]{\lambda^{1/2}\lpar#1,#2,#3\rpar}
\newcommand{\segam}{\Gamma}                             
\newcommand{\egam}[1]{\Gamma\lpar#1\rpar}               
\newcommand{\ebe}[2]{B\lpar#1,#2\rpar}                  
\newcommand{\ddel}[1]{\delta\lpar#1\rpar}               
\newcommand{\drii}[2]{\delta_{#1#2}}                    
\newcommand{\driv}[4]{\delta_{#1#2#3#4}}                
\newcommand{\intmomi}[2]{\int\,d^{#1}#2}
\newcommand{\intmomii}[3]{\int\,d^{#1}#2\,\int\,d^{#1}#3}
\newcommand{\intfx}[1]{\int_{\scriptstyle 0}^{\scriptstyle 1}\,d#1}
\newcommand{\intfxy}[2]{\int_{\scriptstyle 0}^{\scriptstyle 1}\,d#1\,
                        \int_{\scriptstyle 0}^{\scriptstyle #1}\,d#2}
\newcommand{\intfxyz}[3]{\int_{\scriptstyle 0}^{\scriptstyle 1}\,d#1\,
                         \int_{\scriptstyle 0}^{\scriptstyle #1}\,d#2\,
                         \int_{\scriptstyle 0}^{\scriptstyle #2}\,d#3}
\newcommand{\Beta}[2]{{\rm{B}}\lpar #1,#2\rpar}
\newcommand{\sBeta}{\rm{B}}
\newcommand{\sign}[1]{{\rm{sign}}\lpar{#1}\rpar}
%
%
\newcommand{\gn}{\Gamma_{\nu}}
\newcommand{\gel}{\Gamma_{\fe}}
\newcommand{\gmu}{\Gamma_{\mu}}
\newcommand{\gff}{\Gamma_{\ff}}
\newcommand{\gt}{\Gamma_{\tau}}
\newcommand{\gl}{\Gamma_{\fl}}
\newcommand{\gq}{\Gamma_{\fq}}
\newcommand{\gu}{\Gamma_{\fu}}
\newcommand{\gd}{\Gamma_{\fd}}
\newcommand{\gc}{\Gamma_{\fc}}
\newcommand{\gs}{\Gamma_{\fs}}
\newcommand{\gbq}{\Gamma_{\ffb}}
\newcommand{\gz}{\Gamma_{_{\zb}}}
\newcommand{\gw}{\Gamma_{_{\wb}}}
\newcommand{\gh}{\Gamma_{_{h}}}
\newcommand{\ghb}{\Gamma_{_{\hb}}}
\newcommand{\gi}{\Gamma_{\rm{inv}}}
\newcommand{\gzs}{\Gamma^2_{_{\zb}}}
%
%
\newcommand{\tcie}{I^{(3)}_{\fe}}
\newcommand{\tcim}{I^{(3)}_{\flm}}
\newcommand{\tcif}{I^{(3)}_{\ff}}
\newcommand{\tciq}{I^{(3)}_{\fq}}
\newcommand{\tcib}{I^{(3)}_{\ffb}}
\newcommand{\tcih}{I^{(3)}_h}
\newcommand{\tcii}{I^{(3)}_i}
\newcommand{\tcift}{I^{(3)}_{\tilde f}}
\newcommand{\tcifp}{I^{(3)}_{f'}}
\newcommand{\wispt}[1]{I^{(3)}_{#1}}
\newcommand{\ql}{Q_l}
\newcommand{\qe}{Q_e}
\newcommand{\qu}{Q_u}
\newcommand{\qd}{Q_d}
\newcommand{\qb}{Q_b}
\newcommand{\qt}{Q_t}
\newcommand{\qup}{Q'_u}
\newcommand{\qdp}{Q'_d}
\newcommand{\qmu}{Q_{\mu}}
\newcommand{\qes}{Q^2_e}
\newcommand{\qec}{Q^3_e}
\newcommand{\qus}{Q^2_u}
\newcommand{\qds}{Q^2_d}
\newcommand{\qbs}{Q^2_b}
\newcommand{\qts}{Q^2_t}
\newcommand{\qbc}{Q^3_b}
\newcommand{\qf}{Q_f}
\newcommand{\qfs}{Q^2_f}
\newcommand{\qfc}{Q^3_f}
\newcommand{\qff}{Q^4_f}
\newcommand{\qep}{Q_{e'}}
\newcommand{\qfp}{Q_{f'}}
\newcommand{\qfps}{Q^2_{f'}}
\newcommand{\qfpc}{Q^3_{f'}}
\newcommand{\qq}{Q_q}
\newcommand{\qqs}{Q^2_q}
\newcommand{\qi}{Q_i}
\newcommand{\qis}{Q^2_i}
\newcommand{\qj}{Q_j}
\newcommand{\qjs}{Q^2_j}
\newcommand{\QW}{Q_{_\wb}}
\newcommand{\QWs}{Q^2_{_\wb}}
\newcommand{\Qd}{Q_d}
\newcommand{\Qds}{Q^2_d}
\newcommand{\Qu}{Q_u}
\newcommand{\Qus}{Q^2_u}
\newcommand{\vi}{v_i}
\newcommand{\vis}{v^2_i}
\newcommand{\ai}{a_i}
\newcommand{\ais}{a^2_i}
%
%
\newcommand{\piv}{\Pi_{_V}}
\newcommand{\pia}{\Pi_{_A}}
\newcommand{\piva}{\Pi_{_{V,A}}}
\newcommand{\pivi}[1]{\Pi^{({#1})}_{_V}}
\newcommand{\piai}[1]{\Pi^{({#1})}_{_A}}
\newcommand{\pivai}[1]{\Pi^{({#1})}_{_{V,A}}}
\newcommand{\pih}{{\hat\Pi}}
\newcommand{\sgh}{{\hat\Sigma}}
\newcommand{\Pgg}{\Pi_{\ph\ph}}
\newcommand{\Ptg}{\Pi_{_{3Q}}}
\newcommand{\Ptt}{\Pi_{_{33}}}
\newcommand{\Pzg}{\Pi_{_{\zb\ab}}}
\newcommand{\Pzga}[2]{\Pi^{#1}_{_{\zb\ab}}\lpar#2\rpar}
\newcommand{\Pf}{\Pi_{_F}}
\newcommand{\Sgg}{\Sigma_{_{\ab\ab}}}
\newcommand{\Szg}{\Sigma_{_{\zb\ab}}}
\newcommand{\SVV}{\Sigma_{_{\vb\vb}}}
\newcommand{\USvv}{{\hat\Sigma}_{_{\vb\vb}}}
\newcommand{\Sww}{\Sigma_{_{\wb\wb}}}
\newcommand{\Swwg}{\Sigma^{_G}_{_{\wb\wb}}}
\newcommand{\Szz}{\Sigma_{_{\zb\zb}}}
\newcommand{\Shh}{\Sigma_{_{\hb\hb}}}
\newcommand{\Spzz}{\Sigma'_{_{\zb\zb}}}
\newcommand{\Stg}{\Sigma_{_{3Q}}}
\newcommand{\Stt}{\Sigma_{_{33}}}
\newcommand{\bSww}{{\overline\Sigma}_{_{WW}}}
\newcommand{\bStg}{{\overline\Sigma}_{_{3Q}}}
\newcommand{\bStt}{{\overline\Sigma}_{_{33}}}
\newcommand{\Sssn}{\Sigma_{_{\hkn\hkn}}}
\newcommand{\Sssc}{\Sigma_{_{\phi\phi}}}
\newcommand{\Szn}{\Sigma_{_{\zb\hkn}}}
\newcommand{\Swc}{\Sigma_{_{\wb\hkg}}}
\newcommand{\mix}[2]{{\cal{M}}^{#1}\lpar{#2}\rpar}
\newcommand{\bmix}[2]{\Pi^{{#1},F}_{_{\zb\ab}}\lpar{#2}\rpar}
\newcommand{\hPgg}[2]{{\hat{\Pi}^{{#1},F}}_{_{\ph\ph}}\lpar{#2}\rpar}
\newcommand{\hmix}[2]{{\hat{\Pi}^{{#1},F}}_{_{\zb\ab}}\lpar{#2}\rpar}
\newcommand{\Dz}[2]{{\cal{D}}_{_{\zb}}^{#1}\lpar{#2}\rpar}
\newcommand{\bDz}[2]{{\cal{D}}^{{#1},F}_{_{\zb}}\lpar{#2}\rpar}
\newcommand{\hDz}[2]{{\hat{\cal{D}}}^{{#1},F}_{_{\zb}}\lpar{#2}\rpar}
\newcommand{\Szzd}[2]{\Sigma'^{#1}_{_{\zb\zb}}\lpar{#2}\rpar}
\newcommand{\Swwd}[2]{\Sigma'^{#1}_{_{\wb\wb}}\lpar{#2}\rpar}
\newcommand{\Shhd}[2]{\Sigma'^{#1}_{_{\hb\hb}}\lpar{#2}\rpar}
\newcommand{\ZFren}[2]{{\cal{Z}}^{#1}\lpar{#2}\rpar}
\newcommand{\WFren}[2]{{\cal{W}}^{#1}\lpar{#2}\rpar}
\newcommand{\HFren}[2]{{\cal{H}}^{#1}\lpar{#2}\rpar}
\newcommand{\WI}{\cal{W}}
%
%
\newcommand{\cf}{c_f}
\newcommand{\Cf}{C_{_F}}
\newcommand{\Nf}{N_f}
\newcommand{\Nc}{N_c}
\newcommand{\Ncs}{N^2_c}
\newcommand{\nf }{n_f}
\newcommand{\nfs}{n^2_f}
\newcommand{\nfc}{n^3_f}
\newcommand{\MSB}{\overline{MS}}
\newcommand{\LMSB}{\Lambda_{\overline{\mathrm{MS}}}}
\newcommand{\LMSBp}{\Lambda'_{\overline{\mathrm{MS}}}}
\newcommand{\LMSBS}{\Lambda^2_{\overline{\mathrm{MS}}}}
\newcommand{\LMSBv }{\mbox{$\Lambda^{(5)}_{\overline{\mathrm{MS}}}$}}
\newcommand{\LMSBvS}{\mbox{$\left(\Lambda^{(5)}_{\overline{\mathrm{MS}}}\right)^2$}}
\newcommand{\LMSBt }{\mbox{$\Lambda^{(3)}_{\overline{\mathrm{MS}}}$}}
\newcommand{\LMSBtS}{\mbox{$\left(\Lambda^{(3)}_{\overline{\mathrm{MS}}}\right)^2$}}
\newcommand{\LMSBf }{\mbox{$\Lambda^{(4)}_{\overline{\mathrm{MS}}}$}}
\newcommand{\LMSBfS}{\mbox{$\left(\Lambda^{(4)}_{\overline{\mathrm{MS}}}\right)^2$}}
\newcommand{\LMSBn }{\mbox{$\Lambda^{(\nf)}_{\overline{\mathrm{MS}}}$}}
\newcommand{\LMSBnS}{\mbox{$\left(\Lambda^{(\nf)}_{\overline{\mathrm{MS}}}\right)^2$}}
\newcommand{\LMSBnml }{\mbox{$\Lambda^{(\nf-1)}_{\overline{\mathrm{MS}}}$}}
\newcommand{\LMSBnmlS}{\mbox{$\left(\Lambda^{(\nf-1)}_{\overline{\mathrm{MS}}}\right)^2$}}
\newcommand{\Bnf}{\lpar\nf \rpar}
\newcommand{\Bnfm}{\lpar\nf-1 \rpar}
\newcommand{\LuM}{L_{_M}}
\newcommand{\bef}{\beta_{\ff}}
\newcommand{\befs}{\beta^2_{\ff}}
\newcommand{\befc}{\beta^3_{f}}
\newcommand{\alsp}{\alpha'_{_S}}
\newcommand{\api}{\displaystyle \frac{\als(s)}{\pi}}
\newcommand{\alss}{\alpha^2_{_S}}
\newcommand{\ztwo}{\zeta(2)}
\newcommand{\ztri}{\zeta(3)}
\newcommand{\zfor}{\zeta(4)}
\newcommand{\zfiv}{\zeta(5)}
\newcommand{\bi}[1]{b_{#1}}
\newcommand{\ci}[1]{c_{#1}}
\newcommand{\Ci}[1]{C_{#1}}
\newcommand{\bip}[1]{b'_{#1}}
\newcommand{\cip}[1]{c'_{#1}}
%
%
\newcommand{\osps}{16\,\pi^2}
\newcommand{\srt}{\sqrt{2}}
\newcommand{\ospsi}{\displaystyle{\frac{i}{16\,\pi^2}}}
%
%
\newcommand{\tfpromu}{\mbox{$e^+e^-\to \mu^+\mu^-$}}
\newcommand{\tfprotau}{\mbox{$e^+e^-\to \tau^+\tau^-$}}
\newcommand{\tfproe}{\mbox{$e^+e^-\to e^+e^-$}}
\newcommand{\tfpronu}{\mbox{$e^+e^-\to \barnu\nu$}}
\newcommand{\tfproqq}{\mbox{$e^+e^-\to \barq q$}}
\newcommand{\tfprohad}{\mbox{$e^+e^-\to\,$} hadrons}
%
%
\newcommand{\bpromu}{\mbox{$e^+e^-\to \mu^+\mu^-\ph$}}
\newcommand{\bprotau}{\mbox{$e^+e^-\to \tau^+\tau^-\ph$}}
\newcommand{\bproe}{\mbox{$e^+e^-\to e^+e^-\ph$}}
\newcommand{\bpronu}{\mbox{$e^+e^-\to \barnu\nu\ph$}}
\newcommand{\bproqq}{\mbox{$e^+e^-\to \barq q \ph$}}
%
%
\newcommand{\tbprow} {\mbox{$e^+e^-\to \wbp \wbm $}}
\newcommand{\tbproz} {\mbox{$e^+e^-\to \zb  \zb  $}}
\newcommand{\tbproh} {\mbox{$e^+e^-\to \zb  \hb  $}}
\newcommand{\tbprozg}{\mbox{$e^+e^-\to \zb  \ph  $}}
\newcommand{\tbprog} {\mbox{$e^+e^-\to \ph  \ph  $}}
%
%
\newcommand{\Fermionline}[1]{
\vcenter{\hbox{
  \begin{picture}(60,20)(0,{#1})
  \SetScale{2.}
    \ArrowLine(0,5)(30,5)
  \end{picture}}}
}
\newcommand{\AntiFermionline}[1]{
\vcenter{\hbox{
  \begin{picture}(60,20)(0,{#1})
  \SetScale{2.}
    \ArrowLine(30,5)(0,5)
  \end{picture}}}
}
\newcommand{\Photonline}[1]{
\vcenter{\hbox{
  \begin{picture}(60,20)(0,{#1})
  \SetScale{2.}
    \Photon(0,5)(30,5){2}{6.5}
  \end{picture}}}
}
\newcommand{\Gluonline}[1]{
\vcenter{\hbox{
  \begin{picture}(60,20)(0,{#1})
  \SetScale{2.}
    \Gluon(0,5)(30,5){2}{6.5}
  \end{picture}}}
}
\newcommand{\Wbosline}[1]{
\vcenter{\hbox{
  \begin{picture}(60,20)(0,{#1})
  \SetScale{2.}
    \Photon(0,5)(30,5){2}{4}
    \ArrowLine(13.3,3.1)(16.9,7.2)
  \end{picture}}}
}
\newcommand{\Zbosline}[1]{
\vcenter{\hbox{
  \begin{picture}(60,20)(0,{#1})
  \SetScale{2.}
    \Photon(0,5)(30,5){2}{4}
  \end{picture}}}
}
\newcommand{\Philine}[1]{
\vcenter{\hbox{
  \begin{picture}(60,20)(0,{#1})
  \SetScale{2.}
    \DashLine(0,5)(30,5){2}
  \end{picture}}}
}
\newcommand{\Phicline}[1]{
\vcenter{\hbox{
  \begin{picture}(60,20)(0,{#1})
  \SetScale{2.}
    \DashLine(0,5)(30,5){2}
    \ArrowLine(14,5)(16,5)
  \end{picture}}}
}
\newcommand{\Ghostline}[1]{
\vcenter{\hbox{
  \begin{picture}(60,20)(0,{#1})
  \SetScale{2.}
    \DashLine(0,5)(30,5){.5}
    \ArrowLine(14,5)(16,5)
  \end{picture}}}
}
%
%
\newcommand{\gauge}{g}
\newcommand{\gpar}{\xi}
\newcommand{\gparA}{\xi_{_A}}
\newcommand{\gparZ}{\xi_{_Z}}
\newcommand{\gpari}[1]{\gpar_{#1}}
\newcommand{\gparis}[1]{\gpar^2_{#1}}
\newcommand{\gpariq}[1]{\gpar^4_{#1}}
\newcommand{\gpars}{\xi^2}
\newcommand{\dgpar}{\Delta\gpar}
\newcommand{\dgparA}{\Delta\gparA}
\newcommand{\dgparZ}{\Delta\gparZ}
\newcommand{\gparq}{\xi^4}
\newcommand{\gparAs}{\xi^2_{_A}}
\newcommand{\gparAq}{\xi^4_{_A}}
\newcommand{\gparZs}{\xi^2_{_Z}}
\newcommand{\gparZq}{\xi^4_{_Z}}
\newcommand{\Rxi}{R_{\gpar}}
\newcommand{\hxi}{\chi}
%
%
\newcommand{\LSM}{{\cal{L}}_{_{\rm{SM}}}}
\newcommand{\LSMr}{{\cal{L}}^{\rm{ren}}_{_{\rm{SM}}}}
\newcommand{\LYM}{{\cal{L}}_{_{YM}}}
\newcommand{\Lzer}{{\cal{L}}_{_{0}}}
\newcommand{\Lone}{{\cal{L}}^{{\bos},I}}
\newcommand{\Lpro}{{\cal{L}}_{\rm{prop}}}
\newcommand{\Ls  }{{\cal{L}}_{_{S}}}
\newcommand{\Lsi }{{\cal{L}}^{I}_{_{S}}}
\newcommand{\Lgf }{{\cal{L}}_{gf  }}
\newcommand{\Lgfi}{{\cal{L}}^{I}_{gf}}
\newcommand{\Lf  }{{\cal{L}}^{{\fer},I}_{\ssV}}
\newcommand{\LHf }{{\cal{L}}^{\fer}_{\ssS}}
\newcommand{\LHfi}{{\cal{L}}^{{\fer},I}_{\ssS}}
\newcommand{\Lren}{{\cal{L}}_{\rm{ren}}}
\newcommand{\Lct}{{\cal{L}}_{\rm{ct}}}
\newcommand{\Lcti}[1]{{\cal{L}}^{#1}_{\rm{ct}}}
\newcommand{\LctI}{{\cal{L}}^{(2)}_{\rm{ct}}}
\newcommand{\Llone}{{\cal{L}}}
\newcommand{\LQED}{{\cal{L}}_{_{\rm{QED}}}}
\newcommand{\LQEDr}{{\cal{L}}^{\rm{ren}}_{_{\rm{QED}}}}
\newcommand{\FST}[3]{F_{#1#2}^{#3}}
\newcommand{\cD}[1]{D_{#1}}
\newcommand{\pd}[1]{\partial_{#1}}
\newcommand{\tgen}[1]{\tau^{#1}}
\newcommand{\gbl}{g_1}
\newcommand{\lctt}[3]{\varepsilon_{#1#2#3}}
\newcommand{\lctf}[4]{\varepsilon_{#1#2#3#4}}
\newcommand{\lctfb}[4]{\varepsilon\lpar{#1#2#3#4}\rpar}
\newcommand{\slct}{\varepsilon}
\newcommand{\cgfi}[1]{{\cal{C}}^{#1}}
\newcommand{\cgfZ}{{\cal{C}}_{_Z}}
\newcommand{\cgfA}{{\cal{C}}_{_A}}
\newcommand{\cgfZs}{{\cal{C}}^2_{_Z}}
\newcommand{\cgfAs}{{\cal{C}}^2_{_A}}
\newcommand{\hpms}{\mu^2}
\newcommand{\hpal}{\alpha_{_H}}
\newcommand{\hpals}{\alpha^2_{_H}}
\newcommand{\hpbe}{\beta_{_H}}
\newcommand{\hpbep}{\beta^{'}_{_H}}
\newcommand{\hpla}{\lambda}
\newcommand{\hpalf}{\alpha_{f}}
\newcommand{\hpbef}{\beta_{f}}
\newcommand{\tpar}[1]{\Lambda^{#1}}
\newcommand{\Mop}[2]{{\rm{M}}^{#1#2}}
\newcommand{\Lop}[2]{{\rm{L}}^{#1#2}}
\newcommand{\Lgen}[1]{T^{#1}}
\newcommand{\Rgen}[1]{t^{#1}}
\newcommand{\fpari}[1]{\lambda_{#1}}
\newcommand{\fQ}[1]{Q_{#1}}
\newcommand{\unm}{I}
\newcommand{\cDsla}{/\!\!\!\!D}
%
%
\newcommand{\saff}[1]{A_{#1}}                    
\newcommand{\aff}[2]{A_{#1}\lpar #2\rpar}                   
\newcommand{\sbff}[1]{B_{#1}}                    
\newcommand{\sfbff}[1]{B^{F}_{#1}}
\newcommand{\bff}[4]{B_{#1}\lpar #2;#3,#4\rpar}             
\newcommand{\bfft}[3]{B_{#1}\lpar #2,#3\rpar}             
\newcommand{\fbff}[4]{B^{F}_{#1}\lpar #2;#3,#4\rpar}        
\newcommand{\cdbff}[4]{\Delta B_{#1}\lpar #2;#3,#4\rpar}             
\newcommand{\sdbff}[4]{\delta B_{#1}\lpar #2;#3,#4\rpar}             
\newcommand{\cdbfft}[3]{\Delta B_{#1}\lpar #2,#3\rpar}             
\newcommand{\sdbfft}[3]{\delta B_{#1}\lpar #2,#3\rpar}             
\newcommand{\scff}[1]{C_{#1}}                    
\newcommand{\scffo}[2]{C_{#1}\lpar{#2}\rpar}                
\newcommand{\cff}[7]{C_{#1}\lpar #2,#3,#4;#5,#6,#7\rpar}    
\newcommand{\sccff}[5]{c_{#1}\lpar #2;#3,#4,#5\rpar} 
\newcommand{\sdff}[1]{D_{#1}}                    
\newcommand{\dffp}[7]{D_{#1}\lpar #2,#3,#4,#5,#6,#7;}       
\newcommand{\dffm}[4]{#1,#2,#3,#4\rpar}                     
\newcommand{\bzfa}[2]{B^{F}_{_{#2}}\lpar{#1}\rpar}
\newcommand{\bzfaa}[3]{B^{F}_{_{#2#3}}\lpar{#1}\rpar}
\newcommand{\shcff}[4]{C_{_{#2#3#4}}\lpar{#1}\rpar}
\newcommand{\shdff}[6]{D_{_{#3#4#5#6}}\lpar{#1,#2}\rpar}
\newcommand{\scdff}[3]{d_{#1}\lpar #2,#3\rpar} 
\newcommand{\scaldff}[1]{{\cal{D}}^{#1}}
\newcommand{\caldff}[2]{{\cal{D}}^{#1}\lpar{#2}\rpar}
\newcommand{\caldfft}[3]{{\cal{D}}_{#1}^{#2}\lpar{#3}\rpar}
%
%
\newcommand{\slaff}[1]{a_{#1}}                        
\newcommand{\slbff}[1]{b_{#1}}                        
\newcommand{\slbffh}[1]{{\hat{b}}_{#1}}    
\newcommand{\ssldff}[1]{d_{#1}}                        
\newcommand{\sslcff}[1]{c_{#1}}                        
\newcommand{\slcff}[2]{c_{#1}^{(#2)}}                        
\newcommand{\sldff}[2]{d_{#1}^{(#2)}}                        
\newcommand{\lbff}[3]{b_{#1}\lpar #2;#3\rpar}         
\newcommand{\lbffh}[2]{{\hat{b}}_{#1}\lpar #2\rpar}   
\newcommand{\lcff}[8]{c_{#1}^{(#2)}\lpar  #3,#4,#5;#6,#7,#8\rpar}         
\newcommand{\ldffp}[8]{d_{#1}^{(#2)}\lpar #3,#4,#5,#6,#7,#8;}
\newcommand{\ldffm}[4]{#1,#2,#3,#4\rpar}                   
%
%
\newcommand{\Iff}[4]{I_{#1}\lpar #2;#3,#4 \rpar}
\newcommand{\Jff}[4]{J_{#1}\lpar #2;#3,#4 \rpar}
\newcommand{\Jds}[5]{{\bar{J}}_{#1}\lpar #2,#3;#4,#5 \rpar}
%
\newcommand{\nhmt}{\frac{n}{2}-2}
\newcommand{\nhmts}{{n}/{2}-2}
\newcommand{\omnh}{1-\frac{n}{2}}
\newcommand{\nhmo}{\frac{n}{2}-1}
\newcommand{\fmon}{4-n}
\newcommand{\lpi}{\ln\pi}
\newcommand{\lmass}[1]{\ln #1}
\newcommand{\egnh}{\egam{\frac{n}{2}}}
\newcommand{\egomnh}{\egam{1-\frac{n}{2}}}
\newcommand{\egtmnh}{\egam{2-\frac{n}{2}}}
\newcommand{\Ddr}{{\ds\frac{1}{{\bar{\varepsilon}}}}}
\newcommand{\Ddrs}{{\ds\frac{1}{{\bar{\varepsilon}^2}}}}
\newcommand{\Ddrd}{{\bar{\varepsilon}}}
\newcommand{\ept}{\hat\varepsilon}
\newcommand{\Ddrh}{{\ds\frac{1}{\hat{\varepsilon}}}}
\newcommand{\Ddrp}{{\ds\frac{1}{\varepsilon'}}}
\newcommand{\Ddrps}{\lpar{\ds{\frac{1}{\varepsilon'}}}\rpar^2}
\newcommand{\dre}{\varepsilon}
\newcommand{\drei}[1]{\varepsilon_{#1}}
\newcommand{\epp}{\varepsilon'}
\newcommand{\eps}{\varepsilon^*}
\newcommand{\ep}{\epsilon}
\newcommand{\propbt}[6]{{{#1_{#2}#1_{#3}}\over{\lpar #1^2 + #4 
-\ib\ep\rpar\lpar\lpar #5\rpar^2 + #6 -\ib\ep\rpar}}}
\newcommand{\propbo}[5]{{{#1_{#2}}\over{\lpar #1^2 + #3 - \ib\ep\rpar
\lpar\lpar #4\rpar^2 + #5 -\ib\ep\rpar}}}
\newcommand{\propc}[6]{{1\over{\lpar #1^2 + #2 - \ib\ep\rpar
\lpar\lpar #3\rpar^2 + #4 -\ib\ep\rpar
\lpar\lpar #5\rpar^2 + #6 -\ib\ep\rpar}}}
\newcommand{\propa}[2]{{1\over {#1^2 + #2^2 - \ib\ep}}}
\newcommand{\propb}[4]{{1\over {\lpar #1^2 + #2 - \ib\ep\rpar
\lpar\lpar #3\rpar^2 + #4 -\ib\ep\rpar}}}
\newcommand{\propbs}[4]{{1\over {\lpar\lpar #1\rpar^2 + #2 - \ib\ep\rpar
\lpar\lpar #3\rpar^2 + #4 -\ib\ep\rpar}}}
\newcommand{\propat}[4]{{#3_{#1}#3_{#2}\over {#3^2 + #4^2 - \ib\ep}}}
\newcommand{\propaf}[6]{{#5_{#1}#5_{#2}#5_{#3}#5_{#4}\over 
{#5^2 + #6^2 -\ib\ep}}}
\newcommand{\momeps}[1]{#1^2 - \ib\ep}
\newcommand{\mopeps}[1]{#1^2 + \ib\ep}
\newcommand{\propz}[1]{{1\over{#1^2 + \mzs - \ib\ep}}}
\newcommand{\propw}[1]{{1\over{#1^2 + \mws - \ib\ep}}}
\newcommand{\proph}[1]{{1\over{#1^2 + \mhs - \ib\ep}}}
\newcommand{\propf}[2]{{1\over{#1^2 + #2}}}
\newcommand{\propzrg}[3]{{{\delta_{#1#2}}\over{{#3}^2 + \mzs - \ib\ep}}}
\newcommand{\propwrg}[3]{{{\delta_{#1#2}}\over{{#3}^2 + \mws - \ib\ep}}}
\newcommand{\propzug}[3]{{
      {\delta_{#1#2} + \displaystyle{{{#3}^{#1}{#3}^{#2}}\over{\mzs}}}
                         \over{{#3}^2 + \mzs - \ib\ep}}}
\newcommand{\propwug}[3]{{
      {\delta_{#1#2} + \displaystyle{{{#3}^{#1}{#3}^{#2}}\over{\mws}}}
                        \over{{#3}^2 + \mws - \ib\ep}}}
\newcommand{\thf}[1]{\theta\lpar #1\rpar}
\newcommand{\epf}[1]{\varepsilon\lpar #1\rpar}
\newcommand{\singp}{\stackrel{sing}{\rightarrow}}
\newcommand{\aint}[3]{\int_{#1}^{#2}\,d #3}
\newcommand{\aroot}[1]{\sqrt{#1}}
\newcommand{\gramc}{\Delta_3}
\newcommand{\gramd}{\Delta_4}
\newcommand{\pinch}[2]{P^{(#1)}\lpar #2\rpar}
\newcommand{\pinchc}[2]{C^{(#1)}_{#2}}
\newcommand{\pinchd}[2]{D^{(#1)}_{#2}}
\newcommand{\loarg}[1]{\ln\lpar #1\rpar}
\newcommand{\loargr}[1]{\ln\lrbr #1\rrbr}
\newcommand{\lsoarg}[1]{\ln^2\lpar #1\rpar}
\newcommand{\ltarg}[2]{\ln\lpar #1\rpar\lpar #2\rpar}
\newcommand{\rfun}[2]{R\lpar #1,#2\rpar}
\newcommand{\pinchb}[3]{B_{#1}\lpar #2,#3\rpar}
\newcommand{\lga}{\ph}
\newcommand{\lzga}{\ssZ\ph}
%
%
\newcommand{\afa}[5]{A_{#1}^{#2}\lpar #3;#4,#5\rpar}
\newcommand{\bfa}[5]{B_{#1}^{#2}\lpar #3;#4,#5\rpar} 
\newcommand{\hfa}[5]{H_{#1}^{#2}\lpar #3;#4,#5\rpar}
\newcommand{\rfa}[5]{R_{#1}^{#2}\lpar #3;#4,#5\rpar}
\newcommand{\afao}[3]{A_{#1}^{#2}\lpar #3\rpar}
\newcommand{\bfao}[3]{B_{#1}^{#2}\lpar #3\rpar}
\newcommand{\hfao}[3]{H_{#1}^{#2}\lpar #3\rpar}
\newcommand{\rfao}[3]{R_{#1}^{#2}\lpar #3\rpar}
\newcommand{\afas}[2]{A_{#1}^{#2}}
\newcommand{\bfas}[2]{B_{#1}^{#2}}
\newcommand{\hfas}[2]{H_{#1}^{#2}}
\newcommand{\rfas}[2]{R_{#1}^{#2}}
\newcommand{\tfas}[2]{T_{#1}^{#2}}
\newcommand{\afaR}[6]{A_{#1}^{\gpar}\lpar #2;#3,#4,#5,#6 \rpar}
\newcommand{\bfaR}[6]{B_{#1}^{\gpar}\lpar #2;#3,#4,#5,#6 \rpar}
\newcommand{\hfaR}[6]{H_{#1}^{\gpar}\lpar #2;#3,#4,#5,#6 \rpar}
\newcommand{\shfaR}[1]{H_{#1}^{\gpar}}
\newcommand{\rfaR}[6]{R_{#1}^{\gpar}\lpar #2;#3,#4,#5,#6 \rpar}
\newcommand{\srfaR}[1]{R_{#1}^{\gpar}}
\newcommand{\afaRg}[5]{A_{#1 \lga}^{\gpar}\lpar #2;#3,#4,#5 \rpar}
\newcommand{\bfaRg}[5]{B_{#1 \lga}^{\gpar}\lpar #2;#3,#4,#5 \rpar}
\newcommand{\hfaRg}[5]{H_{#1 \lga}^{\gpar}\lpar #2;#3,#4,#5 \rpar}
\newcommand{\shfaRg}[1]{H_{#1\lga}^{\gpar}}
\newcommand{\rfaRg}[5]{R_{#1 \lga}^{\gpar}\lpar #2;#3,#4,#5 \rpar}
\newcommand{\srfaRg}[1]{R_{#1\lga}^{\gpar}}
\newcommand{\afaRt}[3]{A_{#1}^{\gpar}\lpar #2,#3 \rpar}
\newcommand{\hfaRt}[3]{H_{#1}^{\gpar}\lpar #2,#3 \rpar}
\newcommand{\hfaRf}[4]{H_{#1}^{\gpar}\lpar #2,#3,#4 \rpar}
\newcommand{\afasm}[4]{A_{#1}^{\lpar #2,#3,#4 \rpar}}
\newcommand{\bfasm}[4]{B_{#1}^{\lpar #2,#3,#4 \rpar}}
\newcommand{\color}[1]{c_{#1}}
\newcommand{\htf}[2]{H_2\lpar #1,#2\rpar}
\newcommand{\rof}[2]{R_1\lpar #1,#2\rpar}
\newcommand{\rtf}[2]{R_3\lpar #1,#2\rpar}
\newcommand{\rtrans}[2]{R_{#1}^{#2}}
\newcommand{\momf}[2]{#1^2_{#2}}
\newcommand{\Scalvert}[8][70]{
  \vcenter{\hbox{
  \SetScale{0.8}
  \begin{picture}(#1,50)(15,15)
    \Line(25,25)(50,50)      \Text(34,20)[lc]{#6} \Text(11,20)[lc]{#3}
    \Line(50,50)(25,75)      \Text(34,60)[lc]{#7} \Text(11,60)[lc]{#4}
    \Line(50,50)(90,50)      \Text(11,40)[lc]{#2} \Text(55,33)[lc]{#8}
    \GCirc(50,50){10}{1}          \Text(60,48)[lc]{#5} 
  \end{picture}}}
  }
%
%
\newcommand{\tHs}{\mu}
\newcommand{\tHsz}{\mu_{_0}}
\newcommand{\tHss}{\mu^2}
\newcommand{\Reb}{{\rm{Re}}}
\newcommand{\Imb}{{\rm{Im}}}
%
%
\newcommand{\spd}{\partial}
\newcommand{\fun}[1]{f\lpar{#1}\rpar}
\newcommand{\ffun}[2]{F_{#1}\lpar #2\rpar}
\newcommand{\gfun}[2]{G_{#1}\lpar #2\rpar}
\newcommand{\sffun}[1]{F_{#1}}
\newcommand{\csffun}[1]{{\cal{F}}_{#1}}
\newcommand{\sgfun}[1]{G_{#1}}
\newcommand{\tpfi}{\lpar 2\pi\rpar^4\ib}
\newcommand{\ffv}{F_{_V}}
\newcommand{\fga}{G_{_A}}
\newcommand{\ffm}{F_{_M}}
\newcommand{\ffs}{F_{_S}}
\newcommand{\fgp}{G_{_P}}
\newcommand{\fge}{G_{_E}}
\newcommand{\ffa}{F_{_A}}
\newcommand{\ffps}{F_{_P}}
\newcommand{\ffe}{F_{_E}}
\newcommand{\gacom}[2]{\lpar #1 + #2\gfd\rpar}
\newcommand{\mft}{m_{\tilde f}}
\newcommand{\qft}{Q_{f'}}
\newcommand{\vft}{v_{\tilde f}}
\newcommand{\subb}[2]{b_{#1}\lpar #2 \rpar}
\newcommand{\fwfr}[5]{\Sigma\lpar #1,#2,#3;#4,#5 \rpar}
\newcommand{\slim}[2]{\lim_{#1 \to #2}}
\newcommand{\sprop}[3]{
{#1\over {\lpar q^2\rpar^2\lpar \lpar q+ #2\rpar^2+#3^2\rpar }}}
%
%
\newcommand{\xroot}[1]{x_{#1}}
\newcommand{\yroot}[1]{y_{#1}}
\newcommand{\zroot}[1]{z_{#1}}
\newcommand{\lvar}{l}
\newcommand{\rvar}{r}
\newcommand{\tvar}{t}
\newcommand{\uvar}{u}
\newcommand{\vvar}{v}
\newcommand{\xvar}{x}
\newcommand{\yvar}{y}
\newcommand{\zvar}{z}
\newcommand{\yvarp}{y'}
\newcommand{\rvars}{r^2}
\newcommand{\vvars}{v^2}
\newcommand{\xvars}{x^2}
\newcommand{\yvars}{y^2}
\newcommand{\zvars}{z^2}
\newcommand{\rvarc}{r^3}
\newcommand{\xvarc}{x^3}
\newcommand{\yvarc}{y^3}
\newcommand{\zvarc}{z^3}
\newcommand{\rvarq}{r^4}
\newcommand{\xvarq}{x^4}
\newcommand{\yvarq}{y^4}
\newcommand{\zvarq}{z^4}
\newcommand{\avar}{a}
\newcommand{\avars}{a^2}
\newcommand{\avarc}{a^3}
\newcommand{\avari}[1]{a_{#1}}
\newcommand{\avart}[2]{a_{#1}^{#2}}
\newcommand{\delvari}[1]{\delta_{#1}}
\newcommand{\rvari}[1]{r_{#1}}
\newcommand{\xvari}[1]{x_{#1}}
\newcommand{\yvari}[1]{y_{#1}}
\newcommand{\zvari}[1]{z_{#1}}
\newcommand{\rvart}[2]{r_{#1}^{#2}}
\newcommand{\xvart}[2]{x_{#1}^{#2}}
\newcommand{\yvart}[2]{y_{#1}^{#2}}
\newcommand{\zvart}[2]{z_{#1}^{#2}}
\newcommand{\rvaris}[1]{r^2_{#1}}
\newcommand{\xvaris}[1]{x^2_{#1}}
\newcommand{\yvaris}[1]{y^2_{#1}}
\newcommand{\zvaris}[1]{z^2_{#1}}
\newcommand{\Xvar}{X}
\newcommand{\Xvars}{X^2}
\newcommand{\Xvari}[1]{X_{#1}}
\newcommand{\Xvaris}[1]{X^2_{#1}}
\newcommand{\Yvar}{Y}
\newcommand{\Yvars}{Y^2}
\newcommand{\Yvari}[1]{Y_{#1}}
\newcommand{\Yvaris}[1]{Y^2_{#1}}
\newcommand{\lnx}{\ln\xvar}
\newcommand{\lnz}{\ln\zvar}
\newcommand{\lnsx}{\ln^2\xvar}
\newcommand{\lnsz}{\ln^2\zvar}
\newcommand{\lncz}{\ln^3\zvar}
\newcommand{\lnomz}{\ln\lpar 1-\zvar\rpar}
\newcommand{\lnsomz}{\ln^2\lpar 1-\zvar\rpar}
\newcommand{\ccoefi}[1]{c_{#1}}
\newcommand{\ccoeft}[2]{c^{#1}_{#2}}
%
%
\newcommand{\Smat}{{\cal{S}}}
\newcommand{\Mmat}{{\cal{M}}}
\newcommand{\Xmat}[1]{X_{#1}}
\newcommand{\XmatI}[1]{X^{-1}_{#1}}
\newcommand{\unitmat}{I}
\newcommand{\Kmat}{{C}}
\newcommand{\Kmatc}{{C}^{\dagger}}
\newcommand{\Kmati}[1]{{C}_{#1}}
\newcommand{\Kmatci}[1]{{C}^{\dagger}_{#1}}
\newcommand{\ffac}[2]{f_{#1}^{#2}}
\newcommand{\Ffac}[1]{F_{#1}}
\newcommand{\Rvec}[2]{R^{(#1)}_{#2}}
\newcommand{\momfl}[2]{#1_{#2}}
\newcommand{\momfs}[2]{#1^2_{#2}}
\newcommand{\fpseZ}{A^{^{FP,Z}}}
\newcommand{\fpseA}{A^{^{FP,A}}}
\newcommand{\fptZ}{T^{^{FP,Z}}}
\newcommand{\fptA}{T^{^{FP,A}}}
\newcommand{\dprop}{\overline\Delta}
\newcommand{\dpropi}[1]{d_{#1}}
\newcommand{\dpropic}[1]{d^{c}_{#1}}
\newcommand{\dpropii}[2]{d_{#1}\lpar #2\rpar}
\newcommand{\dpropis}[1]{d^2_{#1}}
\newcommand{\dproppi}[1]{d'_{#1}}
\newcommand{\psf}[4]{P\lpar #1;#2,#3,#4\rpar}
\newcommand{\ssf}[5]{S^{(#1)}\lpar #2;#3,#4,#5\rpar}
\newcommand{\csf}[5]{C_{_S}^{(#1)}\lpar #2;#3,#4,#5\rpar}
%
%
\newcommand{\lvec}{l}
\newcommand{\lvecs}{l^2}
\newcommand{\lveci}[1]{l_{#1}}
\newcommand{\mvec}{m}
\newcommand{\mvecs}{m^2}
\newcommand{\mveci}[1]{m_{#1}}
\newcommand{\nvec}{n}
\newcommand{\nvecs}{n^2}
\newcommand{\nveci}[1]{n_{#1}}
\newcommand{\epi}[1]{\epsilon_{#1}}
\newcommand{\phep}[1]{\ep_{#1}}
\newcommand{\sphep}{\ep}
\newcommand{\vbep}[1]{e_{#1}}
\newcommand{\vbepp}[1]{e^{+}_{#1}}
\newcommand{\vbepm}[1]{e^{-}_{#1}}
\newcommand{\svbep}{e}
%
%
\newcommand{\lpol}{\lambda}
\newcommand{\spol}{\sigma}
\newcommand{\rpol}{\rho  }
\newcommand{\kpol}{\kappa}
\newcommand{\lpols}{\lambda^2}
\newcommand{\spols}{\sigma^2}
\newcommand{\rpols}{\rho^2}
\newcommand{\kpols}{\kappa^2}
\newcommand{\lpoli}[1]{\lambda_{#1}}
\newcommand{\spoli}[1]{\sigma_{#1}}
\newcommand{\rpoli}[1]{\rho_{#1}}
\newcommand{\kpoli}[1]{\kappa_{#1}}
%
%
\newcommand{\uvec}{u}
\newcommand{\uveci}[1]{u_{#1}}
%
%
\newcommand{\imom}{q}
\newcommand{\imomi}[1]{q_{#1}}
\newcommand{\imoms}{q^2}
\newcommand{\pmom}{p}
\newcommand{\pmomp}{p'}
\newcommand{\pmoms}{p^2}
\newcommand{\pmomq}{p^4}
\newcommand{\pmomx}{p^6}
\newcommand{\pmomi}[1]{p_{#1}}
\newcommand{\pmomis}[1]{p^2_{#1}}
\newcommand{\Pmom}{P}
\newcommand{\Pmoms}{P^2}
\newcommand{\Pmomi}[1]{P_{#1}}
\newcommand{\Pmomis}[1]{P^2_{#1}}
\newcommand{\Kmom}{K}
\newcommand{\Kmoms}{K^2}
\newcommand{\Kmomi}[1]{K_{#1}}
\newcommand{\Kmomis}[1]{K^2_{#1}}
\newcommand{\kmom}{k}
\newcommand{\kmoms}{k^2}
\newcommand{\kmomi}[1]{k_{#1}}
\newcommand{\lmom}{l}
\newcommand{\lmoms}{l^2}
\newcommand{\lmomi}[1]{l_{#1}}
\newcommand{\qmom}{q}
\newcommand{\qmoms}{q^2}
\newcommand{\qmomi}[1]{q_{#1}}
\newcommand{\qmomis}[1]{q^2_{#1}}
\newcommand{\smom}{s}
\newcommand{\smoms}{s^2}
\newcommand{\smomi}[1]{s_{#1}}
\newcommand{\tmom}{t}
\newcommand{\tmoms}{t^2}
\newcommand{\tmomi}[1]{t_{#1}}
\newcommand{\Trmom}{Q}
\newcommand{\Prmom}{P}
\newcommand{\gmv}{Q^2}
\newcommand{\Trmoms}{Q^2}
\newcommand{\Prmoms}{P^2}
\newcommand{\Ptmoms}{T^2}
\newcommand{\Pumoms}{U^2}
\newcommand{\Trmomq}{Q^4}
\newcommand{\Prmomq}{P^4}
\newcommand{\Ptmomq}{T^4}
\newcommand{\Pumomq}{U^4}
\newcommand{\Trmomx}{Q^6}
\newcommand{\Trmomi}[1]{Q_{#1}}
\newcommand{\Trmomis}[1]{Q^2_{#1}}
\newcommand{\Prmomi}[1]{P_{#1}}
\newcommand{\pone}{p_1}
\newcommand{\ptwo}{p_2}
\newcommand{\ptre}{p_3}
\newcommand{\pfor}{p_4}
\newcommand{\pones}{p_1^2}
\newcommand{\ptwos}{p_2^2}
\newcommand{\ptres}{p_3^2}
\newcommand{\pfors}{p_4^2}
\newcommand{\poneq}{p_1^4}
\newcommand{\ptwoq}{p_2^4}
\newcommand{\ptreq}{p_3^4}
\newcommand{\pforq}{p_4^4}
\newcommand{\qmomit}[2]{q_{#1#2}}
\newcommand{\modmom}[1]{\mid{\vec{#1}}\mid}
\newcommand{\modmomi}[2]{\mid{\vec{#1}}_{#2}\mid}
\newcommand{\vect}[1]{{\vec{#1}}}
\newcommand{\Energ}{E}
\newcommand{\Energp}{E'}
\newcommand{\Energpp}{E''}
\newcommand{\Energs}{E^2}
\newcommand{\Energc}{E^3}
\newcommand{\Energf}{E^4}
\newcommand{\Energv}{E^5}
\newcommand{\Energx}{E^6}
\newcommand{\Energi}[1]{E_{#1}}
\newcommand{\Energt}[2]{E_{#1}^{#2}}
\newcommand{\Energis}[1]{E^2_{#1}}
\newcommand{\energ}{e}
\newcommand{\energp}{e'}
\newcommand{\energpp}{e''}
\newcommand{\energs}{e^2}
\newcommand{\energi}[1]{e_{#1}}
\newcommand{\energt}[2]{e_{#1}^{#2}}
\newcommand{\energis}[1]{e^2_{#1}}
\newcommand{\wenerg}{w}
\newcommand{\wenergs}{w^2}
\newcommand{\wenergi}[1]{w_{#1}}
\newcommand{\wenergp}{w'}
\newcommand{\wenergpp}{w''}
%
%
\newcommand{\ecut}{e}
\newcommand{\ecuts}{e^2}
\newcommand{\ecuti}[1]{e^{#1}}
\newcommand{\ccut}{c_m}
\newcommand{\ccuti}[1]{c_{#1}}
\newcommand{\ccuts}{c^2_m}
\newcommand{\scuts}{s^2_m}
\newcommand{\ccutis}[1]{c^2_{#1}}
\newcommand{\ccutic}[1]{c^3_{#1}}
\newcommand{\ccutc}{c^3_m}
\newcommand{\rcut}{\varrho}
\newcommand{\rcuts}{\varrho^2}
\newcommand{\rcuti}[1]{\varrho_{#1}}
\newcommand{\rcutu}[1]{\varrho^{#1}}
\newcommand{\Dcut}{\Delta}
%
\newcommand{\dwf}{\delta_{_{WF}}}
\newcommand{\gbar}{\overline g}
\newcommand{\PP}{\mbox{PP}}
\newcommand{\mv}{m_{_V}}
\newcommand{\bGv}{{\overline\Gamma}_{_V}}
\newcommand{\Umuv}{\hat{\mu}_\ssV}
\newcommand{\Svv}{{\Sigma}_\ssV}
\newcommand{\muv}{p_\ssV}
\newcommand{\muvb}{\mu_{\ssV_{0}}}
\newcommand{\URPvv}{{P}_\ssV}
\newcommand{\RPvv}{{P}_\ssV}
\newcommand{\Svvrem}{{\Sigma}_\ssV^{\mathrm{rem}}}
\newcommand{\USvvrem}{\hat{\Sigma}_\ssV^{\mathrm{rem}}}
\newcommand{\Gv}{\Gamma_{_V}}
%
%
\newcommand{\param}{p}
\newcommand{\parami}[1]{p^{#1}}
\newcommand{\paramb}{p_{0}}
\newcommand{\Zcon}{Z}
\newcommand{\Zconi}[1]{Z_{#1}}
\newcommand{\zconi}[1]{z_{#1}}
\newcommand{\Zconim}[1]{{Z^-_{#1}}}
\newcommand{\zconim}[1]{{z^-_{#1}}}
\newcommand{\Zcont}[2]{Z_{#1}^{#2}}
\newcommand{\zcont}[2]{z_{#1}^{#2}}
\newcommand{\zcontm}[2]{z_{#1}^{{#2}-}}
\newcommand{\sZconi}[2]{\sqrt{Z_{#1}}^{\;#2}}
\newcommand{\php}[3]{e^{#1}_{#2}\lpar #3 \rpar}
\newcommand{\gacome}[1]{\lpar #1 - \gfd\rpar}
\newcommand{\sPj}[2]{\Lambda^{#1}_{#2}}
\newcommand{\sPjs}[2]{\Lambda_{#1,#2}}
\newcommand{\amos}{\mbox{$M^2_{_1}$}}
\newcommand{\amts}{\mbox{$M^2_{_2}$}}
\newcommand{\er}{e_{_{R}}}
\newcommand{\epr}{e'_{_{R}}}
\newcommand{\ers}{e^2_{_{R}}}
\newcommand{\erc}{e^3_{_{R}}}
\newcommand{\erq}{e^4_{_{R}}}
\newcommand{\erf}{e^5_{_{R}}}
\newcommand{\sour}{J}
\newcommand{\sourb}{\overline J}
\newcommand{\lrm}{M_{_R}}
%
%
\newcommand{\vlami}[1]{\lambda_{#1}}
\newcommand{\vlamis}[1]{\lambda^2_{#1}}
\newcommand{\Vvert}{V}
\newcommand{\Avert}{A}
\newcommand{\Svert}{S}
\newcommand{\Pvert}{P}
\newcommand{\vvert}{F}
\newcommand{\Cvert}{\cal{V}}
\newcommand{\Bvert}{\cal{B}}
\newcommand{\Vveri}[2]{V_{#1}^{#2}}
\newcommand{\Fveri}[1]{{\cal{F}}^{#1}}
\newcommand{\Cveri}[1]{{\cal{V}}\lpar{#1}\rpar}
\newcommand{\Bveri}[1]{{\cal{B}}\lpar{#1}\rpar}
\newcommand{\Vverti}[3]{V_{#1}^{#2}\lpar{#3}\rpar}
\newcommand{\Averti}[3]{A_{#1}^{#2}\lpar{#3}\rpar}
\newcommand{\Gverti}[3]{G_{#1}^{#2}\lpar{#3}\rpar}
\newcommand{\Zverti}[3]{Z_{#1}^{#2}\lpar{#3}\rpar}
\newcommand{\Hverti}[2]{H^{#1}\lpar{#2}\rpar}
\newcommand{\Wverti}[3]{W_{#1}^{#2}\lpar{#3}\rpar}
\newcommand{\Cverti}[2]{{\cal{V}}_{#1}^{#2}}
\newcommand{\vverti}[3]{F^{#1}_{#2}\lpar{#3}\rpar}
\newcommand{\averti}[3]{{\overline{F}}^{#1}_{#2}\lpar{#3}\rpar}
\newcommand{\fveone}[1]{f_{#1}}
\newcommand{\fvetri}[3]{f^{#1}_{#2}\lpar{#3}\rpar}
\newcommand{\gvetri}[3]{g^{#1}_{#2}\lpar{#3}\rpar}
\newcommand{\cvetri}[3]{{\cal{F}}^{#1}_{#2}\lpar{#3}\rpar}
\newcommand{\hvetri}[3]{{\hat{\cal{F}}}^{#1}_{#2}\lpar{#3}\rpar}
\newcommand{\avetri}[3]{{\overline{\cal{F}}}^{#1}_{#2}\lpar{#3}\rpar}
\newcommand{\fverti}[2]{F^{#1}_{#2}}
\newcommand{\cverti}[2]{{\cal{F}}_{#1}^{#2}}
\newcommand{\fV}{f_{_{\Vvert}}}
\newcommand{\gA}{g_{_{\Avert}}}
\newcommand{\fVi}[1]{f^{#1}_{_{\Vvert}}}
\newcommand{\seai}[1]{a_{#1}}
\newcommand{\seapi}[1]{a'_{#1}}
\newcommand{\seAi}[2]{A_{#1}^{#2}}
\newcommand{\sewi}[1]{w_{#1}}
\newcommand{\seWi}[1]{W_{#1}}
\newcommand{\seWsi}[1]{W^{*}_{#1}}
\newcommand{\seWti}[2]{W_{#1}^{#2}}
\newcommand{\sewti}[2]{w_{#1}^{#2}}
\newcommand{\seSig}[1]{\Sigma_{#1}\lpar\sla{\pmom}\rpar}
\newcommand{\ww}{w}
%
%
\newcommand{\bbff}[1]{{\overline B}_{#1}}
\newcommand{\sW}{p_{_W}}
\newcommand{\sZ}{p_{_Z}}
\newcommand{\ssp}{s_p}
\newcommand{\fW}{f_{_W}}
\newcommand{\fZ}{f_{_Z}}
\newcommand{\tabn}[1]{Tab.(\ref{#1})}
\newcommand{\subMSB}[1]{{#1}_{\mbox{$\overline{\scriptscriptstyle MS}$}}}
\newcommand{\supMSB}[1]{{#1}^{\mbox{$\overline{\scriptscriptstyle MS}$}}}
\newcommand{\redMSB}{{\mbox{$\overline{\scriptscriptstyle MS}$}}}
\newcommand{\gpbb}{g'_{0}}
\newcommand{\Zconip}[1]{Z'_{#1}}
\newcommand{\bpff}[4]{B'_{#1}\lpar #2;#3,#4\rpar}             
\newcommand{\xidf}{\xi^2-1}
\newcommand{\tDdr}{1/{\bar{\varepsilon}}}
\newcommand{\cRz}{{\cal R}_{_Z}}
\newcommand{\cRg}{{\cal R}_{\gamma}}
\newcommand{\Sz}{\Sigma_{_Z}}
\newcommand{\alh}{{\hat\alpha}}
\newcommand{\alhz}{\alpha_{_Z}}
\newcommand{\Phzg}{{\hat\Pi}_{_{\zb\ab}}}
\newcommand{\fvvert}{F^{\rm vert}_{_V}}
\newcommand{\gavert}{G^{\rm vert}_{_A}}
\newcommand{\bmv}{{\overline m}_{_V}}
\newcommand{\Sgn}{\Sigma_{\gamma\hkn}}
\newcommand{\tabns}[2]{Tabs.(\ref{#1}--\ref{#2})}
\newcommand{\rmboxd}{{\rm Box}_d\lpar s,t,u;M_1,M_2,M_3,M_4\rpar}
\newcommand{\rmboxc}{{\rm Box}_c\lpar s,t,u;M_1,M_2,M_3,M_4\rpar}
%
%
\newcommand{\Afaci}[1]{A_{#1}}
\newcommand{\Afacis}[1]{A^2_{#1}}
\newcommand{\upar}[1]{u}
\newcommand{\upari}[1]{u_{#1}}
\newcommand{\vpari}[1]{v_{#1}}
\newcommand{\lpari}[1]{l_{#1}}
\newcommand{\Lpari}[1]{l_{#1}}
\newcommand{\Nff}[2]{N^{(#1)}_{#2}}
\newcommand{\Sff}[2]{S^{(#1)}_{#2}}
\newcommand{\sSff}{S}
\newcommand{\FQED}[2]{F_{#1#2}}
\newcommand{\fbpsif}{{\overline{\psi}_f}}
\newcommand{\fpsif}{\psi_f}
\newcommand{\etafd}[2]{\eta_d\lpar#1,#2\rpar}
\newcommand{\sigdu}[2]{\sigma_{#1#2}}
\newcommand{\scalc}[4]{c_{_0}\lpar #1;#2,#3,#4\rpar}
\newcommand{\scald}[2]{d_{_0}\lpar #1,#2\rpar}
\newcommand{\pir}[1]{\Pi^{\rm ren}\lpar #1\rpar}
\newcommand{\sigh}{\sigma_{\rm had}}
\newcommand{\dah}{\Delta\alpha^{(5)}_{\rm had}}
\newcommand{\dat}{\Delta\alpha_{\rm top}}
\newcommand{\Vqed}[3]{V_1^{\rm sub}\lpar#1;#2,#3\rpar}
\newcommand{\thetah}{{\hat\theta}}
\newcommand{\mtsix}{m^6_t}
\newcommand{\smlon}{\frac{\mlones}{s}}
\newcommand{\lntwo}{\ln 2}
\newcommand{\wmin}{w_{\rm min}}
\newcommand{\kmin}{k_{\rm min}}
\newcommand{\scaldi}[3]{d_{_0}^{#1}\lpar #2,#3\rpar}
\newcommand{\mdls}{\Big|}
\newcommand{\smf}{\frac{\mfs}{s}}
\newcommand{\bint}{\beta_{\rm int}}
\newcommand{\IRv}{V_{_{\rm IR}}}
\newcommand{\IRr}{R_{_{\rm IR}}}
\newcommand{\fssts}{\frac{s^2}{t^2}}
\newcommand{\fssus}{\frac{s^2}{u^2}}
\newcommand{\optM}{1+\frac{t}{M^2}}
\newcommand{\opuM}{1+\frac{u}{M^2}}
\newcommand{\ftM}{\lpar -\frac{t}{M^2}\rpar}
\newcommand{\fuM}{\lpar -\frac{u}{M^2}\rpar}
\newcommand{\omsM}{1-\frac{s}{M^2}}
\newcommand{\xsf}{\sigma_{_{\rm F}}}
\newcommand{\xsb}{\sigma_{_{\rm B}}}
\newcommand{\afb}{A_{_{\rm FB}}}
\newcommand{\rsoft}{\rm soft}
\newcommand{\rms}{\rm s}
\newcommand{\rsmx}{\sqrt{s_{\rm max}}}
\newcommand{\rspm}{\sqrt{s_{\pm}}}
\newcommand{\rsp}{\sqrt{s_{+}}}
\newcommand{\rsm}{\sqrt{s_{-}}}
\newcommand{\sigmx}{\sigma_{\rm max}}
\newcommand{\gG}[2]{G_{#1}^{#2}}
\newcommand{\gacomm}[2]{\lpar #1 - #2\gfd\rpar}
\newcommand{\fcsx}{\frac{1}{\ctwsix}}
\newcommand{\fcq}{\frac{1}{\ctwf}}
\newcommand{\fcs}{\frac{1}{\ctws}}
\newcommand{\affs}[2]{{\cal A}_{#1}\lpar #2\rpar}                   
\newcommand{\stwei}{s_{\theta}^8}
\def\mdan{\vspace{1mm}\mpar{\hfil$\downarrow$new\hfil}\vspace{-1mm}
          \ignorespaces}
\def\muan{\vspace{-1mm}\mpar{\hfil$\uparrow$new\hfil}\vspace{1mm}\ignorespaces}
\def\mlan{\vspace{-1mm}\mpar{\hfil$\rightarrow$new\hfil}\vspace{1mm}\ignorespaces}
\def\mnnew{\mpar{\hfil NEWNEW \hfil}\ignorespaces}
%
%
\newcommand{\boxc}[2]{{\cal{B}}_{#1}^{#2}}
\newcommand{\boxct}[3]{{\cal{B}}_{#1}^{#2}\lpar{#3}\rpar}
\newcommand{\hboxc}[3]{\hat{{\cal{B}}}_{#1}^{#2}\lpar{#3}\rpar}
\newcommand{\vev}{\langle v \rangle}
\newcommand{\vevi}[1]{\langle v_{#1}\rangle}
\newcommand{\vevs}{\langle v^2   \rangle}
\newcommand{\fwfrV}[5]{\Sigma_{_V}\lpar #1,#2,#3;#4,#5 \rpar}
\newcommand{\fwfrS}[7]{\Sigma_{_S}\lpar #1,#2,#3;#4,#5;#6,#7 \rpar}
\newcommand{\fSi}[1]{f^{#1}_{_{\Svert}}}
\newcommand{\fPi}[1]{f^{#1}_{_{\Pvert}}}
\newcommand{\mXs}{m_{_X}}
\newcommand{\mXss}{m^2_{_X}}
\newcommand{\mYs}{M^2_{_Y}}
\newcommand{\xik}{\xi_k}
\newcommand{\xiks}{\xi^2_k}
\newcommand{\mpls}{m^2_+}
\newcommand{\mmis}{m^2_-}
%
\newcommand{\SN}{\Sigma_{_N}}
\newcommand{\SC}{\Sigma_{_C}}
\newcommand{\SPN}{\Sigma'_{_N}}
\newcommand{\PFf}{\Pi^{\fer}_{_F}}
\newcommand{\PFb}{\Pi^{\bos}_{_F}}
\newcommand{\dPZ}{\Delta{\hat\Pi}_{_Z}}
\newcommand{\Sfin}{\Sigma_{_F}}
\newcommand{\Sfir}{\Sigma_{_R}}
\newcommand{\Sfinh}{{\hat\Sigma}_{_F}}
\newcommand{\Sfinf}{\Sigma^{\fer}_{_F}}
\newcommand{\Sfinbh}{\Sigma^{\bos}_{_F}}
\newcommand{\alf}{\alpha^{\fer}}
\newcommand{\alhfz}{\alpha^{\fer}\lpar{\ssZ}\rpar}
\newcommand{\alhfs}{\alpha^{\fer}\lpar{\sman}\rpar}
\newcommand{\gfQ}{g^f_{_{Q}}}
\newcommand{\gfL}{g^f_{_{L}}}
\newcommand{\ccf}{\frac{\gbs}{16\,\pi^2}}
\newcommand{\chq}{{\hat c}^4}
\newcommand{\muuq}{m_{u'}}
\newcommand{\muus}{m^2_{u'}}
\newcommand{\mdd}{m_{d'}}
\newcommand{\clf}[2]{\mathrm{Cli}_{_#1}\lpar\displaystyle{#2}\rpar}
\def\stes{\sin^2\theta}
\def\acal{\cal A}
\def\alr{A_{_{\rm{LR}}}}
\newcommand{\barQ}{\overline Q}
\newcommand{\Sptg}{\Sigma'_{_{3Q}}}
\newcommand{\Sptt}{\Sigma'_{_{33}}}
\newcommand{\Ppgg}{\Pi'_{\ph\ph}}
\newcommand{\Pww}{\Pi_{_{\wb\wb}}}
\newcommand{\capV}[2]{{\cal F}^{#2}_{_{#1}}}
\newcommand{\bt}{\beta_t}
\newcommand{\mhsix}{M^6_{_H}}
\newcommand{\topt}{{\cal T}_{33}}
\newcommand{\topq}{{\cal T}_4}
\newcommand{\Phzgf}{{\hat\Pi}^{\fer}_{_{\zb\ab}}}
\newcommand{\Phzgb}{{\hat\Pi}^{\bos}_{_{\zb\ab}}}
\newcommand{\Sfirh}{{\hat\Sigma}_{_R}}
\newcommand{\Szgh}{{\hat\Sigma}_{_{\zb\ab}}}
\newcommand{\Szghb}{{\hat\Sigma}^{\bos}_{_{\zb\ab}}}
\newcommand{\Szghf}{{\hat\Sigma}^{\fer}_{_{\zb\ab}}}
\newcommand{\Szgb}{\Sigma^{\bos}_{_{\zb\ab}}}
\newcommand{\Szgf}{\Sigma^{\fer}_{_{\zb\ab}}}
\newcommand{\chig}{\chi_{_{\ph}}}
\newcommand{\chiz}{\chi_{_{\zb}}}
\newcommand{\Sfih}{{\hat\Sigma}}
\newcommand{\Szzh}{\hat{\Sigma}_{_{\zb\zb}}}
\newcommand{\dPZf}{\Delta{\hat\Pi}^f_{_{\zb}}}
\newcommand{\khZdf}[1]{{\hat\kappa}^{#1}_{_{\zb}}}
\newcommand{\chf}{{\hat c}^4}
\newcommand{\amp}[2]{{\cal{A}}_{_{#1}}^{\rm{#2}}}
\newcommand{\hatvm}[1]{{\hat v}^-_{#1}}
\newcommand{\hatvp}[1]{{\hat v}^+_{#1}}
\newcommand{\hatvpm}[1]{{\hat v}^{\pm}_{#1}}
\newcommand{\kvz}[1]{\kappa^{\zb #1}_{_V}}
\newcommand{\barp}{\overline p}                
\newcommand{\delw}{\Delta_{_{\wb}}}
\newcommand{\bdelw}{{\bar{\Delta}}_{_{\wb}}}
\newcommand{\bdelf}{{\bar{\Delta}}_{\ff}}
\newcommand{\delz}{\Delta_{_\zb}}
\newcommand{\deli}[1]{\Delta\lpar{#1}\rpar}
\newcommand{\chizb}{\chi_{_\zb}}
\newcommand{\Swwp}{\Sigma'_{_{\wb\wb}}}
\newcommand{\epph}{\varepsilon'/2}
\newcommand{\sbffp}[1]{B'_{#1}}                    
\newcommand{\epss}{\varepsilon^*}
\newcommand{\Ddrhs}{{\ds\frac{1}{\hat{\varepsilon}^2}}}
\newcommand{\lnmsb}{L_{_\wb}}
\newcommand{\lnsmsb}{L^2_{_\wb}}
\newcommand{\tpni}{\lpar 2\pi\rpar^n\ib}
\newcommand{\tpn}{2^n\,\pi^{n-2}}
\newcommand{\cmf}{M_f}
\newcommand{\cmfs}{M^2_f}
\newcommand{\toDdr}{{\ds\frac{2}{{\bar{\varepsilon}}}}}
\newcommand{\troDdr}{{\ds\frac{3}{{\bar{\varepsilon}}}}}
\newcommand{\totDdr}{{\ds\frac{3}{{2\,\bar{\varepsilon}}}}}
\newcommand{\foDdr}{{\ds\frac{4}{{\bar{\varepsilon}}}}}
\newcommand{\smh}{m_{_H}}
\newcommand{\smhs}{m^2_{_H}}
\newcommand{\Ph}{\Pi_{_\hb}}
\newcommand{\Sphh}{\Sigma'_{_{\hb\hb}}}
\newcommand{\bh}{\beta}
\newcommand{\alsn}{\alpha^{(n_f)}_{_S}}
\newcommand{\smq}{m_q}
\newcommand{\smqp}{m_{q'}}
\newcommand{\shb}{h}
\newcommand{\hab}{A}
\newcommand{\hbpm}{H^{\pm}}
\newcommand{\hbp}{H^{+}}
\newcommand{\hbm}{H^{-}}
\newcommand{\msh}{M_h}
\newcommand{\mha}{M_{_A}}
\newcommand{\mhc}{M_{_{H^{\pm}}}}
\newcommand{\mshs}{M^2_h}
\newcommand{\mhas}{M^2_{_A}}
\newcommand{\barfp}{\overline{f'}}                
\newcommand{\chiii}{{\hat c}^3}
\newcommand{\chiv}{{\hat c}^4}
\newcommand{\chv}{{\hat c}^5}
\newcommand{\chvi}{{\hat c}^6}
\newcommand{\alsvi}{\alpha^{6}_{_S}}
\newcommand{\tww}{t_{_W}}
\newcommand{\ti}{t_{_1}}
\newcommand{\tii}{t_{_2}}
\newcommand{\tiii}{t_{_3}}
\newcommand{\tiv}{t_{_4}}
\newcommand{\psla}{\hbox{\rlap/p}}
\newcommand{\qsla}{\hbox{\rlap/q}}
\newcommand{\nsla}{\hbox{\rlap/n}}
\newcommand{\lsla}{\hbox{\rlap/l}}
\newcommand{\msla}{\hbox{\rlap/m}}
\newcommand{\cnsla}{\hbox{\rlap/N}}
\newcommand{\clsla}{\hbox{\rlap/L}}
\newcommand{\cmsla}{\hbox{\rlap/M}}
\newcommand{\blmt}{\lrbr - 3\rrbr}
\newcommand{\blfo}{\lrbr 4 1\rrbr}
\newcommand{\bltp}{\lrbr 2 +\rrbr}
\newcommand{\clitwo}[1]{{\rm{Li}}_{2}\lpar{#1}\rpar}
\newcommand{\clitri}[1]{{\rm{Li}}_{3}\lpar{#1}\rpar}
\newcommand{\xt}{x_{\ft}}
\newcommand{\zt}{z_{\ft}}
\newcommand{\Ht}{h_{\ft}}
\newcommand{\xts}{x^2_{\ft}}
\newcommand{\zts}{z^2_{\ft}}
\newcommand{\Hts}{h^2_{\ft}}
\newcommand{\ztc}{z^3_{\ft}}
\newcommand{\Htc}{h^3_{\ft}}
\newcommand{\ztq}{z^4_{\ft}}
\newcommand{\Htq}{h^4_{\ft}}
\newcommand{\ztv}{z^5_{\ft}}
\newcommand{\Htv}{h^5_{\ft}}
\newcommand{\ztx}{z^6_{\ft}}
\newcommand{\Htx}{h^6_{\ft}}
\newcommand{\ztz}{z^7_{\ft}}
\newcommand{\Htz}{h^7_{\ft}}
\newcommand{\sht}{\sqrt{\Ht}}
\newcommand{\atan}[1]{{\rm{arctan}}\lpar{#1}\rpar}
\newcommand{\dbff}[3]{{\hat{B}}_{_{{#2}{#3}}}\lpar{#1}\rpar}
\newcommand{\ztbs}{{\bar{z}}^{2}_{\ft}}
\newcommand{\ztb}{{\bar{z}}_{\ft}}
\newcommand{\Htbs}{{\bar{h}}^{2}_{\ft}}
\newcommand{\Htb}{{\bar{h}}_{\ft}}
\newcommand{\Hztb}{{\bar{hz}}_{\ft}}
\newcommand{\Ln}[1]{{\rm{Ln}}\lpar{#1}\rpar}
\newcommand{\Lns}[1]{{\rm{Ln}}^2\lpar{#1}\rpar}
\newcommand{\wt}{w_{\ft}}
\newcommand{\wts}{w^2_{\ft}}
\newcommand{\wtb}{\overline{w}}
\newcommand{\fra}{\frac{1}{2}}
\newcommand{\frb}{\frac{1}{4}}
\newcommand{\frc}{\frac{3}{2}}
\newcommand{\frd}{\frac{3}{4}}
\newcommand{\fre}{\frac{9}{2}}
\newcommand{\frf}{\frac{9}{4}}
\newcommand{\frg}{\frac{5}{4}}
\newcommand{\frh}{\frac{5}{2}}
\newcommand{\fri}{\frac{1}{8}}
\newcommand{\frj}{\frac{7}{4}}
\newcommand{\frl}{\frac{7}{8}}
\newcommand{\Spzzh}{\hat{\Sigma}'_{_{\zb\zb}}}
\newcommand{\sss}{s\sqrt{s}}
\newcommand{\sqs}{\sqrt{s}}
\newcommand{\Rtg}{R_{_{3Q}}}
\newcommand{\Rtt}{R_{_{33}}}
\newcommand{\Rww}{R_{_{\wb\wb}}}
\newcommand{\ssZ}{{\scriptscriptstyle{\zb}}}
\newcommand{\ssW}{{\scriptscriptstyle{\wb}}}
\newcommand{\ssH}{{\scriptscriptstyle{\hb}}}
\newcommand{\ssV}{{\scriptscriptstyle{\vb}}}
\newcommand{\ssA}{{\scriptscriptstyle{A}}}
\newcommand{\ssB}{{\scriptscriptstyle{B}}}
\newcommand{\ssC}{{\scriptscriptstyle{C}}}
\newcommand{\ssD}{{\scriptscriptstyle{D}}}
\newcommand{\ssF}{{\scriptscriptstyle{F}}}
\newcommand{\ssG}{{\scriptscriptstyle{G}}}
\newcommand{\ssL}{{\scriptscriptstyle{L}}}
\newcommand{\ssM}{{\scriptscriptstyle{M}}}
\newcommand{\ssN}{{\scriptscriptstyle{N}}}
\newcommand{\ssP}{{\scriptscriptstyle{P}}}
\newcommand{\ssQ}{{\scriptscriptstyle{Q}}}
\newcommand{\ssR}{{\scriptscriptstyle{R}}}
\newcommand{\ssS}{{\scriptscriptstyle{S}}}
\newcommand{\ssT}{{\scriptscriptstyle{T}}}
\newcommand{\ssU}{{\scriptscriptstyle{U}}}
\newcommand{\ssX}{{\scriptscriptstyle{X}}}
\newcommand{\ssY}{{\scriptscriptstyle{Y}}}
\newcommand{\ssWF}{{\scriptscriptstyle{WF}}}
\newcommand{\DiagramFermionToBosonFullWithMomenta}[8][70]{
  \vcenter{\hbox{
  \SetScale{0.8}
  \begin{picture}(#1,50)(15,15)
    \put(27,22){$\nearrow$}      
    \put(27,54){$\searrow$}    
    \put(59,29){$\to$}    
    \ArrowLine(25,25)(50,50)      \Text(34,20)[lc]{#6} \Text(11,20)[lc]{#3}
    \ArrowLine(50,50)(25,75)      \Text(34,60)[lc]{#7} \Text(11,60)[lc]{#4}
    \Photon(50,50)(90,50){2}{8}   \Text(80,40)[lc]{#2} \Text(55,33)[ct]{#8}
    \Vertex(50,50){2,5}          \Text(60,48)[cb]{#5} 
    \Vertex(90,50){2}
  \end{picture}}}
  }
\newcommand{\DiagramFermionToBosonPropagator}[4][85]{
  \vcenter{\hbox{
  \SetScale{0.8}
  \begin{picture}(#1,50)(15,15)
    \ArrowLine(25,25)(50,50)
    \ArrowLine(50,50)(25,75)
    \Photon(50,50)(105,50){2}{8}   \Text(90,40)[lc]{#2}
    \Vertex(50,50){0.5}         \Text(80,48)[cb]{#3}
    \GCirc(82,50){8}{1}            \Text(55,48)[cb]{#4}
    \Vertex(105,50){2}
  \end{picture}}}
  }
\newcommand{\DiagramFermionToBosonEffective}[3][70]{
  \vcenter{\hbox{
  \SetScale{0.8}
  \begin{picture}(#1,50)(15,15)
    \ArrowLine(25,25)(50,50)
    \ArrowLine(50,50)(25,75)
    \Photon(50,50)(90,50){2}{8}   \Text(80,40)[lc]{#2}
    \BBoxc(50,50)(5,5)            \Text(55,48)[cb]{#3}
    \Vertex(90,50){2}
  \end{picture}}}
  }
\newcommand{\DiagramFermionToBosonFull}[3][70]{
  \vcenter{\hbox{
  \SetScale{0.8}
  \begin{picture}(#1,50)(15,15)
    \ArrowLine(25,25)(50,50)
    \ArrowLine(50,50)(25,75)
    \Photon(50,50)(90,50){2}{8}   \Text(80,40)[lc]{#2}
    \Vertex(50,50){2.5}          \Text(60,48)[cb]{#3}
    \Vertex(90,50){2}
  \end{picture}}}
  }
\newcommand{\expgw}{\frac{\gf\mws}{2\srt\,\pi^2}}
\newcommand{\expgz}{\frac{\gf\mzs}{2\srt\,\pi^2}}
\newcommand{\Spww}{\Sigma'_{_{\wb\wb}}}
\newcommand{\shf}{{\hat s}^4}
\newcommand{\acz}{\scff{0}}
\newcommand{\acoo}{\scff{11}}
\newcommand{\acod}{\scff{12}}
\newcommand{\acdo}{\scff{21}}
\newcommand{\acdd}{\scff{22}}
\newcommand{\acdt}{\scff{23}}
\newcommand{\acdq}{\scff{24}}
\newcommand{\acto}{\scff{31}}
\newcommand{\actd}{\scff{32}}
\newcommand{\actt}{\scff{33}}
\newcommand{\actq}{\scff{34}}
\newcommand{\actc}{\scff{35}}
\newcommand{\acts}{\scff{36}}
\newcommand{\acoA}{\scff{1A}}
\newcommand{\acdA}{\scff{2A}}
\newcommand{\acdB}{\scff{2B}}
\newcommand{\acdC}{\scff{2C}}
\newcommand{\acdD}{\scff{2D}}
\newcommand{\actA}{\scff{3A}}
\newcommand{\actB}{\scff{3B}}
\newcommand{\actC}{\scff{3C}}
\newcommand{\ada}{\sdff{0}}
\newcommand{\adb}{\sdff{11}}
\newcommand{\adc}{\sdff{12}}
\newcommand{\add}{\sdff{13}}
\newcommand{\ade}{\sdff{21}}
\newcommand{\adf}{\sdff{22}}
\newcommand{\adg}{\sdff{23}}
\newcommand{\adh}{\sdff{24}}
\newcommand{\adi}{\sdff{25}}
\newcommand{\adj}{\sdff{26}}
\newcommand{\adl}{\sdff{27}}
\newcommand{\adm}{\sdff{31}}
\newcommand{\adn}{\sdff{32}}
\newcommand{\ado}{\sdff{33}}
\newcommand{\adp}{\sdff{34}}
\newcommand{\adq}{\sdff{35}}
\newcommand{\adr}{\sdff{36}}
\newcommand{\ads}{\sdff{37}}
\newcommand{\adt}{\sdff{38}}
\newcommand{\adu}{\sdff{39}}
\newcommand{\adw}{\sdff{310}}
\newcommand{\adv}{\sdff{311}}
\newcommand{\ady}{\sdff{312}}
\newcommand{\adz}{\sdff{313}}
\newcommand{\admt}{\frac{\tman}{\sman}}
\newcommand{\admu}{\frac{\uman}{\sman}}
\newcommand{\frm}{\frac{3}{8}}
\newcommand{\frn}{\frac{5}{8}}
\newcommand{\fro}{\frac{15}{8}}
\newcommand{\frp}{\frac{3}{16}}
\newcommand{\frq}{\frac{5}{16}}
\newcommand{\frr}{\frac{1}{16}}
\newcommand{\frs}{\frac{7}{2}}
\newcommand{\frt}{\frac{7}{16}}
\newcommand{\fru}{\frac{1}{3}}
\newcommand{\frw}{\frac{2}{3}}
\newcommand{\frz}{\frac{4}{3}}
\newcommand{\fry}{\frac{13}{3}}
\newcommand{\fraa}{\frac{11}{4}}
\newcommand{\bee}{\beta_{e}}
\newcommand{\beW}{\beta_{_\wb}}
\newcommand{\toDdrh}{{\ds\frac{2}{{\hat{\varepsilon}}}}}
\newcommand{\bqas}{\begin{eqnarray*}}
\newcommand{\eqas}{\end{eqnarray*}}
\newcommand{\mhcub}{M^3_{_H}}
\newcommand{\adComA}{\sdff{A}}
\newcommand{\adComB}{\sdff{B}}
\newcommand{\adComC}{\sdff{C}}
\newcommand{\adComD}{\sdff{D}}
\newcommand{\adComE}{\sdff{E}}
\newcommand{\adComF}{\sdff{F}}
\newcommand{\adComG}{\sdff{G}}
\newcommand{\adComH}{\sdff{H}}
\newcommand{\adComI}{\sdff{I}}
\newcommand{\adComJ}{\sdff{J}}
\newcommand{\adComL}{\sdff{L}}
\newcommand{\adComM}{\sdff{M}}
\newcommand{\adComN}{\sdff{N}}
\newcommand{\adComO}{\sdff{O}}
\newcommand{\adComP}{\sdff{P}}
\newcommand{\adComQ}{\sdff{Q}}
\newcommand{\adComR}{\sdff{R}}
\newcommand{\adComS}{\sdff{S}}
\newcommand{\adComT}{\sdff{T}}
\newcommand{\adComU}{\sdff{U}}
\newcommand{\adComAc}{\sdff{A}^c}
\newcommand{\adComBc}{\sdff{B}^c}
\newcommand{\adComCc}{\sdff{C}^c}
\newcommand{\adComDc}{\sdff{D}^c}
\newcommand{\adComEc}{\sdff{E}^c}
\newcommand{\adComFc}{\sdff{F}^c}
\newcommand{\adComGc}{\sdff{G}^c}
\newcommand{\adComHc}{\sdff{H}^c}
\newcommand{\adComIc}{\sdff{I}^c}
\newcommand{\adComJc}{\sdff{J}^c}
\newcommand{\adComLc}{\sdff{L}^c}
\newcommand{\adComMc}{\sdff{M}^c}
\newcommand{\adComNc}{\sdff{N}^c}
\newcommand{\adComOc}{\sdff{O}^c}
\newcommand{\adComPc}{\sdff{P}^c}
\newcommand{\adComQc}{\sdff{Q}^c}
\newcommand{\adComRc}{\sdff{R}^c}
\newcommand{\adComSc}{\sdff{S}^c}
\newcommand{\adComTc}{\sdff{T}^c}
\newcommand{\adComUc}{\sdff{U}^c}
\newcommand{\adComAf}{\sdff{A}^f}
\newcommand{\adComBf}{\sdff{B}^f}
\newcommand{\adComCf}{\sdff{F}^f}
\newcommand{\adComDf}{\sdff{D}^f}
\newcommand{\adComEf}{\sdff{E}^f}
\newcommand{\adComFf}{\sdff{F}^f}
\newcommand{\adComGf}{\sdff{G}^f}
\newcommand{\adComHf}{\sdff{H}^f}
\newcommand{\adComIf}{\sdff{I}^f}
\newcommand{\adComJf}{\sdff{J}^f}
\newcommand{\adComLf}{\sdff{L}^f}
\newcommand{\adComMf}{\sdff{M}^f}
\newcommand{\adComNf}{\sdff{N}^f}
\newcommand{\adComOf}{\sdff{O}^f}
\newcommand{\adComPf}{\sdff{P}^f}
\newcommand{\adComQf}{\sdff{Q}^f}
\newcommand{\adComRf}{\sdff{R}^f}
\newcommand{\adComSf}{\sdff{S}^f}
\newcommand{\adComTf}{\sdff{T}^f}
\newcommand{\adComUf}{\sdff{U}^f}
\newcommand{\adComBfc}{\sdff{B}^{fc}} 
\newcommand{\adComCfco}{\sdff{C}^{fc1}}
\newcommand{\adComCfcd}{\sdff{C}^{fc2}} 
\newcommand{\adComCfct}{\sdff{C}^{fc3}} 
\newcommand{\adComDfc}{\sdff{D}^{fc}}
\newcommand{\adComEfc}{\sdff{E}^{fc}}
\newcommand{\adComFfc}{\sdff{F}^{fc}}
\newcommand{\adComGfc}{\sdff{G}^{fc}}
\newcommand{\adComHfc}{\sdff{H}^{fc}}
\newcommand{\adComLfc}{\sdff{L}^{fc}}
\newcommand{\afba}[1]{A^{#1}_{_{\rm FB}}}
\newcommand{\alra}[1]{A^{#1}_{_{\rm LR}}}
\newcommand{\adComAt}{\sdff{A}^t}
\newcommand{\adComBt}{\sdff{B}^t}
\newcommand{\adComCt}{\sdff{T}^t}
\newcommand{\adComDt}{\sdff{D}^t}
\newcommand{\adComEt}{\sdff{E}^t}
\newcommand{\adComFt}{\sdff{T}^t}
\newcommand{\adComGt}{\sdff{G}^t}
\newcommand{\adComHt}{\sdff{H}^t}
\newcommand{\adComIt}{\sdff{I}^t}
\newcommand{\adComJt}{\sdff{J}^t}
\newcommand{\adComLt}{\sdff{L}^t}
\newcommand{\adComMt}{\sdff{M}^t}
\newcommand{\adComNt}{\sdff{N}^t}
\newcommand{\adComOt}{\sdff{O}^t}
\newcommand{\adComPt}{\sdff{P}^t}
\newcommand{\adComQt}{\sdff{Q}^t}
\newcommand{\adComRt}{\sdff{R}^t}
\newcommand{\adComSt}{\sdff{S}^t}
\newcommand{\adComTt}{\sdff{T}^t}
\newcommand{\adComUt}{\sdff{U}^t}
\newcommand{\adComAtt}{\sdff{A}^{\tau}}
\newcommand{\adComBtt}{\sdff{B}^{\tau}}
\newcommand{\adComCtt}{\sdff{T}^{\tau}}
\newcommand{\adComDtt}{\sdff{D}^{\tau}}
\newcommand{\adComEtt}{\sdff{E}^{\tau}}
\newcommand{\adComFtt}{\sdff{T}^{\tau}}
\newcommand{\adComGtt}{\sdff{G}^{\tau}}
\newcommand{\adComHtt}{\sdff{H}^{\tau}}
\newcommand{\adComItt}{\sdff{I}^{\tau}}
\newcommand{\adComJtt}{\sdff{J}^{\tau}}
\newcommand{\adComLtt}{\sdff{L}^{\tau}}
\newcommand{\adComMtt}{\sdff{M}^{\tau}}
\newcommand{\adComNtt}{\sdff{N}^{\tau}}
\newcommand{\adComOtt}{\sdff{O}^{\tau}}
\newcommand{\adComPtt}{\sdff{P}^{\tau}}
\newcommand{\adComQtt}{\sdff{Q}^{\tau}}
\newcommand{\adComRtt}{\sdff{R}^{\tau}}
\newcommand{\adComStt}{\sdff{S}^{\tau}}
\newcommand{\adComTtt}{\sdff{T}^{\tau}}
\newcommand{\adComUtt}{\sdff{U}^{\tau}}
\newcommand{\etavz}[1]{\eta^{\zb #1}_{_V}}
\newcommand{\phanst}{$\hphantom{\sigma^{s+t}\ }$}
\newcommand{\phanat}{$\hphantom{A_{FB}^{s+t}\ }$}
\newcommand{\phanss}{$\hphantom{\sigma^{s}\ }$}
\newcommand{\phanas}{$\hphantom{A_{FB}^{s}\ }$} 
\newcommand{\pbb}{\,\mbox{\bf pb}}
\newcommand{\pe}{\,\%\:}
\newcommand{\pc}{\,\%}
\newcommand{\temiv}{10^{-4}}
\newcommand{\temv}{10^{-5}}
\newcommand{\temvi}{10^{-6}}
\newcommand{\di}[1]{d_{#1}}
\newcommand{\delip}[1]{\Delta_+\lpar{#1}\rpar}
\newcommand{\propbb}[5]{{{#1}\over {\lpar #2^2 + #3 - \ib\varepsilon\rpar
\lpar\lpar #4\rpar^2 + #5 -\ib\varepsilon\rpar}}}
\newcommand{\cfft}[5]{C_{#1}\lpar #2;#3,#4,#5\rpar}    
\newcommand{\ppl}[1]{p_{+{#1}}}
\newcommand{\pmi}[1]{p_{-{#1}}}
\newcommand{\bpox}{\beta^2_{\xi}}
\newcommand{\bffdiff}[5]{B_{\rm d}\lpar #1;#2,#3;#4,#5\rpar}             
\newcommand{\cffdiff}[7]{C_{\rm d}\lpar #1;#2,#3,#4;#5,#6,#7\rpar}    
\newcommand{\affdiff}[2]{A_{\rm d}\lpar #1;#2\rpar}             
\newcommand{\Dqf}{\Delta\qf}
\newcommand{\bposx}{\beta^4_{\xi}}
\newcommand{\svverti}[3]{f^{#1}_{#2}\lpar{#3}\rpar}
\newcommand{\Mods}{\mbox{$M^2_{12}$}}
\newcommand{\Mots}{\mbox{$M^2_{13}$}}
\newcommand{\Motq}{\mbox{$M^4_{13}$}}
\newcommand{\Mdts}{\mbox{$M^2_{23}$}}
\newcommand{\Mdos}{\mbox{$M^2_{21}$}}
\newcommand{\Mtds}{\mbox{$M^2_{32}$}}
\newcommand{\dffpt}[3]{D_{#1}\lpar #2,#3;}           
\newcommand{\quu}{Q_{uu}}
\newcommand{\qdd}{Q_{dd}}
\newcommand{\qud}{Q_{ud}}
\newcommand{\qdu}{Q_{du}}
\newcommand{\msPj}[6]{\Lambda^{#1#2#3}_{#4#5#6}}
\newcommand{\bdiff}[4]{B_{\rm d}\lpar #1,#2;#3,#4\rpar}             
\newcommand{\bdifff}[7]{B_{\rm d}\lpar #1;#2;#3;#4,#5;#6,#7\rpar}             
\newcommand{\adiff}[3]{A_{\rm d}\lpar #1;#2;#3\rpar}  
\newcommand{\aw}{a_{_\wb}}
\newcommand{\az}{a_{_\zb}}
\newcommand{\sct}[1]{sect.~\ref{#1}}
\newcommand{\dreim}[1]{\varepsilon^{\rm M}_{#1}}
\newcommand{\drem}{\varepsilon^{\rm M}}
\newcommand{\hcapV}[2]{{\hat{\cal F}}^{#2}_{_{#1}}}
\newcommand{\swww}{{\scriptscriptstyle \wb\wb\wb}}
\newcommand{\szhz}{{\scriptscriptstyle \zb\hb\zb}}
\newcommand{\shzh}{{\scriptscriptstyle \hb\zb\hb}}
\newcommand{\bwith}[3]{\beta^{#3}_{#1}\lpar #2\rpar}
\newcommand{\Shhh}{{\hat\Sigma}_{_{\hb\hb}}}
\newcommand{\Sphhh}{{\hat\Sigma}'_{_{\hb\hb}}}
\newcommand{\seWilc}[1]{w_{#1}}
\newcommand{\seWtilc}[2]{w_{#1}^{#2}}
\newcommand{\eilc}{\gamma}
\newcommand{\eilcs}{\gamma^2}
\newcommand{\eilcc}{\gamma^3}
\newcommand{\eilcb}{{\overline{\gamma}}}
\newcommand{\eilcbs}{{\overline{\gamma}^2}}
\newcommand{\Sttww}{\Sigma_{_{33;\wb\wb}}}
\newcommand{\bSttww}{{\overline\Sigma}_{_{33;\wb\wb}}}
\newcommand{\Pggtg}{\Pi_{\ph\ph;3Q}}
\newcommand{\bDelta}{{\bar\Delta}}
\newcommand{\tDelta}{{\tilde\Delta}}
\newcommand{\tcft}[1]{C_{#1}\lpar t\rpar}
\newcommand{\tcftt}[1]{C_{#1}\lpar tt\rpar}
\newcommand{\tcfp}[1]{C^+_{#1}}
\newcommand{\tcfm}[1]{C^-_{#1}}

%
%
\title{Single-$\wb$ Production and Fermion-Loop Scheme:\\
Numerical Results}

\author{
Giampiero Passarino  \\
{\em Dipartimento di Fisica Teorica, Universit\`a di Torino, Italy} \\
{\em INFN, Sezione di Torino, Italy} \\ \\ \\
}

\date{}
\maketitle

\begin{abstract}
  \normalsize \noindent 
The single-$\wb$ production mechanism is synonymous to the $e^+e^-$
annihilation into $e\nu_e$ and a $\wb$ boson with the outgoing electron lost
in a small cone around the beam direction. It requires a Renormalization
Scheme that preserves gauge invariance and fermion masses cannot be
neglected in the calculation. A recently proposed generalization of the
so-called Fermion-Loop scheme is applied to the evaluation of observables
at LEP~2 energies. The total contribution to single-$\wb$ processes can be 
split, in a gauge invariant manner, into a $s$-channel component and a 
$t$-channel one. The latter is dominated by a regime of low momentum transfer 
of the outgoing electron and any high-energy Renormalization Scheme, as the 
$\gf$-one, fails to give the correct description of the scale. The Fermion-Loop
scheme automatically converts, among other things, all couplings of the theory 
into couplings that are running at the appropriate scale. Therefore, in addition
to represent the only scheme fully justified on a field-theoretical basis, the 
Fermion-Loop is the best starting point to include radiative corrections into
single-$\wb$ production. Numerical results are presented, showing a decrease
in the predictions that can be sizeable. There is no naive and overall
rescaling of $\alpha_{\rm QED}$ in any pragmatic scheme, as the Fixed-Width 
one, that can reproduce the Fermion-Loop results, at the requested accuracy,
for all configurations and for all kinematical cuts.
\end{abstract}

\vskip 2cm
\noindent
Pacs: 11.15.-q, 13.10.-q, 13.38.-b, 13.40.-f, 14.70.Fm

\clearpage

%
%

\section{Introduction}

An interesting process at LEP~2 is the so-called single-$\wb$ production,
$e^+e^- \to \wb e \nu$ which can be seen as a part of the CC20 process,
$e^+e^- \to \barq\,q\,(\mu\,\nu_{\mu},\,\,\tau\,\nu_{\tau})\,e\,\nu_e$, or as
a part of the Mix56 process, $e^+e^- \to e^+\,e^-\,\nu_e\,\barnu_e$. 
For a theoretical review we refer to~\cite{kn:swc} and to~\cite{kn:bd}.
For the experimental aspects we refer to the work of Ref.~\cite{kn:expf}.

The CC20 process is usually considered in two regimes, $|\cos\theta(e^-)|
\ge c$ or LACC20 and $|\cos\theta(e^-)| \le c$ or SACC20.
Strictly speaking the single $\wb$ production is defined by those events that
satisfy $|\cos\theta(e^-)| \ge 0.997$ and, therefore is a SACC20.

The LACC20 cross-section has been computed by many authors and references can 
be found in \cite{kn:lep2ww}. It represents a contribution to the $e^+e^- \to 
\wbp\wbm$ total cross-section, in turn used to derive a value for $\mw$, the 
$\wb$ boson mass.
This point deserves a comment: by $e^+e^- \to \wbp\wbm$ it is meant the
ideal cross-section obtained with the three double-resonant CC03
diagrams, see \fig{fig:CC03}, and therefore the background, i.e. the full
cross-section minus the CC03 one, is evaluated with the help of some 
MonteCarlo, estimating the error on the subtraction by comparing with some 
other MonteCarlo.
Then $\mw$ is derived from a fit to $\sigma$(CC03) with the help of a third 
calculation. From a theoretical point of view the evaluation of LACC20
is free of ambiguity, even in the approximation of massless fermions, as long
as a gauge-preserving scheme is applied and $\theta(e^-)$ is not too small.

For SACC20 instead, one cannot employ the massless approximation anymore
and this fact makes the calculation much more difficult. In other words, 
in addition to double-resonant $\wb$-pair production with one $\wb$
decaying into $e\nu_e$, there are $t$-channel diagrams that give a sizeable
contribution for small values of the polar scattering angle of the
$t$-channel electron. 

Single-$\wb$ processes are sensitive to the breaking of U(1) gauge invariance 
in the collinear limit, as described in Ref.~\cite{kn:bhf1} (see also
\cite{kn:Kurihara}).
The correct way of handling them is based on the so-called 
Fermion-Loop (FL)scheme~\cite{kn:bhf2}, the gauge-invariant treatment of the 
finite-width effects of $\wb$ and $\zb$ bosons in LEP2 processes.
However, till very recently, the Fermion-Loop scheme was available only for the
LACC20 regime.

For $e^+e^- \to e^-\barnu_e f_1\barf_2$, the U(1) gauge 
invariance becomes essential in the region of phase space where the angle 
between the incoming and outgoing electrons is small, see the work of
\cite{kn:bhf1} and also an alternative formulations in \cite{kn:zepp}.
In this limit the superficial $1/Q^4$ divergence of the  
propagator structure is reduced to $1/Q^2$ by U(1) gauge  
invariance. In the presence of light fermion masses this gives raise  
to the familiar $\ln(\mes/s)$ large logarithms.

Furthermore, keeping a finite electron mass through the calculation is not 
enough. One of the main results of~\cite{kn:swc} was to show that
there are remaining subtleties in CC20, associated with the zero mass limit
for the remaining fermions.

In Ref.~\cite{kn:swext} (hereafter I) a generalization of the Fermion-Loop 
scheme has been given (for previous work we refer to~\cite{kn:thosetwo}).
It consists of the re-summation of the fermionic one-loop corrections to the 
vector-boson propagators and of the inclusion of all remaining fermionic 
one-loop corrections, in particular those to the Yang--Mills vertices. 
In the original formulation, the Fermion-Loop scheme requires that vector
bosons couple to conserved currents, i.e., that the masses of all external
fermions can be neglected. There are several examples where fermion masses must
be kept to obtain a reliable prediction. As already stated, this is the case
for the single-$\wb$ production mechanism, where the outgoing electron is 
collinear, within a small cone, with the incoming electron. Therefore, $\me$ 
cannot be neglected.

Furthermore, among the $20$ Feynman diagrams that contribute (for $e\barnu_e 
u\bard$ final states, up to $56$ for $e^+e^-\nu_e\barnu_e$) there are 
multi-peripheral ones that require a non-vanishing mass also for the other
outgoing fermions. In I a generalization of the Fermion-Loop scheme is 
introduced to account for external, non-conserved, currents. Dyson re-summed 
transitions are introduced without neglecting the $p_{\mu}p_{\nu}$-terms and 
including the contributions from the Higgs-Kibble ghosts in the 
't Hooft-Feynman gauge.
In I we have introduced running vector boson masses and studied their relation 
with the corresponding complex poles. Always in I it is shown that any 
$\Smat$-matrix element takes a very simple form when written in terms of 
running masses. Finally, the relevant Ward identity, the 
U(1) Ward identity for single-$\wb$, is derived in the situation of interest,
when all currents are non-conserved and when the top quark mass is not 
neglected inside loops.

For all details concerning the formal construction of the fully massive 
Fermion-Loop scheme we, therefore, refer to~\cite{kn:swext}. Here, instead, 
we concentrate on its implementation within the FORTRAN program 
{\tt WTO}~\cite{kn:wto}, on the corresponding numerical results and on the
comparison, for single-$\wb$, between the Fermion-Loop (hereafter FL) scheme
and the Fixed-Width (hereafter FW) one.

To be specific the name of Fixed-Width scheme is reserved for the following:
the cross-section is computed using the tree-level amplitude. The massive 
gauge-boson propagators are given by $1/(\pmoms+\mlones-\ib\Glone)$.
This gives an unphysical width for $\pmoms>0$, but retains $U(1)$ gauge 
invariance in the CC20 process.

The most recent numerical results produced for single-$\wb$ production are from
the following codes~\cite{kn:groups}: {\tt CompHEP}, {\tt GRC4F}, 
{\tt NEXTCALIBUR},\footnote{
{\tt NEXTCALIBUR} with masses, uses a massive matrix element program called 
{\tt HELAC}, that is a recursive algorithm for electroweak amplitude
calculations based on Dyson-Schwinger equations
developed by A.~Kanaki and C~.G.~Papadopoulos~\cite{kn:wip}.} {\tt PVALPHA}, 
{\tt WPHACT} and {\tt WTO}.

Among these codes, {\tt WPHACT} is the only one to employ the Fermion-Loop 
scheme in its imaginary version~\cite{kn:fli}, where the full imaginary part 
of the Fermion-Loop corrections is used.

In view of a requested, inclusive cross-section, accuracy of $2\%$ we must
include radiative corrections to the best of our knowledge, at least the
bulk of any large effect.
As we know, the correct scale of the couplings and their differentiation between
$s$ and $t$-channel is connected to the real part of the corrections, so that
the imaginary FL is not enough, we need a complete FL for single-$\wb$.
Having all the parts, the tree-level couplings are replaced by running 
couplings at the appropriate momenta and the massive gauge-boson propagators are
modified accordingly. The vertex coefficients, entering through the Yang--Mills 
vertex, contain the lowest order couplings as well as the one-loop fermionic 
vertex corrections. 

Apart from some recent development, each calculation aimed to provide some 
estimate for $e^+e^- \to 4\,$f production is, at least nominally, a tree level 
calculation. Among other things it will require the choice of some Input
Parameter Set (IPS) and of certain relations among the parameters. In the 
literature, although improperly, this is usually referred to as the choice of 
the Renormalization Scheme.

Typically, we have at our disposal four experimental data points
(plus $\alpha_s$), i.e. the measured vector boson masses $\mz, \mw$ and the
coupling constants, $\gf$ and $\alpha$. However, we only 
have three bare parameters at our disposal, the charged vector boson mass, 
the $SU(2)$ coupling constant and the sinus of the weak mixing angle. While 
the inclusion of one loop corrections would allow us to fix at least the 
value of the top quark mass from a consistency relation, this cannot be done 
at the tree level. Thus, different choices of the basic relations among the 
input parameters can lead to different results with deviations which, in some 
case, can be sizeable.

For instance, a possible choice is to fix the coupling constant $g$ as
\bq
g^2 = {{4\pi\alpha}\over {s_{_W}^2}}, \quad
s_{_W}^2 = {{\pi\alpha}\over {{\sqrt 2}\gf\mw^2}},
\eq
where $\gf$ is the Fermi coupling constant. Another possibility would be to use
\bq
g^2 = 4{\sqrt 2}\gf\mw^2, 
\eq
but, in both cases, we miss the correct running of the coupling. Ad hoc 
solutions should be avoided, and the running of the parameters must always 
follow from a fully consistent scheme.
Therefore, the only satisfactory solution can be found in the extension of 
the full Fermion-Loop to having non-zero external masses, or non-conserved 
currents. Unless, of course, one can compute the full set of corrections.

Note that the Fixed-Width scheme behaves properly in the collinear and 
high-energy regions of phase space, to the contrary of the Running-Width
scheme, but it completely misses the running of the couplings, an effect that 
is expected to be above the requested precision tag of $2\%$.
By considering the impact of the FL-scheme on the relevant observables we will 
be able to judge on the goodness of naive rescaling procedures.

\section{The ingredients in the Fermion-Loop scheme.}

There are several building blocks that enter into the construction of
the Fermion-Loop scheme, see for instance~\cite{kn:book}.
In the 't Hooft--Feynman gauge, the $\drii{\mu}{\nu}$ part of the 
vector--vector transitions can be cast in the following form~\cite{kn:book},
where $\stw(\ctw)$ is the sine(cosine) of the weak mixing angle:
\bqa
S_{\ph\ph}  &=& \frac{\gbs\stws}{16\,\pi^2}\,\Pgg(\pmoms)\,\pmoms, 
\quad
S_{\ssZ\ssZ} = \frac{\gbs}{16\pi^2\ctws}\,\Szz(\pmoms),
\nl
S_{\ssZ\ph} &=& \frac{\gbs\stw}{16\pi^2\ctw}\,\Sigma_{_{\ssZ\ph}}(\pmoms),
\quad
S_{\ssW\ssW} = \frac{\gbs}{16\pi^2}\,\Sww(\pmoms).
\label{S_self}
\eqa
Next we have can transform to the $(3,Q)$ basis~\cite{kn:book}, where one
writes
\bqa
\Szz(\pmoms) &=& \Stt(\pmoms)-2\stws\Stg(\pmoms)+\stwf\Pgg(\pmoms)\,\pmoms,
\nl
\Sigma_{_{\ssZ\ph}}(\pmoms) &=& \Stg(\pmoms)-\stws\Pgg(\pmoms)\,\pmoms.
\label{Sig_self}
\eqa
We now consider three parameters, the e.m. coupling constant $\ec$,
the $SU(2)$ coupling constant $\gb$ and the sine of the weak mixing
angle $\stw$. At the tree level they are not independent, but rather
they satisfy the relation $\gbs\stws = \ecs$.
The running of the e.m. coupling constant is easily derived and gives
\bq
\frac{1}{\ecs(\sman)} = \frac{1}{\gbs\stws}-\frac{1}{16\,\pi^2}\,\Pgg(\sman).
\eq
However, we have a natural scale to use since at $\sman = 0$ we have 
the fine structure constant at our disposal. Therefore, the running of 
$\ecs(\sman)$ is completely specified in terms of $\alpha$ by
\bq
\frac{1}{\ecs(\sman)} = \frac{1}{4\,\pi\alpha}\,
\lrbr 1 - \frac{\alpha}{4\,\pi}\,\Pi(\sman)\rrbr,  
\qquad \mbox{with} \quad
\Pi(\sman) = \Pgg(\sman) - \Pgg(0).
\eq
For the running of $\gbs$ we derive a similar equation:
\bq
\frac{1}{\gbs(\sman)} = \frac{1}{\gbs} - \frac{1}{16\,\pi^2}\,\Ptg(\sman).
\eq
The running of the third parameter, $\stws(\sman)$, is now fixed by
\bq
\stws(\sman) = \frac{\ecs(\sman)}{\gbs(\sman)}.
\eq
\begin{figure}[t]
\begin{minipage}[t]{14cm}
{\begin{center}
\vspace*{-2.0cm}
\mbox{\epsfysize=14cm\epsfxsize=15cm\epsffile{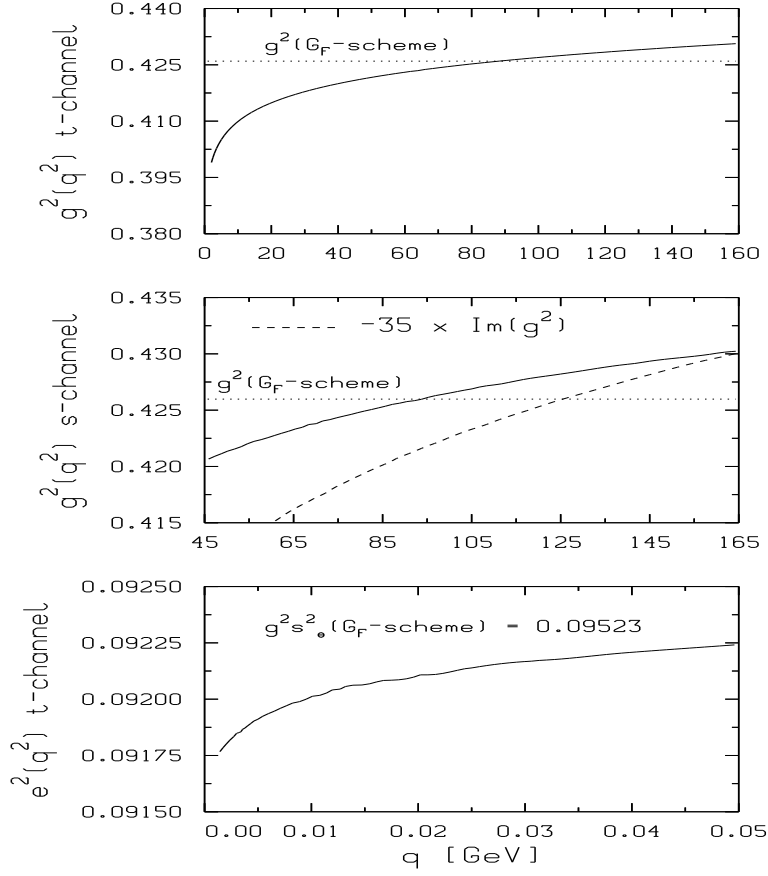}}
\vspace*{-2.7cm}
\end{center}}
\end{minipage}
\vspace*{1.5cm}
\caption[]{Running of coupling constants. The sign of $\Imb g^2(s)$ is reversed
and the corresponding curve is magnified by a factor $35$. Here $q$ is the 
absolute value of the momentum.}
\label{fig:rcoupli}
\end{figure}
In \fig{fig:rcoupli} we show the running of $e^2(q^2)$ for $q^2 \to  0_+$,
compared with the fixed value of $e^2$ that one would use in the $\gf$-scheme.
Furthermore, in the same figure, we show the evolution of $g^2(q^2)$ for
$q^2$ time-like or space-like, again compared with $g^2_{\gf}$. The sizeable
difference that one gets between $e^2$ running in the $t$-channel and
$e^2$ fixed in the $\gf$-scheme is expected to be one of the major 
improvements induced by the full Fermion-Loop scheme.

\begin{figure}[t]
\begin{minipage}[t]{14cm}
{\begin{center}
\vspace*{-2.0cm}
\mbox{\epsfysize=14cm\epsfxsize=15cm\epsffile{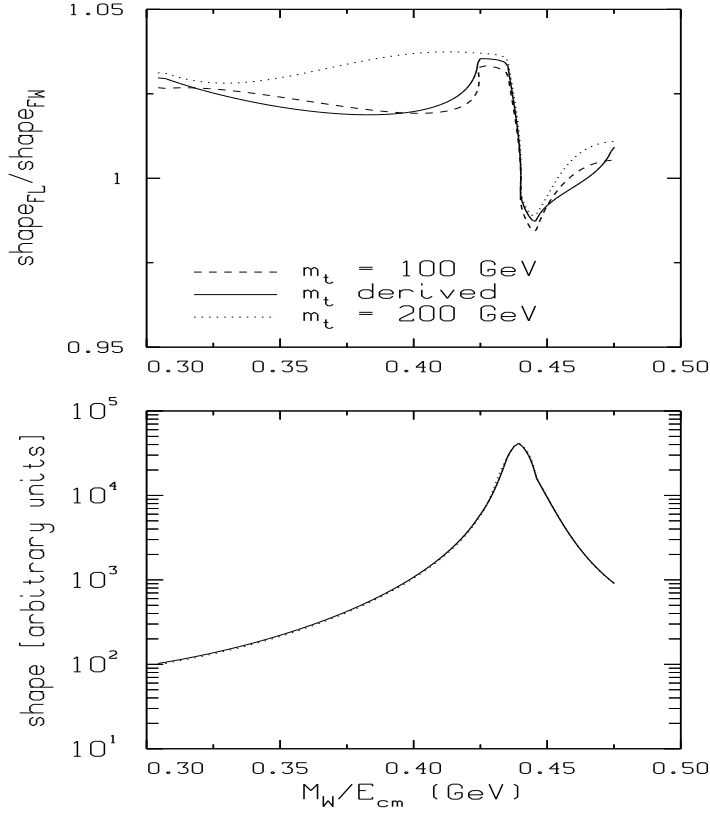}}
\vspace*{-2.7cm}
\end{center}}
\end{minipage}
\vspace*{1.5cm}
\caption[]{Comparison between the Fixed-Width and the Fermion-Loop line-shapes,
\eqn{sLS}.}
\label{fig:LS}
\end{figure}

The re-summed propagators for the vector bosons are:
\bqa
G_{\ph}(\pmoms) &=& \biggl\{\pmoms - S_{\ph\ph}(\pmoms) 
-{{\lrbr S_{\ssZ\ph}(\pmoms)\rrbr^2}\over
{\pmoms + \bzms - S_{\ssZ\ssZ}(\pmoms)}}\biggr\}^{-1},  
\nl
G_{\ssZ\ph}(\pmoms) &=& {{S_{\ssZ\ph}(\pmoms)}\over
{\bigl[\pmoms - S_{\ph\ph}(\pmoms)\bigr]\,
 \bigl[\pmoms + \bzms - S_{\ssZ\ssZ}(\pmoms)\bigr] 
-\bigl[S_{\ssZ\ph}(\pmoms)\bigr]^2}},  
\nl
G_{\ssZ}(\pmoms) &=& \biggl\{\pmoms + \bzms - S_{\ssZ\ssZ} - 
{{\lrbr S_{\ssZ\ph}(\pmoms)\rrbr^2}\over 
{\pmoms - S_{\ph\ph}(\pmoms)}}\biggr\}^{-1},  
\nl
G_{\ssW}(\pmoms) &=& \Bigl[ \pmoms + \LMs - S_{_{\ssW\ssW}}(\pmoms)\Bigr]^{-1}.
\label{defprops}
\eqa
The quantity $\bzm = M/\ctw$ is the bare $\zb$ mass.
An essential ingredient in the construction of the scheme is represented by 
the location of the complex poles.
Substituting the corresponding results into the expressions for the 
propagators, \eqn{defprops}, we see that all ultraviolet divergences not 
proportional to $\pmoms$ cancel. We obtain
\bqa
G_{\ssZ}(\sman) &=& \Bigl[-\sman + \sZ - Z(\sman) + Z(\sZ)\Bigr]^{-1},  
\nl
G_{\ssW}(\sman) &=& \Bigl[-\sman + \sW - S_{_{\ssW\ssW}}(\sman) 
+ S_{_{\ssW\ssW}}(\sW)\Bigr]^{-1},  
\nl
G_{\ssZ\ph}(\sman) &=& -\,{{S_{\ssZ\ph}(\sman)}\over{\sman + S_{\ph\ph}(\sman)}}\,
G_{\ssZ}(\sman),  
\nl
G_{\ph}(\sman) &=& -\,{1\over{\sman + S_{\ph\ph}(\sman)}} 
+\lrbr{{S_{\ssZ\ph}(\sman)}\over{\sman + S_{\ph\ph}(\sman)}}\rrbr^2\,G_{\ssZ}(\sman).
\eqa
The vector boson propagators are now expressed as
\bqa
G_{\ssW}(\sman)&=&-\frac{\gbs(\sman)}{\gbs}\,
\frac{\omega_{\ssW}(\sman)}{\sman},  
\qquad
G_{\ssZ}(\sman) = -\frac{\ctws}{\gbs}\,\frac{\gbs(\sman)}{\canys(\sman)}\,
\frac{\omega_{\ssZ}(\sman)}{\sman},  
\nl
G_{\ssZ\ph}(\sman)&=&\frac{\stw}{\ctw}\lrbr 1 - \frac{\sanys(\sman)}{\stws}
\rrbr\,G_{\ssZ}(\sman),  
\nl
G_{\ph}(\sman) &=& \frac{\ecs(\sman)}{\ecs} + \frac{\stws}{\ctws}
\lrbr 1 - \frac{\sanys(\sman)}{\stws}\rrbr^2\,G_{\ssZ}(\sman),
\label{defG}
\eqa
where the propagation functions are
\bqa
\omega^{-1}_{\ssW}(\sman) &=& 1 - \frac{\gbs(\sman)}{\sman}\,
\lcbr
\frac{\sW}{\gbs(\sW)} - \frac{1}{16\,\pi^2}\,
\Bigl[f_{\ssW}(\sman) - f_{\ssW}(\sW)\Bigr]
\rcbr,  
\nl
\omega^{-1}_{\ssZ}(\sman) &=& 1 - \frac{\gbs(\sman)}{\canys(\sman)\sman}\,
\lcbr
\frac{\canys(\sZ)}{\gbs(\sZ)}\,\sZ - \frac{1}{16\,\pi^2}\,
\Bigl[f_{\ssZ}(\sman) - f_{\ssZ}(\sW)\Bigr]
\rcbr.
\eqa
The explicit form of the $f_{\ssW,\ssZ}$-functions is given in~\cite{kn:swext}.
The complete one-loop re-summation in the
't Hooft-Feynman gauge is equivalent to some {\em effective} unitary-gauge
$\wb$-propagator. The whole amplitude can be written in terms of a $\wb$-boson 
exchange diagram, if we make use of the following effective propagator:
\bq
\Delta^{\mu\nu}_{\rm eff} = \frac{1}{p^2+M^2-S^0_{\ssW}}\,\Big[
\delta^{\mu\nu} + \frac{p^{\mu}p^{\nu}}{M^2(p^2)}\Big].
\eq
The explicit form of $M^2(p^2)$ is, again given in~\cite{kn:swext}. We now 
define the following line-shape functions:
\bqa
L_{\rm FL}(s) &=& {s^2\over{
\lpar s-\Reb\sW\rpar^2 + \lpar \Imb\sW\rpar^2}}\,{1\over {|R(s)|^2}},  \nl
L_{\rm FW}(s) &=& {s^2\over{\lpar s - \mws\rpar^2 + \mws\gw^2}}.
\label{sLS}
\eqa
with a running $\rho$-factor
\bq
R(s) = 1 + \frac{g^2(s)}{16\,\pi^2}\,
{{f_{\ssW}(s)-f_{\ssW}(\sW) + \sW\,\Bigl[ \Ptg(\sW) - \Ptg(s)\Bigr]}\over
{s-\sW}}.
\label{rhof}
\eq
Here $\mw,\gw$ are the on-shell $\wb$ mass and total width, respectively.
The two line-shapes are compared in \fig{fig:LS} for different values of the 
top quark mass.
We are using complex-mass renormalization but we only include fermionic 
corrections. Therefore, we can start with the Fermi coupling constant but also 
with $\mw$ as an input parameter.
Equating the corresponding renormalization conditions yields a relation 
between $\mz$, $\gf$, $\Reb\{\alpha(\mzs)^{-1}\}$, $\mw$, and $\mt$, 
see~\cite{kn:bhf2}. 
This relation can be solved iteratively for $m_t$. For the following input 
parameter set,
\bq
\mw = 80.350\,\GeV, \quad \mz = 91.1867\,\GeV, \quad
\gf= 1.16639\,\times\,10^{-5}\,\GeV^{-2},
\eq
we obtain the following solution:
\bq
\mu_{\ssW} = \sqrt{\Reb\lpar \sW\rpar} =  80.324\,\GeV, \quad
\gamma_{\ssW} = - {{\Imb\lpar \sW\rpar}\over {\mu_{\ssW}}} = 2.0581\,\GeV,
\quad \mt = 148.62\,\GeV.
\eq
The $26\,\MeV$ difference between $\mw$ and $\mu_{\ssW}$ is responsible
for the sharp transition around $80\,$GeV that can be seen in \fig{fig:LS}.
Apart from that, the Fermion-Loop line-shape can be few percents above
the Fixed-Width line shape.
\begin{figure}[t]
\begin{minipage}[t]{14cm}
{\begin{center}
\vspace*{-2.0cm}
\mbox{\epsfysize=13cm\epsfxsize=14cm\epsffile{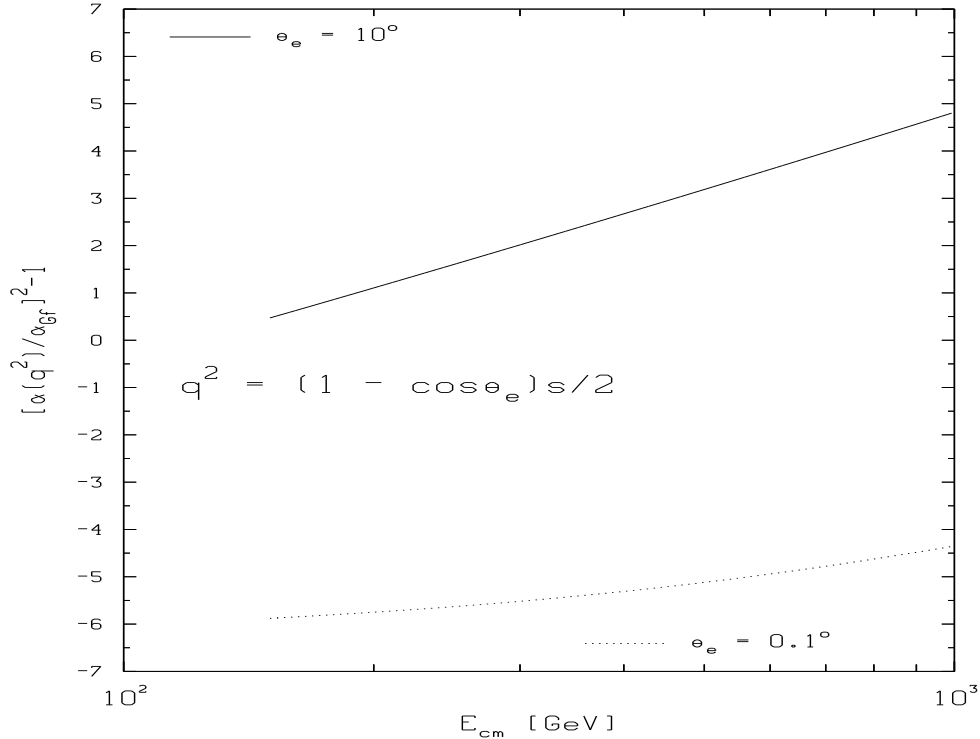}}
\vspace*{-2.7cm}
\end{center}}
\end{minipage}
\vspace*{1.5cm}
\caption[]{Running of $\alpha_{\rm QED}(q^2)$ in the $t$-channel.}
\label{fig:alphaq2}
\end{figure}
The behavior of $e^2(q^2)$ in the $t$-channel and, to some
extent, the ratio between Fermion-Loop and Fixed-Width cross-sections is
crucially dependent on which regime we are considering. A careful examination
of \fig{fig:alphaq2} shows the following: if we choose
\bq
<q^2> = \lpar 1 - <\cos\theta_e> \rpar\, \frac{s}{2},
\eq
and use $10^\circ$ or $0.1^\circ$ as representative for LACC20 or SACC20,
we obtain a ratio
\bqa
\Bigl[ \frac{\alpha(q^2)}{\alpha_{\gf}}\Bigr]^2 -1 &\to& 
-6{\%} \div -4{\%}, \quad <\theta_e> = 0.1^\circ,  \nl
\Bigl[ \frac{\alpha(q^2)}{\alpha_{\gf}}\Bigr]^2 -1 
&\to& 
+0.5{\%} \div +5{\%}, \quad <\theta_e> = 10^\circ
\eqa
for $\sqrt{s}$ ranging from $150\,\GeV$ to $1\,\TeV$.
Therefore, the effect of a running $\alpha_{\rm QED}$ in the $t$-channel is to 
decrease/increase the cross-section, with respect to the Fixed-Width scheme,
for SACC20/LACC20.

A final ingredient, existing in the Fermion-Loop scheme, is represented by
the inclusion of vertices that are $\mt$-dependent.
The top-quark contributions are particularly important for delayed-unitarity 
effects. In this respect also terms involving the totally-antisymmetric 
$\varepsilon$-tensor (originating from vertex corrections) are relevant.
While such terms are absent for complete generations of massless fermions
owing to the anomaly cancellations, they show up for finite fermion masses.

All details concerning vertices can be found in~\cite{kn:swext} where one
can find the explicit expressions for the vertex form-factors in terms
of standard $\scff{}$-functions~\cite{kn:pv}.

\section{The diagrams.}

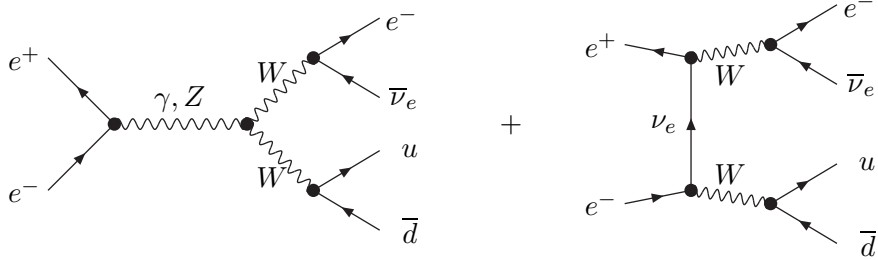
\begin{figure}[h]
\bqas
\ba{ccc}
\vcenter{\hbox{
  \begin{picture}(175,100)(0,0)
    \ArrowLine(25,25)(50,50)        \Text(10,25)[lc]{$\fem$}
    \ArrowLine(50,50)(25,75)        \Text(10,75)[lc]{$\fep$}
    \Vertex(50,50){2.5}
    \Photon(50,50)(100,50){2}{9}    \Text(75,55)[bc]{$\ph,\zb$}
    \Photon(100,50)(125,25){2}{7}   \Text(104,30)[lc]{$\wb$}
    \Photon(100,50)(125,75){2}{7}   \Text(104,70)[lc]{$\wb$}
    \Vertex(100,50){2.5}
    \ArrowLine(150,10)(125,25)      \Text(165,10)[rc]{$\bard$}
    \ArrowLine(125,25)(150,40)      \Text(165,40)[rc]{$u$}
    \Vertex(125,25){2.5}
    \ArrowLine(150,60)(125,75)      \Text(165,60)[rc]{$\barnu_e$}
    \ArrowLine(125,75)(150,90)      \Text(165,90)[rc]{$\fem$}
    \Vertex(125,75){2.5}
  \end{picture}}}
&\quad+&
\vcenter{\hbox{
  \begin{picture}(150,100)(0,0)
    \ArrowLine(50,75)(25,80)        \Text(10,80)[lc]{$\fep$}
    \ArrowLine(50,25)(50,75)        \Text(35,50)[lc]{$\fnue$}
    \ArrowLine(25,20)(50,25)        \Text(10,20)[lc]{$\fem$}
    \Photon(50,75)(80,80){2}{7}     \Text(65,72.5)[tc]{$\wb$}
    \Photon(50,25)(80,20){2}{7}     \Text(65,27.5)[bc]{$\wb$}
    \ArrowLine(105,65)(80,80)       \Text(120,65)[rc]{$\barnu_e$}
    \ArrowLine(80,80)(105,95)       \Text(120,95)[rc]{$\fem$}
    \ArrowLine(105,5)(80,20)        \Text(120,5)[rc]{$\bard$}
    \ArrowLine(80,20)(105,35)       \Text(120,35)[rc]{$u$}
    \Vertex(50,75){2.5}
    \Vertex(50,25){2.5}
    \Vertex(80,80){2.5}
    \Vertex(80,20){2.5}
  \end{picture}}}
\ea
\eqas
\vspace{-2mm}
\caption[]{The CC03 family of diagrams, annihilation $\,\oplus\,$ conversion.}
\label{fig:CC03}
\end{figure}

\begin{figure}[h]
\vspace*{-5mm}
\bqas
\ba{ccc}
\vcenter{\hbox{
  \begin{picture}(165,100)(0,0)
    \ArrowLine(25,25)(50,50)        \Text(10,25)[lc]{$\fem$}
    \ArrowLine(50,50)(25,75)        \Text(10,75)[lc]{$\fep$}
    \Photon(50,50)(75,50){2}{7}     \Text(62.5,55)[cb]{$\ph,\zb$}
    \ArrowLine(95,30)(75,50)        \Text(75,40)[lt]{$\fem$}
    \ArrowLine(115,10)(95,30)       \Text(130,10)[rc]{$\barnu_e$}
    \ArrowLine(75,50)(115,90)       \Text(130,90)[rc]{$\fem$}
    \Photon(95,30)(125,30){2}{7}    \Text(110,35)[cb]{$\wb$}
    \ArrowLine(150,15)(125,30)      \Text(165,15)[rc]{$\bard$}
    \ArrowLine(125,30)(150,45)      \Text(165,45)[rc]{$u$}
    \Vertex(50,50){2.5}
    \Vertex(75,50){2.5}
    \Vertex(95,30){2.5}
    \Vertex(125,30){2.5}
  \end{picture}}}
&\quad+&
\vcenter{\hbox{
  \begin{picture}(165,100)(0,0)
    \ArrowLine(25,25)(50,50)        \Text(10,25)[lc]{$\fem$}
    \ArrowLine(50,50)(25,75)        \Text(10,75)[lc]{$\fep$}
    \Photon(50,50)(75,50){2}{7}     \Text(62.5,55)[bc]{$\ph,\zb$}
    \ArrowLine(115,10)(75,50)       \Text(130,10)[rc]{$\barnu_e$}
    \ArrowLine(75,50)(95,70)        \Text(77,62)[lb]{$\barnu_e$}
    \ArrowLine(95,70)(115,90)       \Text(130,90)[rc]{$\fem$}
    \Photon(95,70)(125,70){2}{7}    \Text(110,65)[ct]{$\wb$}
    \ArrowLine(150,55)(125,70)      \Text(165,55)[rc]{$\bard$}
    \ArrowLine(125,70)(150,85)      \Text(165,95)[rc]{$u$}
    \Vertex(50,50){2.5}
    \Vertex(75,50){2.5}
    \Vertex(95,70){2.5}
    \Vertex(125,70){2.5}
  \end{picture}}}
\ea
\eqas
\bqas
\ba{ccc}
\vcenter{\hbox{
  \begin{picture}(165,100)(0,0)
    \ArrowLine(25,25)(50,50)        \Text(10,25)[lc]{$\fem$}
    \ArrowLine(50,50)(25,75)        \Text(10,75)[lc]{$\fep$}
    \Photon(50,50)(75,50){2}{7}     \Text(62.5,55)[cb]{$\ph,\zb$}
    \ArrowLine(95,30)(75,50)        \Text(75,40)[lt]{$u$}
    \ArrowLine(115,10)(95,30)       \Text(130,10)[rc]{$\bard$}
    \ArrowLine(75,50)(115,90)       \Text(130,90)[rc]{$u$}
    \Photon(95,30)(125,30){2}{7}    \Text(110,35)[cb]{$\wb$}
    \ArrowLine(150,15)(125,30)      \Text(165,15)[rc]{$\barnu_e$}
    \ArrowLine(125,30)(150,45)      \Text(165,45)[rc]{$\fem$}
    \Vertex(50,50){2.5}
    \Vertex(75,50){2.5}
    \Vertex(95,30){2.5}
    \Vertex(125,30){2.5}
  \end{picture}}}
&\quad+&
\vcenter{\hbox{
  \begin{picture}(165,100)(0,0)
    \ArrowLine(25,25)(50,50)        \Text(10,25)[lc]{$\fem$}
    \ArrowLine(50,50)(25,75)        \Text(10,75)[lc]{$\fep$}
    \Photon(50,50)(75,50){2}{7}     \Text(62.5,55)[bc]{$\ph,\zb$}
    \ArrowLine(115,10)(75,50)       \Text(130,10)[rc]{$\bard$}
    \ArrowLine(75,50)(95,70)        \Text(77,62)[lb]{$\bard$}
    \ArrowLine(95,70)(115,90)       \Text(130,90)[rc]{$u$}
    \Photon(95,70)(125,70){2}{7}    \Text(110,65)[ct]{$\wb$}
    \ArrowLine(150,55)(125,70)      \Text(165,55)[rc]{$\barnu_e$}
    \ArrowLine(125,70)(150,85)      \Text(165,95)[rc]{$\fem$}
    \Vertex(50,50){2.5}
    \Vertex(75,50){2.5}
    \Vertex(95,70){2.5}
    \Vertex(125,70){2.5}
  \end{picture}}}
\ea
\eqas
\caption[]{Diagrams belonging to the CC11 $\,-\,$ CC03 family.}
\label{fig:CC11}
\end{figure}
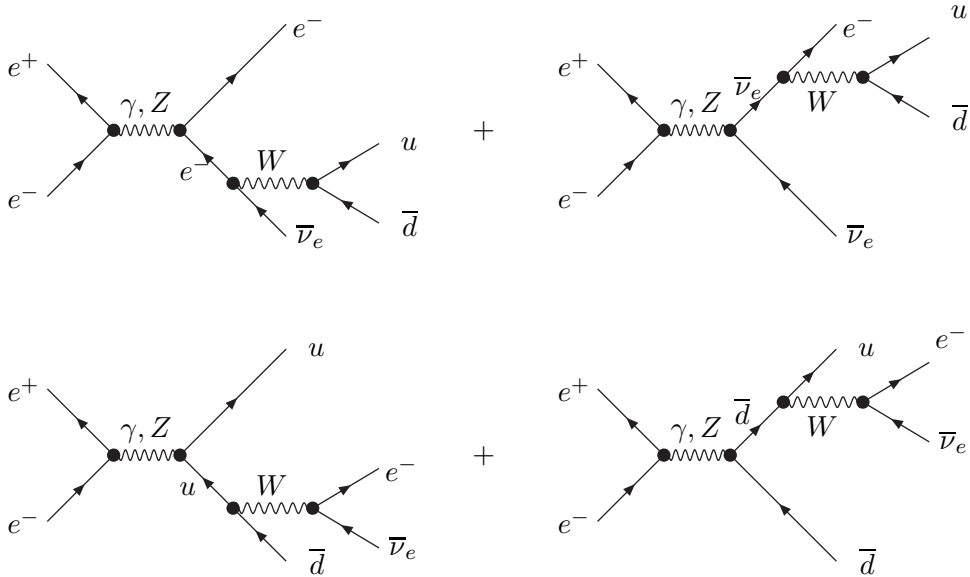

As already mentioned, there are $20$ Feynman diagrams that contribute for 
$e\barnu_e u\bard$ final states, and $56$ for $e^+e^-\nu_e\barnu_e$.
They are well known to the expert working in the field, but we present them
for convenience of the less specialized reader.
For simplicity, we only consider the CC20 family.
The most familiar part of CC20 is the CC03 one, with three diagrams as 
depicted in \fig{fig:CC03},
\vspace*{-2mm}
The next set of diagrams needed to complete the $s$-channel component
shows up with the CC11 class where we add to the
annihilation and conversion CC03 diagrams all topologies corresponding to
pair production of fermions. The CC11 diagrams not in CC03 are 
single-resonant and they are shown in \fig{fig:CC11}
Finally, we have the $10$ diagrams forming the $t$-channel part. In the latter
part, one diagram has a $\wb$-exchange, five a $\zb$-exchange and four a 
$\ph$-exchange. As we have shown in~\cite{kn:swext}, this picture is not 
changed by the inclusion of one-loop fermionic corrections.
\vspace{0.2cm}
\bqas
\ba{ccc}
\vcenter{\hbox{
  \SetScale{0.7}
  \begin{picture}(110,100)(0,0)
  \ArrowLine(50,120)(0,140)
  \ArrowLine(100,140)(50,120)
  \ArrowLine(0,0)(50,20)
  \ArrowLine(50,20)(100,0)
  \ArrowLine(100,110)(80,70)
  \ArrowLine(80,70)(100,30)
  \Photon(50,20)(50,70){2}{7}
  \Photon(50,70)(50,120){2}{7}
  \Photon(50,70)(80,70){2}{7}
  \Text(-14,98)[lc]{$e^+$}
  \Text(77,98)[lc]{$\barnu_e$}
  \Text(-14,0)[lc]{$e^-$}
  \Text(77,0)[lc]{$e^-$}
  \Text(77,77)[lc]{$\bard$}
  \Text(77,21)[lc]{$u$}
  \Text(18,60)[lc]{$\wb$}
  \Text(11,26)[lc]{$\gamma,\zb$}
  \end{picture}}}
&\quad+&
\vcenter{\hbox{
  \SetScale{0.7}
  \begin{picture}(110,100)(0,0)
  \ArrowLine(50,120)(0,140)
  \ArrowLine(65,126)(50,120)
  \ArrowLine(100,140)(65,126)
  \ArrowLine(0,0)(50,20)
  \ArrowLine(50,20)(100,0)
  \ArrowLine(100,110)(80,70)
  \ArrowLine(80,70)(100,30)
  \Photon(50,20)(50,120){2}{7}
  \Photon(65,126)(80,70){2}{7}
  \Text(77,77)[lc]{$\bard$}
  \Text(77,21)[lc]{$u$}
  \Text(54,76)[lc]{$\wb$}
  \Text(11,66)[lc]{$\gamma,\zb$}
  \Text(-14,98)[lc]{$e^+$}
  \Text(77,98)[lc]{$\barnu_e$}
  \Text(-14,0)[lc]{$e^-$}
  \Text(77,0)[lc]{$e^-$}
  \end{picture}}}
\ea
\eqas
\bqas
\ba{ccc}
\vcenter{\hbox{
  \SetScale{0.7}
  \begin{picture}(110,100)(0,0)
  \ArrowLine(50,120)(0,140)
  \ArrowLine(100,140)(50,120)
  \ArrowLine(0,0)(50,20)
  \ArrowLine(50,20)(100,0)
  \ArrowLine(100,90)(50,90)
  \ArrowLine(50,90)(50,50)
  \ArrowLine(50,50)(100,50)
  \Photon(50,20)(50,50){2}{7}
  \Photon(50,90)(50,120){2}{7}
  \Text(77,72)[lc]{$\bard$}
  \Text(77,26)[lc]{$u$}
  \Text(22,52)[lc]{$u$}
  \Text(10,20)[lc]{$\gamma,\zb$}
  \Text(17,77)[lc]{$\wb$}
  \Text(-14,98)[lc]{$e^+$}
  \Text(77,98)[lc]{$\barnu_e$}
  \Text(-14,0)[lc]{$e^-$}
  \Text(77,0)[lc]{$e^-$}
  \end{picture}}}
&\quad+&
\vcenter{\hbox{
  \SetScale{0.7}
  \begin{picture}(110,100)(0,0)
  \ArrowLine(50,120)(0,140)
  \ArrowLine(100,140)(50,120)
  \ArrowLine(0,0)(50,20)
  \ArrowLine(50,20)(100,0)
  \ArrowLine(100,90)(50,50)
  \Line(50,90)(65,78)
  \ArrowLine(90,58)(100,50)
  \ArrowLine(50,50)(50,90)
  \Photon(50,20)(50,50){2}{7}
  \Photon(50,90)(50,120){2}{7}
  \Text(77,72)[lc]{$\bard$}
  \Text(77,21)[lc]{$u$}
  \Text(22,52)[lc]{$d$}
  \Text(10,20)[lc]{$\gamma,\zb$}
  \Text(17,77)[lc]{$\wb$}
  \Text(-14,98)[lc]{$e^+$}
  \Text(77,98)[lc]{$\barnu_e$}
  \Text(-14,0)[lc]{$e^-$}
  \Text(77,0)[lc]{$e^-$}
  \end{picture}}}
\ea
\eqas
\bqas
\ba{ccc}
\vcenter{\hbox{
  \SetScale{0.7}
  \begin{picture}(110,100)(0,0)
  \ArrowLine(50,120)(0,140)
  \ArrowLine(100,140)(50,120)
  \ArrowLine(0,0)(50,20)
  \ArrowLine(50,20)(100,0)
  \ArrowLine(100,110)(80,70)
  \ArrowLine(80,70)(100,30)
  \Photon(50,20)(50,120){2}{7}
  \Photon(18,132)(80,70){2}{7}
  \Text(77,77)[lc]{$\bard$}
  \Text(77,21)[lc]{$u$}
  \Text(48,71)[lc]{$\wb$}
  \Text(21,56)[lc]{$\zb$}
  \Text(-14,98)[lc]{$e^+$}
  \Text(77,98)[lc]{$\barnu_e$}
  \Text(-14,0)[lc]{$e^-$}
  \Text(77,0)[lc]{$e^-$}
  \end{picture}}}
&\quad+&
\vcenter{\hbox{
  \SetScale{0.7}
  \begin{picture}(110,100)(0,0)
  \ArrowLine(50,120)(0,140)
  \ArrowLine(100,140)(50,120)
  \ArrowLine(0,0)(50,20)
  \ArrowLine(50,20)(100,0)
  \ArrowLine(100,110)(80,70)
  \ArrowLine(80,70)(100,30)
  \Photon(50,20)(50,120){2}{7}
  \Photon(64,15)(80,70){2}{7}
  \Text(77,77)[lc]{$\bard$}
  \Text(77,21)[lc]{$u$}
  \Text(40,41)[lc]{$\wb$}
  \Text(19,56)[lc]{$\wb$}
  \Text(-14,98)[lc]{$e^+$}
  \Text(77,98)[lc]{$\barnu_e$}
  \Text(-14,0)[lc]{$e^-$}
  \Text(77,0)[lc]{$e^-$}
  \end{picture}}}
\ea
\eqas
\vspace{-2mm}
\begin{figure}[h]
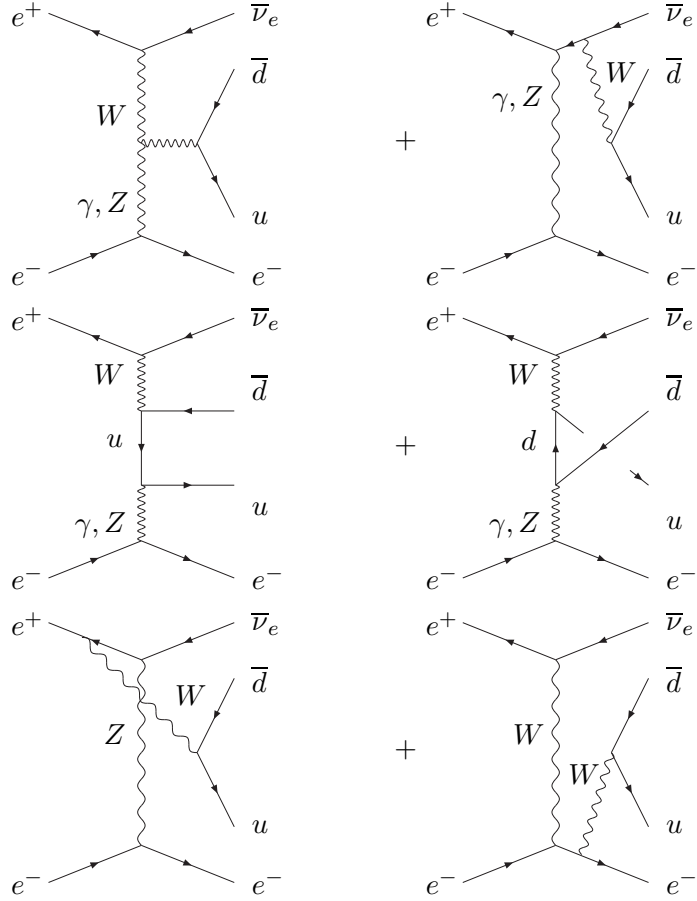

\caption[]{The $t$-channel component of the CC20 family of diagrams: fusion,
bremsstrahlung and multi-peripheral.}
\label{cc20gz}
\end{figure}
\vskip 20pt

\section{Implementing the Fermion-Loop scheme.}

Here we follow closely the spirit of Sect.~6 of ~\cite{kn:swext}:
once gauge invariance is preserved then we are allowed to investigate the 
numerical relevance of masses and to neglect some, if convenient.
For the photon $t$-channel diagrams, the amplitude squared, summed over spin 
and integrated over the phase space of $\nu_e u \bard$, can be written as 
follows:
\bqa
\frac{1}{4}\,L^{\ph\ph}_{\mu\nu}\,W^{\mu\nu}_{\ph\ph} &=&
2\,\frac{\mes}{(Xys)^2}\,W^1_{\ph\ph} 
- \frac{1}{Xys}\,W^1_{\ph\ph}  \nl 
{}&+& \frac{2}{Xy^2s}\,\lpar 1 - \frac{1}{y}\rpar\,W^2_{\ph\ph} 
+ \frac{2}{y^2s}\,\Bigl[\frac{1}{(X+1)^2} + \frac{1}{X+1}\Bigr]\,
\lpar \frac{1}{y} + 1\rpar\,\,W^2_{\ph\ph}  \nl
{}&+& \frac{1}{ys}\,\frac{1}{X+1}\,W^2_{\ph\ph} 
\label{phph}
\eqa
In \eqn{phph} $L$ is the usual leptonic tensor while $W$ is a CC20 tensor, 
pertaining to the specific process under consideration. 
U(1) gauge invariance, proven in~\cite{kn:swext}, allows for the following 
decomposition of $W^{\mu\nu}_{\ph\ph}$,
\bqa
W^{\ph\ph}_{\mu\nu} &=& W^1_{\ph\ph}\, \Big[ - \delta_{\mu\nu}+
\frac{Q_{-\mu}Q_{\nu}}{Q^2_-}\Big] -
     W^2_{\ph\ph}\,\frac{Q^2_-}{\lpar\spro{p_+}{Q_-}\rpar^2}\,
\lpar p_{+\mu}-\frac{\spro{p_+}{Q_-}}{Q^2_-}\,Q_{-\mu} \rpar  \nl
{}&\times&
\lpar p_{+\nu}-\frac{\spro{p_+}{Q_-}}{Q^2_-}\,Q_{-\nu} \rpar.
\label{decom}
\eqa
Moreover, with momenta assignment $e^+(p_+) e^-(p_-) \to e^-(q_-) 
\barnu_e(q_+) u(k) \bard(\kbar)$, $Q_- = p_--q_-$, we have
introduced the variable $y$, equivalent to the fraction of the 
electron energy carried by the photon, and the variable $X$ according to
the following definitions:
\bqa
\spro{p_+}{Q_-} &=& \frac{1}{2}\,\Big[ \mes - \lpar X + 1\rpar\,ys\Big], \nl
Q^2_- &=& X\,ys,  \qquad \lpar p_+ + p_-\rpar^2 = - s.
\label{xydef}
\eqa
Note that our approach is akin to the 
Weizs\"acker-Williams approximation. However, we have gone beyond this 
approximation where one replaces $W^{1,2}_{\ph\ph}(Q^2_-,y)$ of \eqn{decom}
with $W^{1,2}_{\ph\ph}(0,y)$. In our approach, we retain the full 
$Q^2_-$-dependence of the photonic exchange in the $t$-channel, CC20, diagrams.
The sole approximation that we have used is to retain only those parts that, 
after integration, produce logarithmically enhanced terms, 
$\ord{\ln\lpar\mes/s\rpar}$, and constant terms, $\ord{1}$, therefore 
neglecting all terms proportional to $\ord{\lpar\mes/s\rpar^n}$. 
Since $\min\{Q^2_-\} \sim \mes$ this procedure is equivalent to say that
we keep, prior to the integration over $Q^2_-,y$, all terms that are of order
\bq
\ord{\frac{\meq}{Q^6_-}}, \quad \ord{\frac{\mes}{Q^4_-}}, \quad 
\ord{\frac{1}{Q^2_-}}, \quad \ord{1}.
\label{keep}
\eq
By inspection, it is easily seen that the $\ph-\zb$, $\zb-\zb$,
$\wb-\wb$ etc parts of the $t$-channel and the full $s$-channel component
only contain terms of $\ord{1}$.
Therefore, for the non-photonic part of the $t$-channel, for its 
interference with the photonic part and for the $s$-channel we can use
the old version of {\tt WTO}, employing the massless Fermion-Loop scheme.
For the photonic part of the $t$-channel we have introduced in the latest
version of {\tt WTO} the massive Fermion-Loop scheme as described 
in~\cite{kn:swext}.

In contrast with a previous calculation~\cite{kn:swc}, performed with 
the program {\tt WTO} in the Fixed-Width scheme, here we do not have a problem 
of introducing a small cone around the beam axis matching the internal region 
with then external one. Indeed, as noted before, we have been 
able to avoid the Weizs\"acker-Williams approximation. In other words, the
evaluation of the $t$-channel component is now exact in the whole range of 
the photon momentum.

One should not get the wrong impression that our approximation, massive for
the photonic part of the $t$-channel and massless for the rest, is an
inconsistent one. Therefore, it is useful to explain the procedure in more
detail. We construct gauge invariant subsets of amplitudes, consider their
gauge invariant squares and interferences and, after that, we consistently 
neglect all terms of $\ord{\mfs/s}$.

The grand total of $20$ Feynman diagrams is, first of all,
split into a $10$ $s$-channel part and a $10$ $t$-channel part. In the latter
part, one diagram has a $\wb$-exchange, five a $\zb$-exchange and four a 
$\ph$-exchange. As we have shown, this picture is not changed by the inclusion
of one-loop fermionic corrections: when all transitions are properly taken 
into account, we still end up with the $1-5-4$ subdivision
described above, as long as the fermion--anti-fermion-vector--boson couplings
are described in terms of the re-summed expressions.

This $s\,\oplus\,t$ splitting is a gauge invariant one, but we 
can further characterize additional set of diagrams. The argument is as follows:
Take $e^+ \mu^- \to \barnu_e \mu^- u \bard$. Only the CC20 $t$-channel 
diagrams contribute and, since these are all diagrams that we need, this set 
is gauge invariant.
Next, take $e^+ \nu_{\mu} \to \barnu_e \nu_{\mu} u \bard$. It is made, again,
only with $t$-channel parts but the photon does not contribute, so that we have
$10-4 = 6$ diagrams. Moreover, if one writes any $\zb$ current as 
$J = J_{\ssQ} + J_{\ssL}$ (with $J_{\ssL}$ proportional to $\gdp$), only 
$J_{\ssL}$ contributes here, because of the neutrinos, so that we have five 
$J^{\ssZ}_{\ssL}$ diagrams plus one $\wb$ diagram that form a gauge 
invariant set.
Since the whole $t$-channel is gauge invariant, the eight diagrams, four
with photons and four with $J^{\ssZ}_{\ssQ}$, must form a gauge invariant 
sub-set. This remains true for one-loop corrections when writing everything 
in terms of running objects and including vertices. 

Consider, now, the eight $J^{\ph}\,\oplus\,J^{\ssZ}_{\ssQ}$, $t$-channel 
diagrams. They can be decomposed in a product $L_{\mu\nu}\,W^{\mu\nu}$. 
For the $W$-tensors we obtain 
\bqa
W^{\ph\ph}_{\mu\nu} &=& W^1_{\ph\ph}\, \Big[ - \delta_{\mu\nu}+
\frac{Q_{-\mu}Q_{\nu}}{Q^2_-}\Big] -
     W^2_{\ph\ph}\,\frac{Q^2_-}{\lpar\spro{p_+}{Q_-}\rpar^2}\,
\lpar p_{+\mu}-\frac{\spro{p_+}{Q_-}}{Q^2_-}\,Q_{-\mu} \rpar  \nl
{}&\times&
\lpar p_{+\nu}-\frac{\spro{p_+}{Q_-}}{Q^2_-}\,Q_{-\nu} \rpar,  \nl
{}&{}&{}  \nl
W^{\ph\ssZ}_{\mu\nu} &=& W^1_{\ph\ssZ}\, \Big[ - \delta_{\mu\nu}+
\frac{Q_{-\mu}Q_{\nu}}{Q^2_-}\Big] -
     W^2_{\ph\ssZ}\,\frac{Q^2_-}{\lpar\spro{p_+}{Q_-}\rpar^2}\,
\lpar p_{+\mu}-\frac{\spro{p_+}{Q_-}}{Q^2_-}\,Q_{-\mu} \rpar  \nl
{}&\times&
\lpar p_{+\nu}-\frac{\spro{p_+}{Q_-}}{Q^2_-}\,Q_{-\nu} \rpar  \nl
{}&+& {{W^3_{\ph\ssZ}}\over {\spro{p_+}{Q_-}}}\,
\varepsilon\lpar\mu,\nu,Q_-,p_+\rpar + 
{{W^4_{\ph\ssZ}}\over {\spro{p_+}{Q_-}}}\,p_{+\mu}p_{+\nu}  \nl
{}&+&
{{W^5_{\ph\ssZ}}\over {\spro{p_+}{Q_-}}}\,\lpar p_{+\mu}Q_{-\nu} +
p_{+\nu}Q_{-\mu}\rpar  
+ W^6_{\ph\ssZ}\,{{Q_{-\mu}Q_{-\nu}}\over {Q^2_-}}.  \nl
{}&{}&{}  \nl
W^{\ssZ\ssZ}_{\mu\nu} &=& W^1_{\ssZ\ssZ}\, \Big[ - \delta_{\mu\nu}+
\frac{Q_{-\mu}Q_{\nu}}{Q^2_-}\Big] -
     W^2_{\ssZ\ssZ}\,\frac{Q^2_-}{\lpar\spro{p_+}{Q_-}\rpar^2}\,
\lpar p_{+\mu}-\frac{\spro{p_+}{Q_-}}{Q^2_-}\,Q_{-\mu} \rpar  \nl
{}&\times&
\lpar p_{+\nu}-\frac{\spro{p_+}{Q_-}}{Q^2_-}\,Q_{-\nu} \rpar  
+ {{W^3_{\ssZ\ssZ}}\over {\spro{p_+}{Q_-}}}\,
\varepsilon\lpar\mu,\nu,Q_-,p_+\rpar \nl
{}&+& 
{{W^4_{\ssZ\ssZ}}\over {\spro{p_+}{Q_-}}}\,p_{+\mu}p_{+\nu}  
+ {{W^5_{\ssZ\ssZ}}\over {\spro{p_+}{Q_-}}}\,\lpar p_{+\mu}Q_{-\nu} +
p_{+\nu}Q_{-\mu}\rpar  
+ W^6_{\ssZ\ssZ}\,{{Q_{-\mu}Q_{-\nu}}\over {Q^2_-}}.  \nl
\eqa
For $J^{\ssZ}_{\ssQ}$ the $W^3$ form factor does not contribute in the
$L\,\otimes\,W$ contraction.

Strictly speaking, only $J^{\ph}_{\mu}$ satisfies the condition 
$Q^{\mu}_-\,J^{\ph}_{\mu} = 0$. However, by using Ward identities, one can 
easily prove that $Q^{\mu}_-\,J^{\ssZ}_{\mu} = \ord{\mf}$, owing to a direct
coupling of the $\hkn$ with fermions and to the absence of a $\wbp\wbm\hkn$ 
tree-level coupling.

Let's repeat that, inside a gauge invariant subset of diagrams we only keep 
the terms shown in \eqn{keep}.

Let's start with $L^{\mu\nu}\,W^{\ph\ph}_{\mu\nu}$. The result is given in 
\eqn{phph} and, according to our strategy of neglecting terms of
$\ord{\mfs/s}$, we can compute the $W_{\ph\ph}$-tensor in the massless
approximation.

The most delicate case seems to be the interference, $\ph\,\otimes\,\zb$.
Since, schematically, $W_{\mu\nu} = J^{\dagger}_{\mu}\,J_{\nu}$, this 
interference can be written as
\bq
W^{\ph\ssZ}_{\mu\nu} = \lpar J^{\ph}_{\mu}\rpar^{\dagger}\,J^{\ssZ}_{\nu} +
\lpar J^{\ssZ}_{\mu}\rpar^{\dagger}\,J^{\ph}_{\nu}.
\label{currcurr}
\eq
Multiplication by $Q^{\mu}_-$ or by $Q^{\nu}_-$ gives terms of $\ord{\mf}$
since the e.m. current is conserved. Therefore, the form factors
$W^{4,5,6}_{\ph\ssZ}$ are of $\ord{\mf}$. Furthermore, we see from 
\eqn{currcurr} that $Q^{\mu}_-Q^{\nu}_-\,W^{\ph\ssZ}_{\mu\nu} = 0$. It follows:
\bqa
{\cal W}^{\ph\ssZ}_{\mu\nu} &=& 
{{W^4_{\ph\ssZ}}\over {\spro{p_+}{Q_-}}}\,p_{+\mu}p_{+\nu} +
{{W^5_{\ph\ssZ}}\over {\spro{p_+}{Q_-}}}\,\lpar p_{+\mu}Q_{-\nu} +
p_{+\nu}Q_{-\mu}\rpar  
+ W^6_{\ph\ssZ}\,{{Q_{-\mu}Q_{-\nu}}\over {Q^2_-}},  \nl
{}&{}&{}  \nl
Q^{\nu}_-\,{\cal W}^{\ph\ssZ}_{\mu\nu} &=& J^p_{\phi}\,p_{+\mu} +
J^q_{\phi}\,Q_{-\mu},
\eqa
where the $J^{p,q}_{\phi}$ form factors are related to the same set of
diagrams with the replacement $\zb \to \hkn$ and they are $\ord{\mf}$.
Multiplication by $Q^{\mu}_-$ gives zero, with solution
\bqa
J^p_{\phi} &=& - \frac{Q^2_-}{\spro{p_+}{Q_-}}\,J^q_{\phi},  \nl
W^6_{\ph\ssZ} &=& 1, \quad W^5_{\ph\ssZ} = J^q_{\phi} - 1, \quad
W^4_{\ph\ssZ} =  \frac{Q^2_-}{\spro{p_+}{Q_-}}\,\lpar 1 - 2\,J^q_{\phi}\rpar.
\label{propX}
\eqa
In this particular case, the $L\,\otimes\,W$ contraction gives the following 
result:
\bqa
\frac{1}{4}\,L^{\ph\ssZ}_{\mu\nu}\,W^{\mu\nu}_{\ph\ssZ} &=&
\stws\,\Bigl[ -4\,W^1_{\ph\ssZ} + \frac{8}{y\lpar X+1\rpar^2}\,
\lpar 1 - \frac{1}{y}\rpar\,W^2_{\ph\ssZ} \nl
{}&+& 4\,\lpar 1 - \frac{1}{X+1}\rpar\,W^2_{\ph\ssZ} +
\frac{4}{yX}\,\lpar 1 - \frac{1}{y}\rpar\,W^4_{\ph\ssZ} +
\frac{4}{y\lpar X+1\rpar}\,\lpar \frac{1}{y} - 1\rpar\,W^4_{\ph\ssZ}  \nl
{}&+& 2\,\frac{X}{X+1}\,\lpar W^5_{\ph\ssZ} + W^6_{\ph\ssZ}\rpar\Bigr].
\eqa
The only term which is proportional to $1/X$ multiplies $W^4_{\ph\ssZ}$. 
Note, however, that this form factor is proportional to X by virtue of
\eqn{propX}.
Therefore, the contraction does not contain any term that is divergent in the 
massless limit and we can neglect fermion masses.

The $\zb\zb$ terms do not have any divergence to start with so that, in our
procedure, we can neglect all fermion masses. The same is true for the subset
formed by the $J^{\ssZ}_{\ssL}$ diagrams plus one $\wb$ diagram, for the 
$s$-channel diagrams, for their mutual interferences and for their
interference with the photonic $t$-channel part. The latter follows from
the same line of argument used in studying the $J^{\ph}-J^{\ssZ}_{\ssQ}$ 
interference and by recalling that a cross-section is a sum of cut diagrams: 
infrared power counting shows that there are no mass singular contributions 
in the interferences. As far as the $s\,\otimes\,t$ interference is concerned
this confirms previous numerical findings, see Fig. 4 of the first reference
in~\cite{kn:bd}.

\section{Numerical results.}

In this paper we will consider two examples belonging to the single-$\wb$ 
family:
\begin{enumerate}
\item $e^+e^- \to u\bard e\nu$, $|\cos\theta_e| > 0.997$, 
  $M(u\bard) > 45\,$GeV;
\item $e^+e^- \to e\nu\mu\nu(\ph)$,
  $|\cos\theta_e| > 0.997$, $E_{\mu} > 15\,$GeV,
  $|\cos\theta_{\mu}| < 0.95$.
\end{enumerate}
Our Input Parameter Set is specified by
\bq
\sqrt{s} = 183/189/200\,\GeV, \quad   \mw = 80.350\,\GeV, \quad
\mz = 91.1867\,\GeV,
\eq
Moreover, we are mostly interested in the variations induced by the 
Fermion-Loop scheme with respect to the Fixed-Width scheme. Therefore, we will 
not include other effects that are nevertheless needed in any realistic
evaluation of observables:
\begin{enumerate}
\item QED Initial State Radiation;
\item QCD naive corrections;
\item Final State Coulomb correction factor;
\item Final State QED radiation 
\item Anomalous couplings.
\end{enumerate}
Furthermore, we use trivial CKM matrix element, lepton masses according to 
PDG~\cite{kn:pdg}, light quark masses $5\,\MeV$ (up) and $10\,\MeV$ (down),
$\gf= 1.16637\,10^{-5}$ (recent Fermi coupling).

When working in the Fixed-Width approach we use the $\gf$-scheme, that is
\bq
\stws = 1 - \frac{\mws}{\mzs}, \qquad
\alpha \equiv \alpha_{\gf}= 4 \sqrt{2}\,\frac{\gf\mws\stws}{4\,\pi},
\eq
Numerical (integration) errors returned for the cross-section computed
with $0 < \theta_e < \theta^{\rm max}_e$ and $\theta^{\rm max}_e = 0.1^\circ 
\div 0.4^\circ$ are, typically, of order $0.05 \div 0.1\%$ for $u\bard$ and
$\nu_{\mu}\mu^+$ final states, respectively. We start in \tabn{tab0}
by comparing FL and FW cross-sections at $\sqrt{s} = 183\,$GeV. 
Clearly, the bulk of the cross-section is for very small scattering angle of 
the electron, dominated by the first bin. The differences between FL-scheme
and FW-scheme are decreasing when $\theta^{\rm max}_e$ becomes larger and,
from the shape of the angular distribution, we infer smaller differences
in the higher bins.
 \begin{table}[htbp]\centering
 \begin{tabular}{|c||c|c|c|}
 \hline
  & \multicolumn{3}{c|}{} \\
  & \multicolumn{3}{c|}{$e^+e^- \to e^- \barnu_e u \bard$} \\
  & \multicolumn{3}{c|}{} \\
 \hline
                           &    &    &                   \\
$\theta^{\rm max}_e$ [deg] & FL & FW & FL/FW-1 (percent) \\
                           &    &    &                   \\
 \hline
                           &    &    &                   \\
$0.1^\circ$   & 0.04478(3) & 0.04841(3) & -7.50(8) \\
$0.2^\circ$   & 0.05139(3) & 0.05543(3) & -7.29(7) \\
$0.3^\circ$   & 0.05525(3) & 0.05953(3) & -7.19(7) \\
$0.4^\circ$   & 0.05798(3) & 0.06243(3) & -7.13(7) \\
                           &    &    &                   \\
 \hline
  & \multicolumn{3}{c|}{} \\
  & \multicolumn{3}{c|}{$e^+e^- \to e^- \barnu_e \nu_{\mu} \mu^+$} \\
  & \multicolumn{3}{c|}{} \\
 \hline
                           &    &    &                   \\
$\theta^{\rm max}_e$ [deg] & FL & FW & FL/FW-1 (percent) \\
                           &    &    &                   \\
 \hline
                           &    &    &                   \\
$0.1^\circ$   & 0.01345(3) & 0.01415(1) & -4.9(2) \\   
$0.2^\circ$   & 0.01548(3) & 0.01627(2) & -4.9(2) \\   
$0.3^\circ$   & 0.01667(3) & 0.01751(2) & -4.8(2) \\   
$0.4^\circ$   & 0.01752(3) & 0.01839(2) & -4.7(2) \\   
                           &    &    &                   \\
 \hline
 \end{tabular}
\vspace*{3mm}
 \caption[
 ]{
$\sigma$ in pb for the processes $e^+e^- \to e^- \barnu_e 
u \bard (\nu_{\mu} \mu^+)$, for $\sqrt{s} = 183\,\GeV$ and for 
$\theta_e < \theta^{\rm max}_e$. Kinematical cuts are as
described in the main text. The number in parenthesis shows the statistical 
error of the numerical integration on the last digit.}
 \label{tab0}
 \end{table}
 \normalsize

\begin{figure}[t]
\begin{minipage}[t]{14cm}
{\begin{center}
\vspace*{-2.0cm}
\mbox{\epsfysize=14cm\epsfxsize=15cm\epsffile{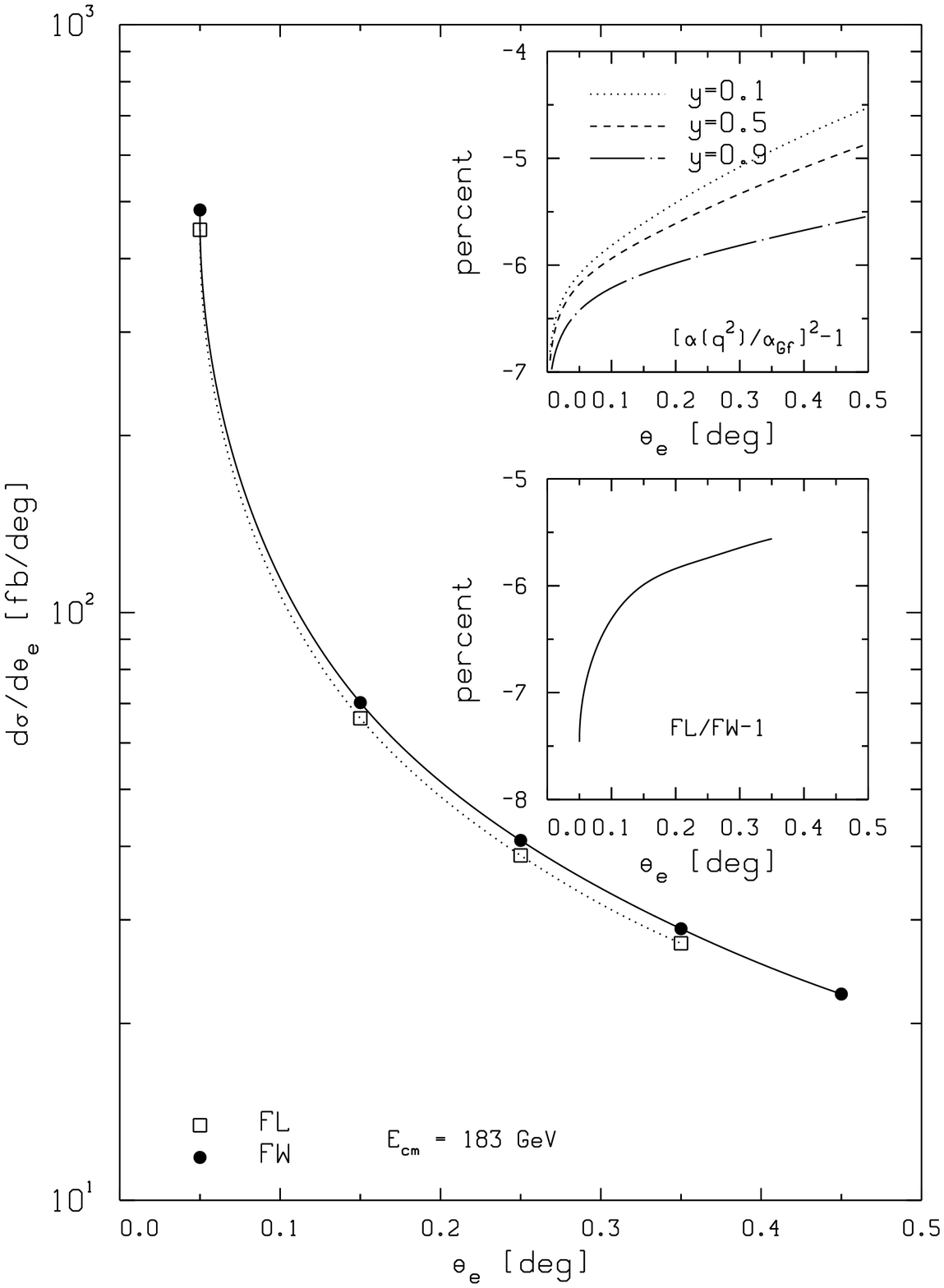}}
\vspace*{-2.7cm}
\end{center}}
\end{minipage}
\vspace*{1.5cm}
\caption[]{$d\sigma/d\cos\theta_e$ [fb/degrees] for $e^+e^- \to u \bard e^-
\barnu_e$ with $M(u\bard) > 45\,$GeV and $\sqrt{s} = 183\,\GeV$.}
\label{fig:rs183}
\end{figure}

To continue our discussion of the numerical results we stay at $\sqrt{s} = 
183\,$GeV and consider now the angular distribution, $d\sigma/d\theta_e$ for 
the $u \bard e^- \barnu_e$ final states. 
The results are shown in \fig{fig:rs183}.
From \fig{fig:rs183} we see that the FL prediction is lower than the FW one,
from $-7.46\%$ in the bin $0^\circ - 0.1^\circ$ to $-5.56\%$ in the bin
$0.3^\circ - 0.4^\circ$. Correspondingly, the first bin is $6.78$ higher
than the second one, $11.60(16.37)$ than the third(fourth) one. This is not
a surprise, since the first bin represents $50\%$ of the total single-$\wb$
cross-section. 

Always in the same figure, we have reported the behavior
of $\Bigl[\alpha(q^2)/\alpha_{\gf}-1\Bigr]^2$ as a function of $\theta_e$ for
three values of $y$, \eqn{xydef}, using the appropriate relation: 
$q^2 = q^2(\theta_e,y)$. 
The behavior of FL/FW-1, when we vary $\theta_e$, is very similar to the one
given by the ratio of coupling constants, indicating that the bulk of the 
effect is in the running of the e.m. coupling constant.

\begin{figure}[t]
\begin{minipage}[t]{14cm}
{\begin{center}
\vspace*{-2.0cm}
\mbox{\epsfysize=14cm\epsfxsize=15cm\epsffile{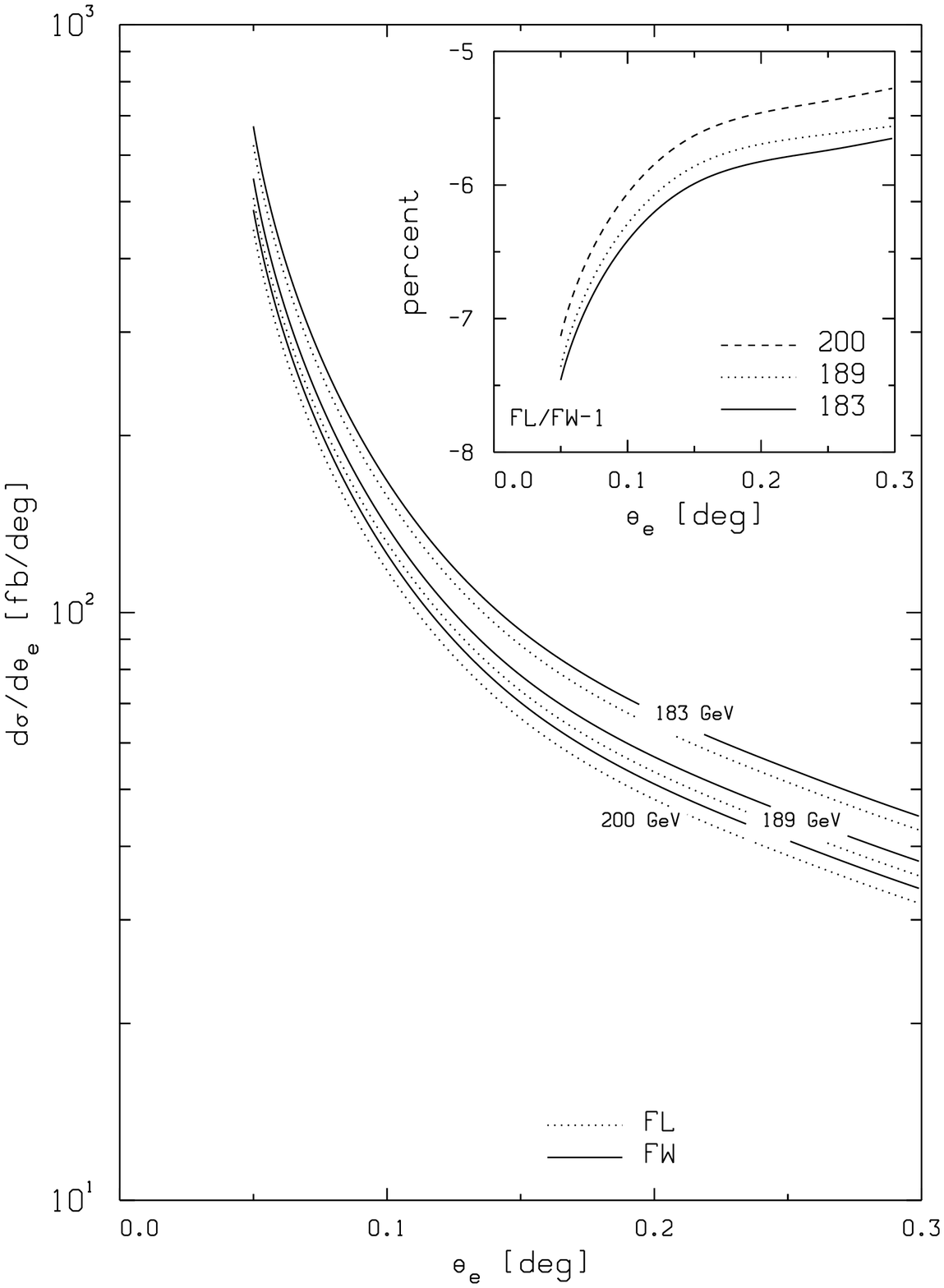}}
\vspace*{-2.7cm}
\end{center}}
\end{minipage}
\vspace*{1.5cm}
\caption[]{$d\sigma/d\cos\theta_e$ [fb/degrees] for $e^+e^- \to u \bard e^-
\barnu_e$ with $M(u\bard) > 45\,$GeV and $\sqrt{s} = 183, 189, 200\,$GeV.}
\label{fig:rsall}
\end{figure}

We have repeated the calculation for all the three energies, finding similar 
results that are shown in \fig{fig:rsall}.
For completeness we have reported the numerical results for the three energies
in \tabn{tab1}, where the first entry is Fixed-Width distribution and the 
second entry is Fermion-Loop one. Only the first four bins are shown, owing
to the fact that they are the most significant in the distribution. The third
entry in \tabn{tab1} gives FL/FW-1 in percent.
 \begin{table}[htbp]\centering
 \begin{tabular}{|c||c|c|c|}
 \hline
 $\theta_e\,$[Deg] &  $\sqrt{s} = 183\,\GeV$  & $\sqrt{s} = 189\,\GeV$  & 
$\sqrt{s} = 200\,\GeV$ \\
 \hline
                                &           &           &               \\
   $0.0^\circ \div 0.1^\circ$   &  0.48395  & 0.54721   & 0.67147       \\
                                &  0.44784  & 0.50695   & 0.62357       \\
                                &  -7.46    & -7.36     & -7.13         \\
                                &           &           &               \\
 \hline
                                &           &           &               \\
   $0.1^\circ \div 0.2^\circ$   &  0.07026  & 0.07815   & 0.09323       \\
                                &  0.06605  & 0.07357   & 0.08798       \\
                                &  -5.99    & -5.86     & -5.63         \\
                                &           &           &               \\
 \hline
                                &           &           &               \\
   $0.2^\circ \div 0.3^\circ$   &  0.04095  & 0.04554   & 0.05433       \\
                                &  0.03860  & 0.04298   & 0.05141       \\
                                &  -5.74    & -5.62     & -5.37         \\
                                &           &           &               \\
 \hline
                                &           &           &               \\
   $0.3^\circ \div 0.4^\circ$   &  0.02897  & 0.03223   & 0.03845       \\
                                &  0.02736  & 0.03045   & 0.03646       \\
                                &  -5.56    & -5.52     & -5.18         \\
                                &           &           &               \\
 \hline
 \end{tabular}
\vspace*{3mm}
 \caption[
 ]{
$d\sigma/d\theta_e$ in [pb/degrees] for the process $e^+e^- \to e^- \barnu_e 
u \bard$, for $M(u\bard) > 45\,$GeV. First entry is Fixed-Width distribution,
second entry is Fermion-Loop one and third entry if FL/FW-1 in percent.}
 \label{tab1}
 \end{table}
 \normalsize
After discussing the angular distribution we consider the total -- 
single-$\wb$ -- cross-section, defined by an acceptance cut of 
$|\cos\theta_e| > 0.997$, again for the high-mass case, i.e.
$M(u\bard) > 45\,$GeV. It is reported in \tabn{tab2}, where one sees that
differences between schemes are of the order of $5\%$ for all energies, 
supporting the use of some effective $\alpha(<q^2>)$ with $<q^2>= 
(1-cos<\theta_e>)\,s/2$ where $<\theta_e> \approx 0.1^\circ \div 0.2^\circ$.
 \begin{table}[htbp]\centering
 \begin{tabular}{|c||c|c|c|}
 \hline
 $\sqrt{s}$ &  FW  & FL & FL/FW-1 (percent)  \\
 \hline
            &      &    &                     \\
 $183\,\GeV$ & 88.17(44)  & 83.28(6)  & -5.5(5)  \\
            &      &    &                     \\
 $189\,\GeV$ & 98.45(25)  & 93.79(7)  & -4.7(3)  \\
            &      &    &                     \\
 $200\,\GeV$ & 119.77(67) & 113.67(8) & -5.1(5)  \\
            &      &    &                     \\
 \hline
 \end{tabular}
\vspace*{3mm}
 \caption[
 ]{
Total single-$\wb$ cross-section in fb for the process $e^+e^- \to e^- \barnu_e 
u \bard$, for $M(u\bard) > 45\,$GeV and $|\cos\theta_e| > 0.997$.
The number in parenthesis shows the statistical 
error of the numerical integration on the last digit.}
 \label{tab2}
 \end{table}
 \normalsize

We have also computed the single-$\wb$ cross-sections for a fixed mass of
the top quark, $\mt = 173.8\,$GeV, without finding any significative difference
with the previous case where $\mt$ is fixed by a consistency relation: they 
give $83.29(6)\,$fb, $93.80(7)\,$fb and $113.68(8)\,$fb for $\sqrt{s} = 183\,
\GeV, 189\,\GeV$ and $200\,$GeV.

Next we consider $e^+e^- \to e\nu\mu\nu$, with $|\cos\theta_e| > 0.997$, 
$E_{\mu} > 15\,$GeV, and $|\cos\theta_{\mu}| < 0.95$.
The angular cut on the muon direction is chosen to prevent the logarithmic
enhancement that would, otherwise, show up in the multi-peripheral terms.
Indeed, when the electron is lost in the beam pipe, and the
quasi-massless fermion is emitted parallel to the (quasi-real) photon
then the internal fermion propagator will produce an enhancement in the cross 
section. 
In \tabn{tab3} we report the comparison between the FL distribution and the
FW one for $\sqrt{s} = 183\,$GeV. As before, only the most significant bins 
are shown ($0.0^\circ \div 0.4^\circ$).
 \begin{table}[htbp]\centering
 \begin{tabular}{|c||c|c|c|}
 \hline
 $\theta_e\,$[Deg] &  FW  & FL  & FL/FW-1 (percent)  \\
 \hline
                                &           &           &               \\
   $0.0^\circ \div 0.1^\circ$   &  0.14154  & 0.13448   &  -4.99      \\
                                &           &           &               \\
 \hline
                                &           &           &               \\
   $0.1^\circ \div 0.2^\circ$   &  0.02113  & 0.02031   &  -3.88      \\
                                &           &           &               \\
 \hline
                                &           &           &               \\
   $0.2^\circ \div 0.3^\circ$   &  0.01238  & 0.01194   &  -3.55      \\
                                &           &           &               \\
 \hline
                                &           &           &               \\
   $0.3^\circ \div 0.4^\circ$   &  0.00880  & 0.00851   &  -3.30      \\
                                &           &           &               \\
 \hline
 \end{tabular}
\vspace*{3mm}
 \caption[
 ]{
$d\sigma/d\theta_e$ in [pb/degrees] for the process $e^+e^- \to e^- \barnu_e 
\nu_{\mu} \mu^+$, for $|\cos\theta_e| > 0.997$, 
$E_{\mu} > 15\,$GeV, and $|\cos\theta_{\mu}| < 0.95$. Furthermore, 
$\sqrt{s} = 183\,$GeV.}
 \label{tab3}
 \end{table}
 \normalsize

As for the hadronic case, the FL prediction is considerably lower than the
FW one, although the percentage difference between the two is 
approximately $2.2\% \div 2.4\%$ smaller than in the previous case.

This fact is reflected in the total single-$\wb$ cross-section for 
$e^+e^- \to e^- \barnu_e \mu^+ \nu_{\mu}$ that we report in \tabn{tab4}.
 \begin{table}[htbp]\centering
 \begin{tabular}{|c||c|c|c|}
 \hline
 $\sqrt{s}$ &  FW  & FL & FL/FW-1 (percent)  \\
 \hline
            &      &    &                     \\
 $183\,\GeV$ & 26.77(14) & 25.53(4)  & -4.6(5)  \\
            &      &    &                     \\
 $189\,\GeV$ & 29.73(14) & 28.78(4)  & -3.2(5)  \\
            &      &    &                     \\
 $200\,\GeV$ & 36.45(23) & 34.97(6)  & -4.1(6)  \\
            &      &    &                     \\
 \hline
 \end{tabular}
\vspace*{3mm}
 \caption[
 ]{
Total single-$\wb$ cross-section in fb for the process $e^+e^- \to e^- \barnu_e 
\mu^+ \nu_{\mu}$, for $|\cos\theta_e| > 0.997$, $E_{\mu} > 15\,$GeV, and 
$|\cos\theta_{\mu}| < 0.95$.
The number in parenthesis shows the statistical 
error of the numerical integration on the last digit.}
 \label{tab4}
 \end{table}
 \normalsize

When we want to improve upon the Fixed-Width scheme, the main accent has to
be put on the correct evaluation of the scale in the running of 
$\alpha_{\rm QED}$. The latter is particularly important for the $t$-channel 
diagrams, dominated by a scale $q^2 \approx 0$ and not $q^2 \approx \mws$. 
However, a consistent implementation of radiative corrections does more than
evolving $\alpha_{\rm QED}$ to the correct scale, other couplings are also 
running, propagators are modified and vertices are included. 

We have performed a simple exercise, namely to take
the Fixed-Width scheme augmented by a rescaling $\bigl[\alpha(q^2)/\alpha_{\gf}
\bigr]^2$. Since the largest part of the effect is, in any case, confined in
the region of very small electron scattering angle we only compute
$d\sigma/d\theta_e$ for $0.0^\circ < \theta_e < 0.4^\circ$. Results of our
computation are reported in \tabn{tab5}.
 \begin{table}[htbp]\centering
 \begin{tabular}{|c||c|c|c|c|c|c|}
 \hline
 bin &  FL  & FWE & FL/FWE-1 (percent) &  FL  & FWE & FL/FWE-1 (percent)  \\
 \hline
     &      &    &                   &      &    &                    \\
$0.0^\circ \div 0.1^\circ$ & 0.44784 & 0.45166 & +0.85 & 
0.13448 & 0.13207 & -1.79  \\
     &      &    &                   &      &    &                    \\
\hline
     &      &    &                   &      &    &                    \\
$0.1^\circ \div 0.2^\circ$ & 0.06605 & 0.06615 & +0.15 &
0.02031 & 0.01989 & -2.07 \\
     &      &    &                   &      &    &                    \\
\hline
     &      &    &                   &      &    &                    \\
$0.2^\circ \div 0.3^\circ$ & 0.03860 & 0.03867 & +0.18 &
0.01194 & 0.01169 & -2.09 \\
     &      &    &                   &      &    &                    \\
\hline
     &      &    &                   &      &    &                    \\
$0.3^\circ \div 0.4^\circ$ & 0.02736 & 0.02742 & +0.22 &
0.00851 & 0.00832 & -2.23 \\
     &      &    &                   &      &    &                    \\
 \hline
 \end{tabular}
\vspace*{3mm}
 \caption[
 ]{
Comparison between FL-scheme and effective FW-scheme, i.e. FW augmented by 
the running of $\alpha_{\rm QED}$ in the $t$-channel. First set of results 
refers to $u \bard$ final states. Second set to the $\mu^+\nu_{\mu}$ final
state. Acceptance cuts are as in the main text and $\sqrt{s} = 183\,\GeV$.}
 \label{tab5}
 \end{table}
 \normalsize
From \tabn{tab5} we see that the effective FW-scheme describes considerably well
the hadronic final state with a cut of $M(u\bard) > 45\,$GeV, with a maximal
difference of $0.85\%$ in the first bin. However, the diminution induced by
$\alpha_{\rm QED}(q^2)$ is too large for the leptonic final state. The latter is
a clear sign that other effects are relevant and a naive rescaling does not
suffice in reproducing a realistic approximation in all situations, at least 
not within the $2\%$ level of requested theoretical accuracy.

In order to understand this fact we have computed $d\sigma/d\theta_e$, in
the first bin (largest effect), both for $u\bard$ and for $\mu^+\nu_{\mu}$ 
final states with varying acceptance cuts. The results of this investigation
are shown in \tabn{tab6} and \tabn{tab7}, respectively.
 \begin{table}[htbp]\centering
 \begin{tabular}{|c||c|c|c|}
 \hline
                             &    &    &                   \\
$M_{\rm min}(u\bard)\,$[GeV] & FL & FW & FL/FW-1 (percent) \\
                             &    &    &                   \\
 \hline
                             &    &    &                   \\
45 & 0.04478(3) & 0.04841(3) & -7.5(1) \\
                             &    &    &                   \\
35 & 0.04711(6) & 0.5104(7)  & -7.7(1) \\
                             &    &    &                   \\
25 & 0.504(1)   & 0.0546(1)  & -7.7(2) \\
                             &    &    &                   \\
15 & 0.0552(1)  & 0.0595(1)  & -7.2(2) \\
                             &    &    &                   \\
10 & 0.0582(1)  & 0.0626(1)  & -7.0(2) \\
                             &    &    &                   \\
5  & 0.0615(1)  & 0.0659(1)  & -6.7(2) \\
                             &    &    &                   \\
1  & 0.0637(1)  & 0.0682(1)  & -6.6(2) \\
                             &    &    &                   \\
 \hline
 \end{tabular}
\vspace*{3mm}
 \caption[
 ]{
Comparison between FL-scheme and FW-scheme for the cross-section in pb
and $0.0^\circ < \theta_e < 0.1^\circ$. Furthermore, $M(u\bard) \ge 
M_{\rm min}$ and $\sqrt{s} = 183\,\GeV$.
The number in parenthesis shows the statistical 
error of the numerical integration on the last digit.}
 \label{tab6}
 \end{table}
 \normalsize
From the numbers in \tabn{tab6} we see the growth in the cross-section,
for all schemes, due to the multi-peripheral peak.
The percentage difference between FL-scheme and FW-scheme decreases, although
slowly, for decreasing $M_{\rm min}(u\bard)$. There is more, the difference
between the two schemes goes from $-12.91\%$ for $35\,\GeV < M(u\bard) <
45\,\GeV$ to $-6.96\%$ for $25\,\GeV < M(u\bard) < 35\,\GeV$, to $-3.58\%$
for $15\,\GeV < M(u\bard) < 25\,\GeV$ and reaches $-1.20\%$ for
$5\,\GeV < M(u\bard) < 10\,\GeV$. With the above cuts we are removing
the contribution from the single-resonant $\wb$-exchange and the
multi-peripheral diagrams start dominating. 
 \begin{table}[htbp]\centering
 \begin{tabular}{|c||c|c|c|}
 \hline
$\cos\theta_{\mu,\rm min}$ & FL & FW & FL/FW-1 (percent) \\
 \hline
                             &    &    &                   \\
0.95   & 0.01345(1)   & 0.01415(1) & -4.9(1) \\
                             &    &    &                   \\
0.995  & 0.01485(9)   & 0.01538(4) & -3.4(6) \\
                             &    &    &                   \\
0.9995 & 0.01543(13)  & 0.01571(9) & -1.8(9) \\
                             &    &    &                   \\
 \hline
 \end{tabular}
\vspace*{3mm}
 \caption[
 ]{
Comparison between FL-scheme and FW-scheme for the cross-section in pb,
$0.0^\circ < \theta_e < 0.1^\circ, |\cos\theta_{\mu}| < 
\cos\theta_{\mu,\rm min}$. Furthermore $\sqrt{s} = 183\,GeV$.
The number in parenthesis shows the statistical 
error of the numerical integration on the last digit.}
 \label{tab7}
 \end{table}
 \normalsize
The same trend is clearly visible in \tabn{tab7}, despite the somewhat larger
errors. By increasing $\cos\theta_{\mu,\rm min}$ we allow the muon to become 
more and more collinear to the quasi-real photon, therefore the 
multi-peripheral contribution counts more and more.

\begin{figure}[t]
\begin{minipage}[t]{14cm}
{\begin{center}
\vspace*{-2.0cm}
\mbox{\epsfysize=14cm\epsfxsize=15cm\epsffile{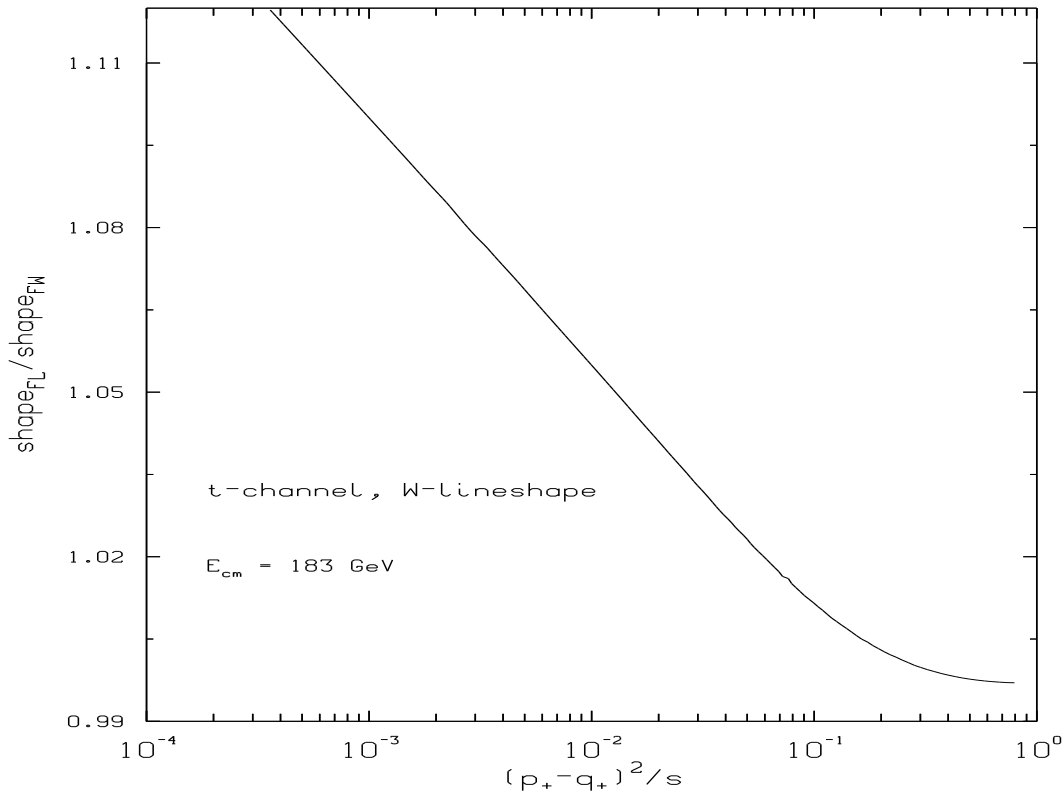}}
\vspace*{-2.7cm}
\end{center}}
\end{minipage}
\vspace*{1.5cm}
\caption[]{Comparison between the Fixed-Width and the Fermion-Loop 
line-shape functions in the $t$-channel, \eqn{tshape}.}
\label{fig:tshape}
\end{figure}

\begin{figure}[t]
\begin{minipage}[t]{14cm}
{\begin{center}
\vspace*{-2.0cm}
\mbox{\epsfysize=14cm\epsfxsize=15cm\epsffile{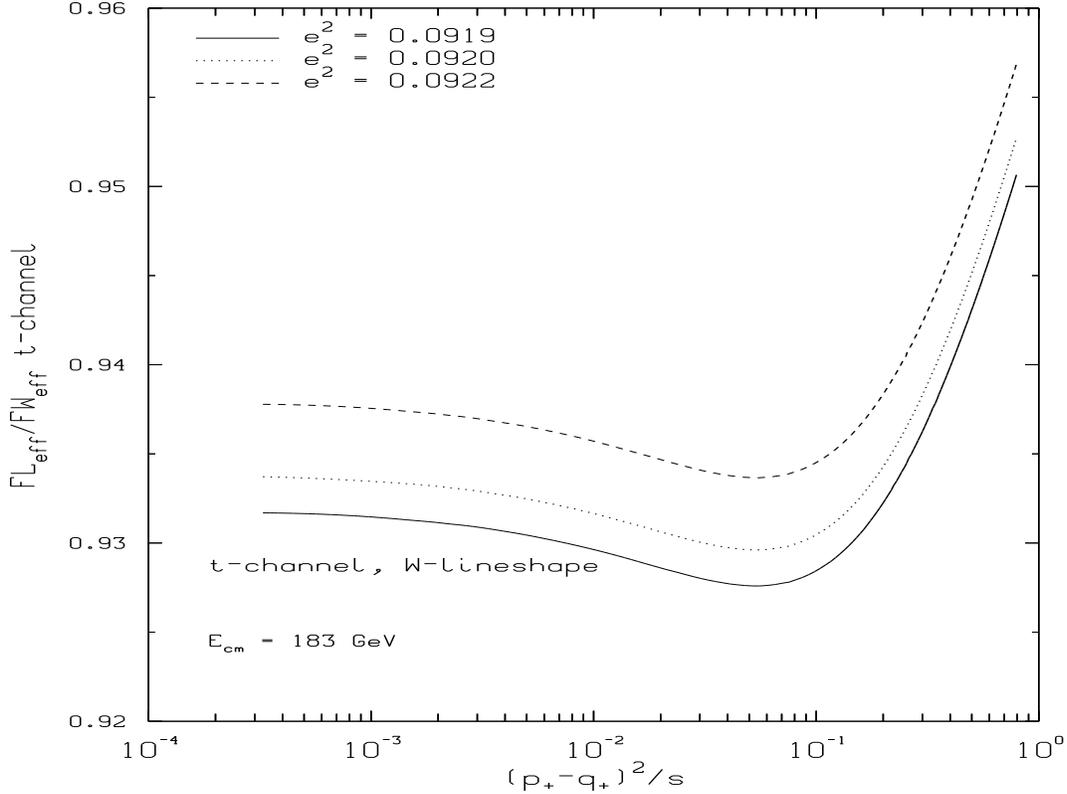}}
\vspace*{-2.7cm}
\end{center}}
\end{minipage}
\vspace*{1.5cm}
\caption[]{Comparison between the fixed-width and the fermion-loop 
line-shape functions in the $t$-channel, \eqn{tshapec}. Three values of
$e^2(T)$ are reported.}
\label{fig:tshapec}
\end{figure}

For the multi-peripheral diagrams alone, the inclusion of fermionic corrections
changes $\alpha \to \alpha((p-_ - q_-)^2)$ in the $t$-channel photon
exchange, $g^2 \to g^2((p_+ - q^2_+)^2)$ in the $t$-channel $\wb$
exchange and modifies the $t$-channel $\wb$-propagator. The latter is a large 
effect. To see it, let us define $t$-channel line-shape functions,
\bqa
L_{\rm FL}(t) &=& {s^2\over{
\lpar t-\Reb\sW\rpar^2 + \lpar \Imb\sW\rpar^2}}\,{1\over {|R(t)|^2}},  \nl
L_{\rm FW}(t) &=& {s^2\over{\lpar t - \mws\rpar^2 + \mws\gw^2}}.
\label{tshape}
\eqa
and compare them, as shown in \fig{fig:tshape}.
The FL $t$-channel line-shape can be considerably larger than the FW one, due
to the $\rho$-factor, $R(t)$, of \eqn{rhof}. In a circumstance where vertices 
and $s$-channel components count very little we can simulate the three effects
induced by the FL-scheme in a simple way. To this end, we compare the 
following objects:
\bqa
{\rm FL}_{\rm eff}(t,T) &=& \alpha^2(T)\,|g^2(t)|^2\,
{s^2\over{\lpar t-\Reb\sW\rpar^2 + \lpar \Imb\sW\rpar^2}}\,
{1\over {|R(t)|^2}},  \nl
{\rm FW}_{\rm eff}(t) &=& g^8\,\stwf\,
{s^2\over{\lpar t - \mws\rpar^2 + \mws\gw^2}},
\label{tshapec}
\eqa
where $T= - (p_- - q_-)^2$ and $t = - (p_+ - q_+)^2$. The comparison is shown,
for fixed $T$ and as a function of $t$, in \fig{fig:tshapec}.
The difference, especially for non-vanishing values of $t$ (remember that
$\mws/s = 0.19$ at $\sqrt{s} = 183\,$GeV), is considerably smaller than the one 
caused by the running of $\alpha_{\rm QED}$. This confirms our
previous findings and gives clear evidence that a naive rescaling of
the Fixed-Width results cannot cover all possible cases, with different
processes and with kinematical cuts of different kinds, at least not below the 
threshold of $2\%$ for the theoretical accuracy. 

Modifications induced by the fermionic loops are sensitive
to the relative weight of single-resonant terms and of multi-peripheral
peaks. Furthermore, the effect of radiative corrections inside the 
$\wb$-propagators ($\rho$-factors) is far from being negligible and tends to 
compensate the change due to the running of $\alpha_{\rm QED}$. 

\section{Conclusions.}

In this paper we have analyzed the numerical impact of introducing the
Fermion-Loop scheme for evaluating cross-sections and distributions in
single-$\wb$ production.

A description of single-$\wb$ processes by means of the Fermion-Loop scheme is
mandatory from, at least, two points of view. FL is the only 
field-theoretically consistent scheme that preserves gauge invariance in
processes including unstable vector-bosons coupled to e.m. currents.
Furthermore, single-$\wb$ production is a process that depends on several
scales, the single-resonant $s$-channel exchange of $\wb$-bosons, the
exchange of $\wb$-bosons in $t$-channel, the small scattering angle peak
of outgoing electrons. 

A correct treatment of the multi-scale problem can only be achieved when we
include radiative corrections in the calculation, not only one-loop terms
but also the re-summation of leading higher-order terms. Recent months have
shown that this project can be brought to a very satisfactory level by
identifying the correct approximation, process-by-process.

In particular, the $\wb-\wb$ configuration, dominated by double-resonant
terms, can be treated within the so-called double-pole-approximation or
DPA~\cite{kn:dpa}. As a consequence, the theoretical uncertainty associated 
with the determination of the $\wb\wb$ cross-section is sizably decreased. In
principle, the same procedure applies to the determination of the $\zb\zb$
cross-section, where one develops a NC02-DPA approximation instead of 
the CC03-DPA one.

 \begin{table}[htbp]\centering
 \begin{tabular}{|c||c|c|c|}
 \hline
 & & & \\
Process & $\sqrt{s} = 183\,\GeV$ & $\sqrt{s} = 189\,\GeV$ & 
$\sqrt{s} = 200\,\GeV$ \\
 & & & \\
 \hline
 & & & \\
$e^+e^- \to e^- \barnu_e u \bard$ &
83.237(36) & 93.797(75) & 113.71(5)  \\
 & & & \\
 \hline
 & & & \\
$e^+e^- \to e^- \barnu_e \nu_{\mu} \mu^+$ &
25.476(40) & 28.741(44) & 34.900(50) \\
 & & & \\
 \hline
 & & & \\
$e^+e^- \to e^- \barnu_e \nu_{\tau} \tau^+$ &
25.442(38) & 28.702(41) & 34.856(48) \\
 & & & \\
 \hline
 \end{tabular}
\vspace*{3mm}
 \caption[
 ]{
Single-$\wb$ cross-sections for $\sqrt{s} = 183(189,200)\,\GeV$ and
$|\cos\theta_e| > 0.997$ in the Fermion-Loop scheme. Additional cuts are:
$M(u\bard) > 45\,$GeV; $E_{\mu(\tau)} > 15\,$GeV;  
$|\cos\theta_{\mu(\tau)}| < 0.95$. The number in parenthesis shows the 
statistical error of the numerical integration on the last two digits.}
 \label{tab8}
 \end{table}
 \normalsize

In this paper we have considered the single-$\wb$ production, i.e.
$e^+e^- \to \wb e \nu$. Fermionic loops have been included, following
the generalization of the Fermion-Loop scheme worked out in~\cite{kn:swext}.
Within this context theoretical predictions for single-$\wb$ production become 
smaller, as compared to other pragmatic, gauge-preserving, schemes. We have 
found that all the modifications introduced via the Fermion-Loop scheme are 
relevant: running of the couplings, $\rho$-factors and vertices, not only the
change $\alpha_{\rm QED}({\rm fixed}) \to \alpha_{\rm QED}({\rm running})$.
Therefore, a naive rescaling cannot reproduce the Fermion-Loop answers for
all situations, all kinematical cuts. 

The high-energy Input Parameter Set used in all calculations that are
presently available -- we quote, among the various schemes, the Fixed-Width 
scheme, the Overall scheme and the Imaginary Fermion-Loop one -- is
based on $\gf, \mw$ and $\mz$ with $\alpha_{\rm QED}({\rm fixed}) =
1/131.95798$. It allows for the inclusion of part of higher order effects in 
the Born cross-sections but, it fails to give a correct and accurate 
description of the $q^2 \sim 0$ dominated processes. 

A naive, overall, rescaling would lower the single-$\wb$ cross-section of 
about $7\%$. We have found, with the complete Fermion-Loop, that this decrease 
is process and cut dependent.
Moreover, the effect is larger in the first bin for $\theta_e$ -- $0.0^\circ 
\div 0.01^\circ$ -- in the distribution $d\sigma/d\theta_e$ and tends to 
become less pronounced for larger scattering angles of the electron. However, 
the first bin represents almost $50\%$ of the total single-$\wb$ cross-section,
so that, in general, the compensations that occur among several effects never 
bring the Fl/FW ratio to one. We obtain a maximal decrease of about $7\%$ in
the result but, on average, the effect is smaller. We have also found that
the effect is rather sensitive to the relative weight of multi-peripheral
contributions.

A final comment is needed to quantify the theoretical accuracy of single-$\wb$
production. Bosonic corrections are still missing and, very often, our 
experience has shown, especially at LEP1, that bosonic corrections may become 
sizeable~ \cite{kn:LEPreport_95}.
A large part of the bosonic corrections, as e.g.\ the leading-logarithmic 
corrections, factorize and can be treated by a convolution. Nevertheless
the remaining bosonic corrections can still be non-negligible, i.e.,
of the order of one percent at LEP2~\cite{kn:LEP2WWreport}. For the Born 
cross-sections $1\%$ should, therefore, be understood as the present limit
for the theoretical uncertainty. This will have to be improved, soon or later,
since bosonic corrections are even larger at higher energies \cite{kn:LCreport}
and the single-$\wb$ cross-section will cross over the $\wb\wb$ one at $500\,$
GeV.
Single-$\wb$ will be one of the major processes at LC to measure not only 
triple gauge couplings but also the $\wb$-mass without color reconnection.

To summarize our findings we present a collection of single-$\wb$ 
cross-sections obtained in the Fermion-Loop scheme with {\tt WTO} in a high
precision run. They are shown in \tabn{tab8} where we also report the
$\nu_{\tau}\tau$ final state. Numerical precision ranges from $0.05\% \div
0.08\%$ for $u\bard$ to $0.14\% \div 0.16\%$ for $\nu_{\mu}\mu(\nu_{\tau}\tau)$.

\section{Acknowledgements.}

Over the last year I had a lot of fruitful discussions with many
experimental colleagues and I profited a lot from their experience. In
particular I want to mention the work of Martin Gr\"unewald, Hywel Phillips,
Alexander Schmidt-Kaerst, Reisaburo Tanaka and Marco Verzocchi.
I also acknowledge many discussions with the participants in the LEP2/MC
workshop, in particular Alessandro Ballestrero, Misha Dubinin, Roberto Pittau, 
Yoshimasa Kurihara and Maciej Skrzypek.

\end{document}